\documentclass[aip,jcp,floatfix,showkeys,preprint]{revtex4-1}
\usepackage{graphicx}
\usepackage[tbtags]{amsmath}
\usepackage{amssymb}
\usepackage{longtable}
\makeatletter

\@addtoreset{equation}{section}
\makeatother
\begin{document}
\title{Polarization effects on intermolecular vibrational energy transfer analyzed by 2DIR spectroscopy}

\author{A.A. Villaeys}
\email{albert.villaeys@wanadoo.fr}
\affiliation{Universit\'e de Strasbourg et Institut de Physique et Chimie des Mat\'eriaux de Strasbourg, France}
\affiliation{Research Center for Applied Sciences, Academia Sinica, Taipei 115, Taiwan}
\author{M. Zouari}
\email{mabroukzy@yahoo.fr}
\affiliation{Universit\'e de Carthage, Facult\'e des Sciences de Bizerte, D\'epartement de Physique, 7021 Zarzouna, Tunisie}
\author{Kuo Kan Liang}
\email{kkliang@sinica.edu.tw}
\affiliation{Research Center for Applied Sciences, Academia Sinica, Taipei 115, Taiwan}
\affiliation{Department of Biochemical Science and Technology, National Taiwan University, Taipei 106, Taiwan}

\begin{abstract}
In the present work, we analyze the influence of the polarization effects taking place during the course of a 2DIR spectroscopy experiment performed on a molecular system undergoing an intermolecular vibrational energy transfer process. When both donor and acceptor molecules participating in the vibrational energy transfer are embedded in a host solvent, they face rotational diffusion that strongly distorts the resulting 2DIR spectra. 
It could be expected that the difference between rotational diffusion constants will be of particular interest.
For this purpose, the polarization effects are discussed according to the different orderings of the laser-molecule interactions. Next, we study the distortions of the spectra as a function of the rotational diffusion constants of the individual molecules. The knowledge of these polarization effects are relevant to the interpretation of the spectra. 
Finally, the conclusions reached in this work for a vibrational energy transfer are valid for any other type of third-order optical process performed on the same molecular system.
\end{abstract}

\keywords{coherent nonlinear optical spectroscopy, intermolecular vibrational energy transfer, polarization effect, rotational dynamics.}
%\PACS: 78.47.jh, 34.50.ez, 45.20.dc

\maketitle

\section{Introduction}

Measurements and subsequent interpretations of the spectra obtained from 
two-dimensional (2D) or higher multidimensional spectroscopy experiments, are difficult tasks.
Under certain circumstances, polarization effects can be an additional and quite effective tool to overcome these difficulties
\cite{hamm.zanni.concepts_methods_2dir.11,cho.two_dim_opt_spectr.09}.
In fact, polarization has long been used as a spectroscopic tool, but it is especially useful in 2D spectroscopy because the diagonal and cross peaks characterizing 2D spectra arise from a well-defined number and ordering of the pulsed laser fields interacting with the molecular transition dipoles. Therefore, taking advantage of particular combinations of the laser fields, specific processes can be emphasized in the 2D spectra.
It can be expected that polarization will play an even more important role in 3D spectroscopy
\cite{ding.zanni.cp341.07.95}.

Today, these polarization effects became very important in 2D infrared (2DIR) spectroscopy where sequences of polarized laser pulses are used to 
detect angles between molecular transition dipoles
\cite{golonska.tokmakoff.jcp115.01.297,zanni.hochstrasser.jpcb105.01.6520,krummel.zanni.jpcb107.03.9165}.
This emerges as an important tool to monitor the molecular structures. Indeed, these effects have played a major role in developing polarization-selective 2DIR spectroscopy which provides a unique way to analyze rotational dynamics. Among the large number of applications, we can mention the determination of the anisotropy dynamics of both intraconfigurational hydroxyl groups and those hydroxyl groups undergoing interconfigurational H-bond exchange in six-molar aqueous sodium perchlorate\cite{ji.gaffney.jcp134.11.044516}.
Of course, these investigations of H-bond exchange rely on previous developments in the theoretical description of rotational dynamics in H-bonded solutions, as well as theoretical models of 2DIR spectroscopy
\cite{golonska.tokmakoff.jcp115.01.297,khalil.tokmakoff.jpca107.03.5258,kwak.fayer.jpcb110.06.19998}.
Theoretical models for rotational dynamics originating from librational, diffusive and angular jump motions have been developed and applied to experimental studies of rotational dynamics in water and aqueous solutions
\cite{szabo.jcp81.84.150,fleming.chem_appl_ultra_spect.86, myers.hochstrasser.ieeejqeqe22.86.1482, purucker.laubereau.jrs24.93.453, wan.johnson.jcp99.93.7602, lim.anfinrud.nsmb4.97.209}.
Besides, polarization effects are useful for supressing background signals
\cite{woutersen.hamm.jpcb104.00.11316,xiong.zanni.ol33.08.1371}
and even for enhancing cross peaks at the expense of diagonal peaks
\cite{zanni.hochstrasser.pnasusa98.01.11265}.
This is of particular interest because large diagonal peaks quite often mask weaker cross peaks and
suppressing the diagonal peaks provides better resolution of the cross peaks.

From a more fundamental point of view, control over polarization generally leads to improved spectral resolution by suppressing or enhancing the contributions associated with specific pathways in the Liouvillian space according to the relative orientations of the laser field polarizations and dipole moments of the molecular states involved in a particular chronological pathway
\cite{woutersen.hamm.jpcb104.00.11316,zanni.hochstrasser.jpcb105.01.6520,golonska.tokmakoff.jcp115.01.297,golonska.tokmakoff.jcp115.01.10814,villaeys.liang.cp450.15.12}.
Therefore, taking advantage of adequate polarization schemes, physical insights can be extracted 
by enhancing or suppressing the contributions associated with some elementary processes that can possibly exhibit or cover the relevant information on which we want to focus
\cite{voronine.mukamel.jcp124.06.034104,strasfeld.zanni.njp11.09.105046}.

Orientation as well as rotational relaxation dynamics are among the most fundamental characteristics of molecules at surfaces and interfaces. They play an important role in many physical, chemical, and biological phenomena, including molecular mechanisms of energy relaxation, solvation, electron transfer, and many others.
What is more relevant to the present work is that when we perform a 2DIR experiment on molecules embedded in a host solvent, for example upon investigating the intermolecular vibrational energy transfer, these effects make the spectra more intricate because rotational diffusion can strongly perturb the optical response of the molecular system under investigation.
When rotational motions of the molecules are much slower than the longest time delays required to measure the spectra, rotational motions can be neglected. However, for small molecules, rotational motions can contributes significantly to homogeneous dephasing. Here, the polarization effects can drastically alter the resulting spectra. Consequently, any interpretation of the 2D spectra will require a proper account of the rotational motion undergone by the molecules participating in the energy transfer.
Therefore, an average has to be performed over the orientations of the dipoles of all molecules.
The resulting 2DIR spectra will be dependent on the magnitude of the rotational diffusion constants of all molecules involved.
This is exactly what we want to analyze in the present work.

The paper is organized as follows. Sec.~2 reviews briefly the general expression of the signal intensity valid for the rephasing and non-rephasing directions. Next, rotational diffusion is taken into account to calculate the orientational average of the above-mentioned signal.
In Sec.~3, the vibrational molecular model supporting intermolecular vibrational energy transfer is introduced to evaluate the internal dynamics as well as the orientational evolution of the molecules undergoing the energy transfer. Finally, in Sec.~4, numerical simulations are performed to give a quantitative analysis of the polarization effects on the 2DIR spectra of a system undergoing intermolecular vibrational energy transfer.

\section{Polarization effects on intermolecular vibrational energy transfer}

The purpose of this section is to describe an intermolecular vibrational energy transfer taking place between molecules undergoing rotational diffusion. Due to their respective shapes, the molecules involved in the energy transfer process generally experience different frictions from the host solvent. 
Since we are interested in the intermolecular energy transfer, the 2DIR experiment involves the dipole moments of two different molecules, which is a typical situation in 2DIR studies.
The total vibrational system is therefore composed of vibrational modes $Q_{\rm A}$ and $Q_{\rm B}$ pertaining to molecules A and B, respectively. The laser-system interaction Hamiltonian is given by
\begin{equation}
\boldsymbol{V}\left(t\right)=-\sum_{\text{p}=a,b,c} \mathcal{A}_{\rm p}\left(t-T_{\rm p}\right)\left[\vec{\boldsymbol{\mu}}\cdot\vec{\mathcal{E}}_{\rm p}e^{-i\omega_{\rm p}\left(t-T_{\rm p}\right)+i\vec{k}_{\rm p}\cdot\vec{r}}+ {\rm C.C.}\right],
\label{2.1}
\end{equation}
where the notation C.C.\ stands for the complex conjugate. The symbol $\mathcal{A}_{\rm p}\left(t-T_{\rm p}\right)$ stands for the normalized envelops of the three laser pulses defined as $\mathcal{A}_{\rm p}\left(t-T_{\rm p}\right)=\sqrt{\gamma_{\rm p}}\exp\left(-\gamma_{\rm p}\left\vert t-T_{\rm p}\right\vert\right)$ with $\text{p}=a, b\text{ or }c$. Any theoretical description of the 2DIR spectroscopy experiment requires the third-order perturbation term of the density matrix with respect to the laser-molecule interaction $\boldsymbol{V}\left(t\right)$
\cite{cho.mukamel.jcp96.92.5618,grimberg.mukamel.jpca106.02.697,lozovoy.dantus.cp267.01.99,khalil.tokmakoff.cp266.01.213,khalil.tokmakoff.prl90.03.047401,
cho.two_dim_opt_spectr.09,hamm.zanni.concepts_methods_2dir.11}.
The contribution to the third-order term of the density matrix, $\boldsymbol{\rho}^{\left(3\right)}\left(t\right)$, relevant to our purpose, takes the form
\begin{multline}
\boldsymbol{\rho}^{\left(3\right)}\left(t\right)=\frac{i}{\hbar^3}\int_{t_0}^{t}\!d\tau_3\int_{t_0}^{\tau_3}\!d\tau_2\int_{t_0}^{\tau_2}\!d\tau_1
\\ \times
\boldsymbol{G}\left(t-\tau_3\right)\boldsymbol{L}_{\rm v}\left(\tau_3\right)\boldsymbol{G}\left(\tau_3-\tau_2\right)
\boldsymbol{L}_{\rm v}\left(\tau_2\right)\boldsymbol{G}\left(\tau_2-\tau_1\right) \boldsymbol{L}_{\rm v}\left(\tau_1\right)\boldsymbol{\rho}\left(t_0\right)\;,
\label{2.2}
\end{multline}
where the interaction Liouvillian is defined by $\boldsymbol{L}_{\rm v}\!\left(\tau_{\alpha}\right)=\left[\boldsymbol{V}\!\left(\tau_{\alpha}\right),\cdots\right]$.
Also involved are the propagators
$\boldsymbol{G}\!\left(\tau_{\alpha}-\tau_{\beta}\right)=
\exp\left\{-\frac{i}{\hbar}\boldsymbol{L}\!\left(\tau_{\alpha}-\tau_{\beta}\right)\right\}$.
They depend on another Liouvillian $\boldsymbol{L}=\left[\boldsymbol{H},\cdots\right]$ with $\boldsymbol{H}$ the free vibrational system Hamiltonian.
The matrix elements of the propagator with specific form of the indices
$\boldsymbol{G}_{iijj}\!\left(\tau_{\alpha}-\tau_{\beta}\right)$ account for free population evolution if $i=j$ and population
transfer if $j\ne i$, respectively. Another set of matrix elements, namely $\boldsymbol{G}_{ijij}\!\left(\tau_{\alpha}-\tau_{\beta}\right)$ with $i\ne j$ account for the evolution of the coherences. The emitted radiation signal along the rephasing phase-matched direction $\vec{k}_{\text{re}}=-\vec{k}_a+\vec{k}_b+\vec{k}_c$ and the non-rephasing direction $\vec{k}_{\text{nre}}=\vec{k}_a-\vec{k}_b+\vec{k}_c$
are deduced from the third-order term of the polarization
\begin{equation}
\vec{P}^{\left(3\right)}_{\vec{k}_s}\left(T_a,T_b,T_c,t\right)=2\Re\left[\sum_{i}\sum_{j<i}\boldsymbol{\rho}^{\left(3\right)}_{\vec{k}_{\rm s},ij}\left(t\right)\vec{\boldsymbol{\mu}}_{ji}\right]
\label{2.3}
\end{equation}
where $\vec{k}_s$ stands for $\vec{k}_{\text{re}}$ or $\vec{k}_{\text{nre}}$.

Using a local field $\vec{E}_{\text{lo}}\left(t\right)=\mathcal{A}_{\text{lo}}\left(t-T_{\text{lo}}\right)\left[\vec{\mathcal{E}}_{\text{lo}}e^{-i\omega_{\text{lo}}\left(t-T_{\text{lo}}\right)+i\vec{k}_{\text{lo}}\cdot\vec{r}-i\Psi}+{\rm C.C.}\right]$ with $\Psi$ an additional phase, the heterodyne detection is introduced
\cite{lepetit.joffre.ol21.96.564,likforman.thierrymieg.ol22.97.1104,belabas.joffre.ol27.02.2043,cho.two_dim_opt_spectr.09,hamm.zanni.concepts_methods_2dir.11}.
As usual, the signal field is deduced from the polarization through the expression $\vec{E}_{\vec{k}_s}\left(T_a,T_b,T_c,t\right)\propto i \vec{P}^{\left(3\right)}_{\vec{k}_s}\left(T_a,T_b,T_c,t\right)$. With typical experimental conditions of weak signal field intensity and after subtracting the intensity of the local field, the temporal Fourier transform of the total intensity is given by
\begin{equation}
I_{{\vec{k}_s}}\left(T_a,T_b,T_c,\omega_t\right)=
2\Re\left[\left(\int_{-\infty}^{+\infty}dt \vec{E}_{\text{lo}}
\left(t-T_{\text{lo}}\right)e^{i\omega_t t}\right)^*
\left(\int_{-\infty}^{+\infty} dt \vec{E}_{\vec{k}_s}\left(T_a,T_b,T_c,t\right)e^{i\omega_t t}\right) \right]\;.
\label{2.4}
\end{equation}
Then, with the time origin chosen at the center of the last pulse and by assuming clean fast leading edges for the pulses
\cite{lepetit.joffre.josab12.95.2467}
the time integration can be reduced to $t\in\left[0,+\infty\right)$. Finally, the 2D spectra is obtained by performing a second Fourier transform over the delay time $\tau=T_b-T_a$ so that
\begin{equation}
I_{{\vec{k}_{\rm s}}}\left(\omega_d,\omega_t,T\right)=\int_{-\infty}^{+\infty}d\tau e^{-i\omega_d\tau}I_{{\vec{k}_{\rm s}}}\left(\tau,T,\omega_t\right)
\label{2.5}
\end{equation}
The experimental waiting time is defined as $T=\min\left(\left\vert T_a\right\vert,\left\vert T_b\right\vert\right)$ according to the prescription by Jonas
\cite{gallagherfaeder.jonas.jpca103.99.10489}.
The time origin is chosen at the center of the last pulse.

The donor and acceptor of the vibrational energy transfer process are both dissolved in liquid solvent. Therefore, a proper account of the polarization effects requires the introduction of orientational dynamics of the molecular motion. For simplicity,
we assume that both the donor and acceptor can be approximated by spherically symmetric molecules having a homogeneous moment of inertia. The collisions between the donor, the acceptor, and the solvent molecules induce angular random walks of the dipole moments which can be modeled by a rotational diffusion equation of motion of the
angular probability distribution function
$\boldsymbol{\mathcal G}\left(\Omega,t\right)$
\cite{berne.pecora.1990}: % Sec. 7.3
\begin{equation}
\frac{\partial \boldsymbol{\mathcal G}\left(\Omega,t\right)}{\partial t}=-D\hat{\boldsymbol{J}}^2 \boldsymbol{\mathcal G}\left(\Omega,t\right)
\label{2.6}
\end{equation}
whose general solution for a molecule rotating from one direction $\Omega_{\alpha}=\left(\theta_{\alpha},\phi_{\alpha}\right)$ to another one $\Omega_{\beta}=\left(\theta_{\beta},\phi_{\beta}\right)$
during the time interval $\delta t$ is given by the propagator
\begin{equation}
\boldsymbol{\mathcal G}\left(\Omega_{\beta},\delta t\left\vert\Omega_{\alpha}\right.\right)=e^{-D\hat{\boldsymbol{J}}^2 \delta t}\boldsymbol{\mathcal G}\left(\Omega_{\alpha},0\left\vert\Omega_{\alpha}\right.\right)
\label{2.7}
\end{equation}
if we neglect the rotation during the laser pulse durations. The dipole propagators can be expanded in terms of spherical harmonics $Y_{\ell,m}\left(\theta,\phi\right)$, abbreviated as $Y_{\ell,m}\left(\Omega\right)$,
satisfying the well-known eigenvalue equation
\begin{equation}
\boldsymbol{J}^2 Y_{\ell,m}\left(\Omega\right)=\ell\left(\ell+1\right)Y_{\ell,m}\left(\Omega\right)
\label{2.8}
\end{equation}
as well as the orthonormalization and closure relations
\begin{equation}\left\{\begin{aligned}
&\int d\Omega Y_{\ell',m'}\left(\Omega\right)Y^{\star}_{\ell,m}\left(\Omega\right)=\delta_{\ell,\ell'}\delta_{m,m'} \\
&\delta\left(\Omega-\Omega_0\right)=\sum_{\ell=0}^{\infty}\sum_{m=-\ell}^{\ell}Y_{\ell,m}\left(\Omega_0\right)Y^{\star}_{\ell,m}\left(\Omega\right)\equiv\sum_{\ell,m}Y_{\ell,m}\left(\Omega_0\right)Y^{\star}_{\ell,m}\left(\Omega\right)
\end{aligned}\right.
\label{2.9}
\end{equation}
If the initial condition $\boldsymbol{\mathcal G}\left(\Omega,0\left\vert\Omega_0\right.\right)=\delta\left(\Omega-\Omega_0\right)$ is satisfied, Eq.\eqref{2.7} becomes
\begin{equation}
\boldsymbol{\mathcal G}\left(\Omega,t\vert\Omega_0\right)=e^{-D\boldsymbol{J}^2t}\sum_{\ell,m}Y_{\ell,m}\left(\Omega_0\right)Y^{\star}_{\ell,m}\left(\Omega\right)=\sum_{\ell,m}e^{-\ell\left(\ell+1\right)Dt}Y_{\ell,m}\left(\Omega_0\right)Y^{\star}_{\ell,m}\left(\Omega\right)
\label{2.10}
\end{equation}
This is the dipole propagator required in the following. Then, the complete polarization dependence of the 
overall evolution of molecules A and B participating in the energy transfer is 
included in the term $\hat{E}_{\text{lo}}\cdot\vec{\boldsymbol{\mu}}\,\boldsymbol{\rho}^{\left(3\right)}\left(t\right)$ where $\hat{E}_{\text{lo}}$ stands for the unitary vector along the oscillating local field. We shall expediently call this heterodyne-detected signal using the local field the \emph{heterodyne signal}. The average over molecular orientations of the heterodyne signal, denoted as $\left\langle\Pi\left(t\right)\right\rangle_{\text{or}}=\left\langle\hat{E}_{\text{lo}}\cdot\vec{\boldsymbol{\mu}}\boldsymbol{\rho}^{\left(3\right)}\left(t\right)\right\rangle_{\text{or}}$, can be
expressed as
\begin{multline}
\left\langle\Pi\left(t\right)\right\rangle_{\text{or}} =\frac{i}{\hbar^3}\left[\prod_{s=\text{A,B}}\int d\Omega^{\left(s\right)}_3\int d\Omega^{\left(s\right)}_2\int d\Omega^{\left(s\right)}_1\int d\Omega^{\left(s\right)}_0\right]
\int_{t_0}^{t}\!d\tau_3\int_{t_0}^{\tau_3}\!d\tau_2\int_{t_0}^{\tau_2}\!d\tau_1 \vec{E}_{\text{lo}}\cdot\vec{\boldsymbol{\mu}}                       \\
\begin{aligned}[b]
&\times\boldsymbol{\mathcal G}\left(\Omega^{\left({\rm A}\right)}_3\Omega^{\left({\rm B}\right)}_3,t-\tau_3\left\vert\Omega^{\left({\rm A}\right)}_2\Omega^{\left({\rm B}\right)}_2\right.\right)\boldsymbol{G}\left(t-\tau_3\right)\boldsymbol{L}_{\rm v}\left(\tau_3\right) \\
&\times\boldsymbol{\mathcal G}\left(\Omega^{\left({\rm A}\right)}_2\Omega^{\left({\rm B}\right)}_2,\tau_3-\tau_2\left\vert\Omega^{\left({\rm A}\right)}_1\Omega^{\left({\rm B}\right)}_1\right.\right)\boldsymbol{G}\left(\tau_3-\tau_2\right)\boldsymbol{L}_{\rm v}\left(\tau_2\right)\\
&\times\boldsymbol{\mathcal G}\left(\Omega^{\left({\rm A}\right)}_1\Omega^{\left({\rm B}\right)}_1,\tau_2-\tau_1\left\vert\Omega^{\left({\rm A}\right)}_0\Omega^{\left({\rm B}\right)}_0\right.\right) \boldsymbol{G}\left(\tau_2-\tau_1\right) \boldsymbol{L}_{\rm v}\left(\tau_1\right)\\
&\times{\mathcal P}^{\left({\rm A}\right)}_0\left(\Omega^{\left({\rm A}\right)}_0\right){\mathcal P}^{\left({\rm B}\right)}_0\left(\Omega^{\left({\rm B}\right)}_0\right)\boldsymbol{\rho}\left(t_0\right)
\end{aligned}
\label{2.11}
\end{multline}
Assuming random initial dipole orientations we have
${\mathcal P}^{\left(\rm A\right)}_0={\mathcal P}^{\left(\rm B\right)}_0=1/4\pi$.
If there is no correlation between the orientations of the molecules, the propagator can be factorized so that
$\boldsymbol{\mathcal G}\left(\Omega^{\left({\rm A}\right)}_n\Omega^{\left({\rm B}\right)}_n,\tau_n-\tau_m\left\vert\Omega^{\left({\rm A}\right)}_m\Omega^{\left({\rm B}\right)}_m\right.\right)
=\boldsymbol{\mathcal G}\left(\Omega^{\left({\rm A}\right)}_n,\tau_n-\tau_m\left\vert\Omega^{\left({\rm A}\right)}_m\right.\right)\boldsymbol{\mathcal G}\left(\Omega^{\left({\rm B}\right)}_n,\tau_n-\tau_m\left\vert\Omega^{\left({\rm B}\right)}_m\right.\right)$. Also, owing to the perturbative approach, only one molecule participates in the laser-molecule interaction at a given time and we have if $j\ne k$:
\begin{equation}
\boldsymbol{\mathcal G}\left(\Omega^{\left(j\right)}_q,\tau_r-\tau_q\left\vert\Omega^{\left(j\right)}_p\right.\right)L^{\left(k\right)}_{\rm v}\left(\tau_q\right)\boldsymbol{\mathcal G}\left(\Omega^{\left(j\right)}_p,\tau_q-\tau_p\left\vert\Omega^{\left(j\right)}_m\right.\right)
=\boldsymbol{\mathcal G}\left(\Omega^{\left(j\right)}_q,\tau_r-\tau_p\left\vert\Omega^{\left(j\right)}_m\right.\right)L^{\left(k\right)}_{\rm v}\left(\tau_q\right)
\label{2.12}
\end{equation}
where $L^{\left(k\right)}_{\rm v}\left(\tau_q\right)$ stands for $L_{\rm v}\left(\tau_q\right)$ if molecule $k$ interacts with the laser pulse at time $\tau_q$.

In the three-photon process considered in the theory of 2DIR spectroscopy, three photons interact with either of the two molecules involved respectively in different chronological order.
It is important to recall that the procedure of averaging over molecular dipole orientations depends on this chronological ordering of the photon-molecule interactions.
Accordingly, we divide all of the processes into four groups so that the
$\left\langle\Pi\left(t\right)\right\rangle_{\text{or}}=\sum_{M=\rm I\rightarrow IV}\left\langle\Pi^{M}\left(t\right)\right\rangle_{\text{or}}$.
The integral in Eq.~\eqref{2.11} can be simplified in these four cases in respective ways.

In the first case, the three photons interact with the same molecule.
The corresponding heterodyne signal takes the form
\begin{multline}
\left\langle\Pi^{\rm I}\left(t\right)\right\rangle_{\text{or}}=\frac{i}{\hbar^3}\sum_{j={\rm A,B}}\int d\Omega^{\left(j\right)}_3\int d\Omega^{\left(j\right)}_2\int d\Omega^{\left(j\right)}_1\int d\Omega^{\left(j\right)}_0 \int_{t_0}^{t}\!d\tau_3\int_{t_0}^{\tau_3}\!d\tau_2\int_{t_0}^{\tau_2}\!d\tau_1 \vec{E}_{\text{lo}}\cdot\vec{\boldsymbol{\mu}}\\
\begin{aligned}[b]
&\times
\boldsymbol{\mathcal G}\left(\Omega^{\left(j\right)}_3,t-\tau_3\left\vert\Omega^{\left(j\right)}_2\right.\right)
\boldsymbol{G}\left(t-\tau_3\right)\boldsymbol{L}_{\rm v}\left(\tau_3\right)
\boldsymbol{\mathcal G}\left(\Omega^{\left(j\right)}_2,\tau_3-\tau_2\left\vert\Omega^{\left(j\right)}_1\right.\right)
\boldsymbol{G}\left(\tau_3-\tau_2\right)\boldsymbol{L}_{\rm v}\left(\tau_2\right) \\
&\times
\boldsymbol{\mathcal G}\left(\Omega^{\left(j\right)}_1,\tau_2-\tau_1\left\vert\Omega^{\left(j\right)}_0\right.\right)
\boldsymbol{G}\left(\tau_2-\tau_1\right) \boldsymbol{L}_{\rm v}\left(\tau_1\right){\mathcal P}^{\left(j\right)}_0\left(\Omega^{\left(j\right)}_0\right)\boldsymbol{\rho}\left(t_0\right)
\end{aligned}
\label{2.13}
\end{multline}
The second case corresponds to two photons interacting with the same molecule first, followed by the third photon interacting with the other molecule. Then, we have
\begin{multline}
\left\langle\Pi^{\rm II}\left(t\right)\right\rangle_{\text{or}}=\frac{i}{\hbar^3}\sum_{j,k={\rm A,B}}^{k\not=j}\int d\Omega^{\left(j\right)}_2\int d\Omega^{\left(j\right)}_1\int d\Omega^{\left(j\right)}_0\int d\Omega^{\left(k\right)}_1\int d\Omega^{\left(k\right)}_0 \int_{t_0}^{t}\!d\tau_3\int_{t_0}^{\tau_3}\!d\tau_2\int_{t_0}^{\tau_2}\!d\tau_1 \\
\begin{aligned}[b]
&\times\vec{E}_{\text{lo}}\cdot\vec{\boldsymbol{\mu}}\,\boldsymbol{\mathcal G}\left(\Omega^{\left(k\right)}_1,t-\tau_3\left\vert\Omega^{\left(k\right)}_0\right.\right)\boldsymbol{G}\left(t-\tau_3\right)\boldsymbol{L}_{\rm v}\left(\tau_3\right) \\
&\times\boldsymbol{\mathcal G}\left(\Omega^{\left(j\right)}_2,t-\tau_2\left\vert\Omega^{\left(j\right)}_1\right.\right)\boldsymbol{G}\left(\tau_3-\tau_2\right)\boldsymbol{L}_{\rm v}\left(\tau_2\right) \\
&\times\boldsymbol{\mathcal G}\left(\Omega^{\left(j\right)}_1,\tau_2-\tau_1\left\vert\Omega^{\left(j\right)}_0\right.\right) \boldsymbol{G}\left(\tau_2-\tau_1\right) \boldsymbol{L}_{\rm v}\left(\tau_1\right) {\mathcal P}^{\left(j\right)}_0\left(\Omega^{\left(j\right)}_0\right){\mathcal P}^{\left(k\right)}_0\left(\Omega^{\left(k\right)}_0\right)\boldsymbol{\rho}\left(t_0\right)
\end{aligned}
\label{2.14}
\end{multline}
The third case involves one photon interacting with one molecule followed by two photons interacting with the other. It turns out
\begin{multline}
\left\langle\Pi^{\rm III}\left(t\right)\right\rangle_{\text{or}}=\frac{i}{\hbar^3}\sum_{j,k={\rm A,B}}^{k\not=j}\int d\Omega^{\left(k\right)}_2\int d\Omega^{\left(k\right)}_1\int d\Omega^{\left(k\right)}_0\int d\Omega^{\left(j\right)}_1\int d\Omega^{\left(j\right)}_0 \int_{t_0}^{t}\!d\tau_3\int_{t_0}^{\tau_3}\!d\tau_2\int_{t_0}^{\tau_2}\!d\tau_1 \\
\begin{aligned}[b]
&\times\vec{E}_{\text{lo}}\cdot\vec{\boldsymbol{\mu}}\,\boldsymbol{\mathcal G}\left(\Omega^{\left(k\right)}_2,t-\tau_3\left\vert\Omega^{\left(k\right)}_1\right.\right)\boldsymbol{G}\left(t-\tau_3\right)\boldsymbol{L}_{\rm v}\left(\tau_3\right) \\
&\times\boldsymbol{\mathcal G}\left(\Omega^{\left(k\right)}_1,\tau_3-\tau_2\left\vert\Omega^{\left(k\right)}_0\right.\right)\boldsymbol{G}\left(\tau_3-\tau_2\right)\boldsymbol{L}_{\rm v}\left(\tau_2\right) \\
&\times\boldsymbol{\mathcal G}\left(\Omega^{\left(j\right)}_1,t-\tau_1\left\vert\Omega^{\left(j\right)}_0\right.\right) \boldsymbol{G}\left(\tau_2-\tau_1\right) \boldsymbol{L}_{\rm v}\left(\tau_1\right) {\mathcal P}^{\left(j\right)}_0\left(\Omega^{\left(j\right)}_0\right){\mathcal P}^{\left(k\right)}_0\left(\Omega^{\left(k\right)}_0\right)\boldsymbol{\rho}\left(t_0\right)
\end{aligned}
\label{2.15}
\end{multline}
Finally, in the last case, three photons interact chronologically with different molecules in turns. In other words, the order is either ${\text{A-B-A}}$ or ${\text{B-A-B}}$. Thus,
\begin{multline}
\left\langle\Pi^{\rm IV}\left(t\right)\right\rangle_{\text{or}}=\frac{i}{\hbar^3}\sum_{j,k={\rm A,B}}^{k\not=j}\int d\Omega^{\left(j\right)}_2\int d\Omega^{\left(j\right)}_1\int d\Omega^{\left(j\right)}_0\int d\Omega^{\left(k\right)}_1\int d\Omega^{\left(k\right)}_0 \int_{t_0}^{t}\!d\tau_3\int_{t_0}^{\tau_3}\!d\tau_2\int_{t_0}^{\tau_2}\!d\tau_1 \\
\begin{aligned}[b]
&\times\vec{E}_{\text{lo}}\cdot\vec{\boldsymbol{\mu}}\,\boldsymbol{\mathcal G}\left(\Omega^{\left(j\right)}_2,t-\tau_3\left\vert\Omega^{\left(j\right)}_1\right.\right)\boldsymbol{G}\left(t-\tau_3\right)\boldsymbol{L}_{\rm v}\left(\tau_3\right) \\
&\times\boldsymbol{\mathcal G}\left(\Omega^{\left(k\right)}_1,t-\tau_2\left\vert\Omega^{\left(k\right)}_0\right.\right)\boldsymbol{G}\left(\tau_3-\tau_2\right)\boldsymbol{L}_{\rm v}\left(\tau_2\right) \\
&\times\boldsymbol{\mathcal G}\left(\Omega^{\left(j\right)}_1,\tau_3-\tau_1\left\vert\Omega^{\left(j\right)}_0\right.\right) \boldsymbol{G}\left(\tau_2-\tau_1\right) \boldsymbol{L}_{\rm v}\left(\tau_1\right) {\mathcal P}^{\left(j\right)}_0\left(\Omega^{\left(j\right)}_0\right){\mathcal P}^{\left(k\right)}_0\left(\Omega^{\left(k\right)}_0\right)\boldsymbol{\rho}\left(t_0\right)
\end{aligned}
\label{2.16}
\end{multline}
All these quantities are required to evaluate the spectra associated with the energy transfer. The first one, namely $\left\langle\Pi^{\rm I}\left(t\right)\right\rangle_{\text{or}}$, is the simplest because only one molecule is involved in the averaging procedure. The scalar product involved in the interaction term can be extracted, so that $L_{\rm v}\left(\tau\right)=\hat{\boldsymbol{\mu}}\cdot\hat{E}\tilde{L}_{\rm v}\left(\tau\right)$.  
Applying the spherical-harmonics expansion given by relation \eqref{2.10}
and with the help of the orthogonality relation of the spherical harmonics we get
\begin{multline}
\left\langle\Pi^{\rm I}\left(t\right)\right\rangle_{\text{or}}=\frac{i}{\hbar^3}\sqrt{\frac{1}{12\pi}}\sum_{j={\rm A,B}}\int d\Omega^{\left(j\right)}_3\int d\Omega^{\left(j\right)}_2\int d\Omega^{\left(j\right)}_1 \int_{t_0}^{t}\!d\tau_3\int_{t_0}^{\tau_3}\!d\tau_2\int_{t_0}^{\tau_2}\!d\tau_1  \\
\begin{aligned}[b]
&\times \vec{E}_{\text{lo}}\cdot\vec{\boldsymbol{\mu}}^{\left(j\right)}e^{-D^{\left(j\right)}\boldsymbol{J}^2\left(t-\tau_3\right)}\\
&\times \sum_{\ell,m}Y_{\ell,m}\left(\Omega_2^{\left(j\right)}\right)Y^{\star}_{\ell,m}\left(\Omega_3^{\left(j\right)}\right)\hat{\boldsymbol{\mu}}^{\left(j\right)}\cdot\hat{E}_p\boldsymbol{G}\left(t-\tau_3\right)\tilde{\boldsymbol{L}}_{\rm v}\left(\tau_3\right)e^{-D^{\left(j\right)}\boldsymbol{J}^2\left(\tau_3-\tau_2\right)} \\
&\times\sum_{\ell',m'}Y_{\ell',m'}\left(\Omega_1^{\left(j\right)}\right)Y^{\star}_{\ell',m'}\left(\Omega_2^{\left(j\right)}\right)\hat{\boldsymbol{\mu}}^{\left(j\right)}\cdot\hat{E}_q\boldsymbol{G}\left(\tau_3-\tau_2\right)\tilde{\boldsymbol{L}}_{\rm v}\left(\tau_2\right)e^{-2D^{\left(j\right)}\left(\tau_2-\tau_1\right)} \\
&\times Y^{\star}_{1,0}\left(\Omega_1^{\left(j\right)}\right)\boldsymbol{G}\left(\tau_2-\tau_1\right)\tilde{\boldsymbol{L}}_{\rm v}\left(\tau_1\right)\boldsymbol{\rho}\left(t_0\right)
\end{aligned}
\label{2.17}
\end{multline}
In Eq.~\eqref{2.17}, we have also assumed a uniformly distributed initial dipole orientation, ${\mathcal P}^{\left(j\right)}_0\left(\Omega^{\left(j\right)}_0\right)=1/4\pi$.
The polarization of the laser field is chosen to be along the Z-axis. Therefore, $\hat{\boldsymbol{\mu}}^{\left(j\right)}\cdot\hat{E}_r=\cos\theta=\sqrt{4\pi/3}\,Y_{1,0}\!\left(\Omega_0^{\left(j\right)}\right)$.
Similarly, $\hat{\boldsymbol{\mu}}^{\left(j\right)}\cdot\hat{E}_q=\sqrt{4\pi/3}\,Y_{1,0}\!\left(\Omega_1^{\left(j\right)}\right)$ and $\hat{\boldsymbol{\mu}}^{\left(j\right)}\cdot\hat{E}_p=\sqrt{4\pi/3}\,Y_{1,0}\!\left(\Omega_2^{\left(j\right)}\right)$.
With these explicit expressions, we shall take advantage of the following integral relations between spherical harmonics:
\begin{equation}
\begin{aligned}
&\int d\Omega\, Y_{1,0}^{\star}\left(\Omega\right)Y_{1,0}\left(\Omega\right)Y_{\ell,m}\left(\Omega\right) =\sqrt{\frac{1}{4\pi}}\;\delta_{\ell,0}\delta_{m,0}+\sqrt{\frac{1}{5\pi}}\;\delta_{\ell,2}\delta_{m,0} \\
&\int d\Omega\, Y_{0,0}^{\star}\left(\Omega\right)Y_{1,0}\left(\Omega\right)Y_{\ell,m}\left(\Omega\right) =\sqrt{\frac{1}{4\pi}}\;\delta_{\ell,1}\delta_{m,0} \\
&\int d\Omega\, Y_{2,0}^{\star}\left(\Omega\right)Y_{1,0}\left(\Omega\right)Y_{\ell,m}\left(\Omega\right) =\sqrt{\frac{1}{5\pi}}\;\delta_{\ell,1}\delta_{m,0}+\frac{3}{2}\sqrt{\frac{3}{35\pi}}\;\delta_{\ell,3}\delta_{m,0}
\end{aligned}
\label{2.18}
\end{equation}
we obtain, for the first term, the final result
\begin{multline}
\left\langle\Pi^{\rm I}\left(t\right)\right\rangle_{\text{or}}=\frac{i}{\hbar^3}\frac{4\pi}{9}E_{\text{lo}}\mu^{\left(j\right)}\sum_{j={\text{A,B}}} \int_{t_0}^{t}\!d\tau_3\int_{t_0}^{\tau_3}\!d\tau_2\int_{t_0}^{\tau_2}\!d\tau_1 e^{-2D^{\left(j\right)}\left(t-\tau_3\right)}\\
\begin{aligned}[b]
\times&\left[ \frac{1}{4\pi}\boldsymbol{G}\left(t-\tau_3\right)\tilde{\boldsymbol{L}}_{\rm v}\left(\tau_3\right)\boldsymbol{G}\left(\tau_3-\tau_2\right)\tilde{\boldsymbol{L}}_{\rm v}\left(\tau_2\right) e^{-2D^{\left(j\right)}\left(\tau_2-\tau_1\right)}\boldsymbol{G}\left(\tau_2-\tau_1\right)\tilde{\boldsymbol{L}}_{\rm v}\left(\tau_1\right)\boldsymbol{\rho}\left(t_0\right)         \right. \\
&\phantom\lbrack +\frac{1}{5\pi}\boldsymbol{G}\left(t-\tau_3\right)\tilde{\boldsymbol{L}}_{\rm v}\left(\tau_3\right)e^{-6D^{\left(j\right)}\left(\tau_3-\tau_2\right)}  \boldsymbol{G}\left(\tau_3-\tau_2\right)\tilde{\boldsymbol{L}}_{\rm v}\left(\tau_2\right)e^{-2D^{\left(j\right)}\left(\tau_2-\tau_1\right)}\\
&\left.\phantom{\lbrack +}\times\boldsymbol{G}\left(\tau_2-\tau_1\right)\tilde{\boldsymbol{L}}_{\rm v}\left(\tau_1\right)\boldsymbol{\rho}\left(t_0\right) \right]
\end{aligned}
\label{2.19}
\end{multline}
where the notations $E_p=\vert\vec{E}_p\vert$ and $\mu=\vert\vec{\mu}\vert$ have been introduced.

Next we evaluate $\left\langle\Pi^{\rm II}\left(t\right)\right\rangle_{\text{or}}$. Using the previous properties and definitions, integrations over $\Omega^{\left(j\right)}_0$ and $\Omega^{\left(k\right)}_0$ can be performed. Then, expression \eqref{2.14} reduces to
\begin{multline}
\left\langle\Pi^{\rm II}\left(t\right)\right\rangle_{\text{or}}=\frac{i}{\hbar^3}\frac{1}{3\sqrt{12\pi}}\sum_{j,k={\text{A,B}}}^{k\not=j}\int d\Omega^{\left(j\right)}_2\int d\Omega^{\left(j\right)}_1\int d\Omega^{\left(k\right)}_1 \int_{t_0}^{t}\!d\tau_3\int_{t_0}^{\tau_3}\!d\tau_2\int_{t_0}^{\tau_2}\!d\tau_1 \vec{E}_{\text{lo}}\cdot\vec{\boldsymbol{\mu}} \\
\begin{aligned}[b]
&\times e^{-2D^{\left(k\right)}\left(t-\tau_3\right)}Y^{\star}_{1,0}\left(\Omega_1^{\left(k\right)}\right)\boldsymbol{G}\left(t-\tau_3\right)\tilde{\boldsymbol{L}}_v\left(\tau_3\right)e^{-D^{\left(j\right)}\boldsymbol{J}^2\left(t-\tau_2\right)} \\
&\times\sum_{\ell',m'}Y_{\ell',m'}\left(\Omega_1^{\left(j\right)}\right)Y^{\star}_{\ell',m'}\left(\Omega_2^{\left(j\right)}\right)Y_{1,0}\left(\Omega_1^{\left(j\right)}\right)\boldsymbol{G}\left(\tau_3-\tau_2\right)\tilde{\boldsymbol{L}}_v\left(\tau_2\right)
e^{-2D^{\left(j\right)}\left(\tau_2-\tau_1\right)}
\\ &\times
Y^{\star}_{1,0}\left(\Omega_1^{\left(j\right)}\right)\boldsymbol{G}\left(\tau_2-\tau_1\right)\tilde{\boldsymbol{L}}_v\left(\tau_1\right) \boldsymbol{\rho}\left(t_0\right)
\end{aligned}
\label{2.20}
\end{multline}
where relations
\begin{align}
&{\mathcal P}^{\left(j\right)}_0\left(\Omega^{\left(j\right)}_0\right)={\mathcal P}^{\left(k\right)}_0\left(\Omega^{\left(k\right)}_0\right)   \nonumber\\
&\hat{\boldsymbol{\mu}}^{\left(j\right)}\cdot \hat{E}_{r\{q\}}=\sqrt{\frac{4\pi}{3}}Y_{1,0}\left(\Omega^{\left(j\right)}_{0\{1\}}\right)
\text{\quad and \quad}\hat{\boldsymbol{\mu}}^{\left(k\right)}\cdot \hat{E}_p=\sqrt{\frac{4\pi}{3}}Y_{1,0}\left(\Omega^{\left(k\right)}_0\right)
\label{2.21}
\end{align}
have been introduced. Now, we perform the integration over $\Omega_1^{\left(j\right)}$ using the integral relation
\begin{equation}
\int d\Omega_1^{\left(j\right)} Y_{1,0}^{\star}\left(\Omega_1^{\left(j\right)}\right)Y_{1,0}\left(\Omega_1^{\left(j\right)}\right)Y_{\ell,m}\left(\Omega_1^{\left(j\right)}\right) =\left(\frac{1}{4\pi}\right)^{1/2}\delta_{l,0}\delta_{m,0}+\left(\frac{1}{5\pi}\right)^{1/2}\delta_{l,2}\delta_{m,0} \;.
\label{2.22}
\end{equation}
to get
\begin{multline}
\left\langle\Pi^{\rm II}\left(t\right)\right\rangle_{\text{or}}=\frac{i}{\hbar^3}\frac{1}{3\sqrt{12\pi}}\sum_{j,k={\text{A,B}}}^{k\not=j}\int d\Omega^{\left(j\right)}_2\int d\Omega^{\left(k\right)}_1 \int_{t_0}^{t}\!d\tau_3\int_{t_0}^{\tau_3}\!d\tau_2\int_{t_0}^{\tau_2}\!d\tau_1 \vec{E}_{\text{lo}}\cdot\vec{\boldsymbol{\mu}} \\
\begin{aligned}[b]
\times & e^{-2D^{\left(k\right)}\left(t-\tau_3\right)}Y^{\star}_{1,0}\left(\Omega_1^{\left(k\right)}\right)\boldsymbol{G}\left(t-\tau_3\right)\tilde{\boldsymbol{L}}_{\rm v}\left(\tau_3\right)
\left[\tfrac{1}{\sqrt{4\pi}}Y^{\star}_{0,0}\left(\Omega_2^{\left(j\right)}\right)
+\tfrac{1}{\sqrt{5\pi}}e^{-6D^{\left(j\right)}\left(t-\tau_2\right)}Y^{\star}_{2,0}\left(\Omega_2^{\left(j\right)}\right)
\right]
\\ \times &
\boldsymbol{G}\left(\tau_3-\tau_2\right)\tilde{\boldsymbol{L}}_{\rm v}\left(\tau_2\right)
e^{-2D^{\left(j\right)}\left(\tau_2-\tau_1\right)}
\boldsymbol{G}\left(\tau_2-\tau_1\right) \tilde{\boldsymbol{L}}_{\rm v}\left(\tau_1\right) \boldsymbol{\rho}\left(t_0\right)
\end{aligned}
\label{2.23}
\end{multline}
We still have to integrate over $\Omega_1^{\left(k\right)}$ and $\Omega_2^{\left(j\right)}$. To this end, we must specify the molecular transition contributing to the heterodyne signal. If we assumed that the detection focus on a molecular transition of molecule $j$, we have $\vec{E}_{\text{lo}}\cdot\vec{\boldsymbol{\mu}^{\left(j\right)}}=\sqrt{4\pi/3}\;Y_{1,0}\left(\Omega_2^{\left(j\right)}\right) $ and
\begin{equation}
\int d\Omega^{\left(j\right)}_2\sqrt{\frac{4\pi}{3}}\;Y_{1,0}\left(\Omega_2^{\left(j\right)}\right)Y^{\star}_{0,0}\left(\Omega_2^{\left(j\right)}\right)
=\int d\Omega^{\left(j\right)}_2\sqrt{\frac{4\pi}{3}}\;Y_{1,0}\left(\Omega_2^{\left(j\right)}\right)Y^{\star}_{2,0}\left(\Omega_2^{\left(j\right)}\right)=0
\label{2.24}
\end{equation}
which implies in turn that $\left\langle\Pi^{\rm II}\left(t\right)\right\rangle_{\text{or}}=0$ if the local field acts on the $j$-molecule. This result is not surprising at all because it is well established that the second-order optical response cancels for an isotropic system, which is the case here since the laser field interacts twice with the same molecule. Otherwise, if detection focus on a particular transition of molecule $k$, $\vec{E}_{\text{lo}}\cdot\vec{\boldsymbol{\mu}^{\left(k\right)}}=\sqrt{4\pi/3\,}Y_{1,0}\left(\Omega_1^{\left(k\right)}\right)$ and performing the integrations over $\Omega_1^{\left(k\right)}$ and $\Omega_2^{\left(j\right)}$ we get
\begin{multline}
\left\langle\Pi^{\rm II}\left(t\right)\right\rangle_{\text{or}}=\frac{i}{9\hbar^3}E_{\text{lo}}\mu^{\left(k\right)}\sum_{j,k={\text{A,B}}}^{k\not=j} \int_{t_0}^{t}\!d\tau_3\int_{t_0}^{\tau_3}\!d\tau_2
\int_{t_0}^{\tau_2}\!d\tau_1 e^{-2D^{\left(k\right)}\left(t-\tau_3\right)}\boldsymbol{G}\left(t-\tau_3\right)\tilde{\boldsymbol{L}}_{\rm v}\left(\tau_3\right) \\
\times\boldsymbol{G}\left(\tau_3-\tau_2\right)\tilde{\boldsymbol{L}}_{\rm v}\left(\tau_2\right)e^{-2D^{\left(j\right)}\left(\tau_2-\tau_1\right)}\boldsymbol{G}\left(\tau_2-\tau_1\right)\tilde{\boldsymbol{L}}_{\rm v}\left(\tau_1\right)\boldsymbol{\rho}\left(t_0\right)
\label{2.25}
\end{multline}
because $\int d\Omega^{\left(j\right)}_2Y^{\star}_{2,0}\left(\Omega_2^{\left(j\right)}\right)=2\pi\int_{0}^{\pi}d\theta\sin\theta\sqrt{\frac{5}{{16\pi}}}\left(3\cos^2\theta-1\right)=0$.

Then we calculate the third term $\left\langle\Pi^{\rm III}\left(t\right)\right\rangle_{\text{or}}$. It takes the form
\begin{multline}
\left\langle\Pi^{\rm III}\left(t\right)\right\rangle_{\text{or}}=\frac{i}{\hbar^3}\frac{1}{6\sqrt{2\pi}}\sum_{j,k={\text{A,B}}}^{k\not=j} \\
\begin{aligned}[b]
&\times\int d\Omega^{\left(k\right)}_2\int d\Omega^{\left(k\right)}_1\int d\Omega^{\left(k\right)}_0
\int d\Omega^{\left(j\right)}_1\int d\Omega^{\left(j\right)}_0 \int_{t_0}^{t}\!d\tau_3\int_{t_0}^{\tau_3}\!d\tau_2\int_{t_0}^{\tau_2}\!d\tau_1 \vec{E}_{\text{lo}}\cdot\vec{\boldsymbol{\mu}} \\
&\times e^{-D^{\left(k\right)}\boldsymbol{J}^2\left(t-\tau_3\right)}\sum_{\ell,m}Y_{\ell,m}\left(\Omega_1^{\left(k\right)}\right)Y^{\star}_{\ell,m}\left(\Omega_2^{\left(k\right)}\right)Y_{1,0}\left(\Omega_1^{\left(k\right)}\right)\boldsymbol{G}\left(t-\tau_3\right)\tilde{\boldsymbol{L}}_{\rm v}\left(\tau_3\right) \\ &\times
e^{-D^{\left(k\right)}\boldsymbol{J}^2\left(\tau_3-\tau_2\right)} \sum_{\ell',m'}Y_{\ell',m'}\left(\Omega_0^{\left(k\right)}\right) Y^{\star}_{\ell',m'}\left(\Omega_1^{\left(k\right)}\right)Y_{1,0}\left(\Omega_0^{\left(k\right)}\rm \right)\boldsymbol{G}\left(\tau_3-\tau_2\right)\tilde{\boldsymbol{L}}_{\rm v}\left(\tau_2\right)\\
&\times e^{-D^{\left(j\right)}\boldsymbol{J}^2\left(t-\tau_1\right)} \sum_{\ell'',m''}Y_{\ell'',m''}\left(\Omega_0^{\left(j\right)}\right)Y^{\star}_{\ell'',m''}\left(\Omega_1^{\left(j\right)}\right)Y_{1,0}\left(\Omega_0^{\left(j\right)}\right)  \boldsymbol{G}\left(\tau_2-\tau_1\right)\tilde{\boldsymbol{L}}_{\rm v}\left(\tau_1\right)\boldsymbol{\rho}\left(t_0\right)
\end{aligned}
\label{2.26}
\end{multline}
Here, we need to introduce the definitions of the polarization propagators and interaction terms. First, we integrate over $\Omega^{\left(j\right)}_0$ and $\Omega^{\left(k\right)}_0$. Next, taking advantage of the first relation in \eqref{2.18}, we integrate over $\Omega_1^{\left(k\right)}$. The following result is obtained:
\begin{multline}
\left\langle\Pi^{\rm III}\left(t\right)\right\rangle_{\text{or}}=\frac{i}{\hbar^3}\frac{1}{6\sqrt{2\pi}}\sum_{j,k={\text{A,B}}}^{k\not=j}\int d\Omega^{\left(k\right)}_2\int d\Omega^{\left(j\right)}_1 \int_{t_0}^{t}\!d\tau_3\int_{t_0}^{\tau_3}\!d\tau_2\int_{t_0}^{\tau_2}\!d\tau_1 \vec{E}_{\text{lo}}\cdot\vec{\boldsymbol{\mu}}e^{-D^{\left(k\right)}\boldsymbol{J}^2\left(t-\tau_3\right)} \\
\begin{aligned}[b]
&\times\left[\frac{1}{\sqrt{4\pi}}Y^{\star}_{0,0}\left(\Omega_2^{\left(k\right)}\right)+\frac{1}{\sqrt{5\pi}} Y^{\star}_{2,0}\left(\Omega_2^{\left(k\right)}\right)\right]
\boldsymbol{G}\left(t-\tau_3\right)\tilde{\boldsymbol{L}}_{\rm v}\left(\tau_3\right)e^{-2D^{\left(k\right)}\left(\tau_3-\tau_2\right)}
\\ &\times
\boldsymbol{G}\left(\tau_3-\tau_2\right)\tilde{\boldsymbol{L}}_{\rm v}\left(\tau_2\right)e^{-2D^{\left(j\right)}\left(t-\tau_1\right)}
Y^{\star}_{1,0}\left(\Omega_1^{\left(j\right)}\right) \boldsymbol{G}\left(\tau_2-\tau_1\right)\tilde{\boldsymbol{L}}_{\rm v}\left(\tau_1\right)\boldsymbol{\rho}\left(t_0\right)
\end{aligned}
\label{2.27}
\end{multline}
Similar to the case of $\left\langle\Pi^{\rm II}\left(t\right)\right\rangle_{\text{or}}$, further simplification is possible.
Due to the orthogonality relations of the spherical harmonics, the term $\vec{E}_{\text{lo}}\cdot\vec{\boldsymbol{\mu}^{\left(k\right)}}$ does not contribute, while the term $\vec{E}_{\text{lo}}\cdot\vec{\boldsymbol{\mu}^{\left(j\right)}}$ gives
\begin{multline}
\left\langle\Pi^{\rm III}\left(t\right)\right\rangle_{\text{or}}=\frac{i}{9\hbar^3}E_{\text{lo}}\mu^{\left(j\right)}\sum_{j,k={\text{A,B}}}^{k\not=j} \int_{t_0}^{t}\!d\tau_3\int_{t_0}^{\tau_3}\!d\tau_2
\int_{t_0}^{\tau_2}\!d\tau_1\boldsymbol{G}\left(t-\tau_3\right)\tilde{\boldsymbol{L}}_{\rm v}\left(\tau_3\right)e^{-2D^{\left(k\right)}\left(\tau_3-\tau_2\right)} \\
\times\boldsymbol{G}\left(\tau_3-\tau_2\right)\tilde{\boldsymbol{L}}_{\rm v}\left(\tau_2\right)e^{-2D^{\left(j\right)}\left(t-\tau_1\right)}\boldsymbol{G}\left(\tau_2-\tau_1\right)\tilde{\boldsymbol{L}}_{\rm v}\left(\tau_1\right)\boldsymbol{\rho}\left(t_0\right)
\label{2.28}
\end{multline}

Finally, we evaluate the last term $\left\langle\Pi^{\rm IV}\left(t\right)\right\rangle_{\text{or}}$. It can be written as
\begin{multline}
\left\langle\Pi^{\rm IV}\left(t\right)\right\rangle_{\text{or}}=\frac{i}{\hbar^3}\frac{1}{6\sqrt{2\pi}}\sum_{j,k={\text{A,B}}}^{k\not=j} \\
\begin{aligned}[b]
&\times\int d\Omega^{\left(j\right)}_2\int d\Omega^{\left(j\right)}_1\int d\Omega^{\left(j\right)}_0 \int d\Omega^{\left(k\right)}_1\int d\Omega^{\left(k\right)}_0 \int_{t_0}^{t}\!d\tau_3\int_{t_0}^{\tau_3}\!d\tau_2\int_{t_0}^{\tau_2}\!d\tau_1 \vec{E}_{\text{lo}}\cdot\vec{\boldsymbol{\mu}} \\
&\times e^{-D^{\left(j\right)}\boldsymbol{J}^2\left(t-\tau_3\right)}\sum_{\ell,m}Y_{\ell,m}\left(\Omega_1^{\left(j\right)}\right)Y^{\star}_{\ell,m}\left(\Omega_2^{\left(j\right)}\right)Y_{1,0}\left(\Omega_1^{\left(j\right)}\right)\boldsymbol{G}\left(t-\tau_3\right)\tilde{\boldsymbol{L}}_{\rm v}\left(\tau_3\right) \\
&\times e^{-D^{\left(k\right)}\boldsymbol{J}^2\left(t-\tau_2\right)}\sum_{\ell',m'}Y_{\ell',m'}\left(\Omega_0^{\left(k\right)}\right) Y^{\star}_{\ell',m'}\left(\Omega_1^{\left(k\right)}\right)Y_{1,0}\left(\Omega_0^{\left(k\right)}\right)\boldsymbol{G}\left(\tau_3-\tau_2\right)\tilde{\boldsymbol{L}}_{\rm v}\left(\tau_2\right) \\
&\times e^{-D^{\left(j\right)}\boldsymbol{J}^2\left(\tau_3-\tau_1\right)}\sum_{\ell'',m''}Y_{\ell''m''}\left(\Omega_0^{\left(j\right)}\right)Y^{\star}_{\ell'',m''}\left(\Omega_1^{\left(j\right)}\right)Y_{1,0}\left(\Omega_0^{\left(j\right)}\right) \boldsymbol{G}\left(\tau_2-\tau_1\right)\tilde{\boldsymbol{L}}_{\rm v}\left(\tau_1\right)\boldsymbol{\rho}\left(t_0\right)
\end{aligned}
\label{2.29}
\end{multline}
Like in the case of calculating $\left\langle\Pi^{\rm III}\left(t\right)\right\rangle_{\text{or}}$, we integrate over $\Omega^{\left(j\right)}_0$ and $\Omega^{\left(k\right)}_0$ first. Then, using the definitions and properties of the spherical harmonics we obtain the expression
\begin{multline}
\left\langle\Pi^{\rm IV}\left(t\right)\right\rangle_{\text{or}}=\frac{i}{\hbar^3}\frac{1}{6\sqrt{2\pi}}\sum_{j,k={\text{A,B}}}^{k\not=j}\int d\Omega^{\left(j\right)}_2\int d\Omega^{\left(k\right)}_1 \int_{t_0}^{t}\!
d\tau_3 \int_{t_0}^{\tau_3}\!d\tau_2\int_{t_0}^{\tau_2}\!d\tau_1 \vec{E}_{\text{lo}}\cdot\vec{\boldsymbol{\mu}} \\
\begin{aligned}[b]
&\times\left[\frac{1}{\sqrt{4\pi}}Y^{\star}_{0,0}\left(\Omega_2^{\left(j\right)}\right)+\frac{1}{\sqrt{5\pi}}e^{-6D^{\left(j\right)}\left(t-\tau_3\right)}Y^{\star}_{2,0}\left(\Omega_2^{\left(j\right)}\right)\right]
\boldsymbol{G}\left(t-\tau_3\right)\tilde{\boldsymbol{L}}_{\rm v}\left(\tau_3\right)e^{-2D^{\left(k\right)}\left(t-\tau_2\right)} \\
&\times Y^{\star}_{1,0}\left(\Omega_1^{\left(k\right)}\right) \boldsymbol{G}\left(\tau_3-\tau_2\right)\tilde{\boldsymbol{L}}_{\rm v}\left(\tau_2\right)e^{-2D^{\left(j\right)}\left(\tau_3-\tau_1\right)}\boldsymbol{G}\left(\tau_2-\tau_1\right)\tilde{\boldsymbol{L}}_{\rm v}\left(\tau_1\right)\boldsymbol{\rho}\left(t_0\right)
\end{aligned}
\label{2.30}
\end{multline}
Here, only $\vec{E}_{\text{lo}}\cdot\vec{\boldsymbol{\mu}}^{\left(k\right)}$ contributes. Integrating over $\Omega^{\left(k\right)}_1$ and $\Omega^{\left(j\right)}_2$ we have
\begin{multline}
\left\langle\Pi^{\rm IV}\left(t\right)\right\rangle_{\text{or}}=\frac{i}{9\hbar^3}E_{\text{lo}}\mu^{\left(k\right)}\sum_{j,k={\text{A,B}}}^{k\not=j}\int_{t_0}^{t}\!\!\!d\tau_3\int_{t_0}^{\tau_3}\!d\tau_2\int_{t_0}^{\tau_2}\!d\tau_1 \boldsymbol{G}\left(t-\tau_3\right)\tilde{\boldsymbol{L}}_{\rm v}\left(\tau_3\right) e^{-2D^{\left(k\right)}\left(t-\tau_2\right)} \\
\times \boldsymbol{G}\left(\tau_3-\tau_2\right)\tilde{\boldsymbol{L}}_{\rm v}\left(\tau_2\right)e^{-2D^{\left(j\right)}\left(\tau_3-\tau_1\right)}\boldsymbol{G}\left(\tau_2-\tau_1\right)\tilde{\boldsymbol{L}}_{\rm v}\left(\tau_1\right)\boldsymbol{\rho}\left(t_0\right)
\label{2.31}
\end{multline}
Equations \eqref{2.19}, \eqref{2.25}, \eqref{2.28}, and \eqref{2.31} conclude the influence of the dipolar orientational average when an energy transfer process occurs between two molecules embedded in a liquid solvent. In the following, these results will be applied to emphasize the peculiar role of the polarization on vibrational energy transfer.

\section{Vibrational molecular model and dynamics}

\begin{figure}[ht]
\includegraphics[clip]{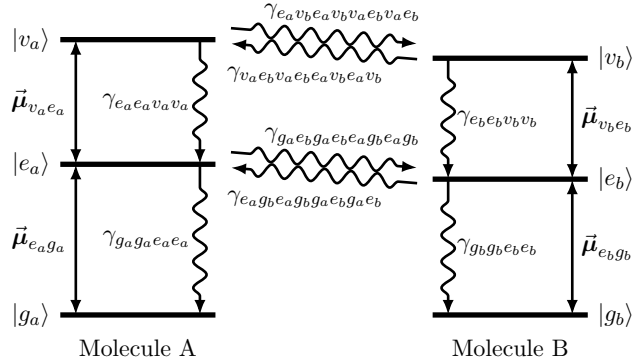}
\caption{
\label{fig1}
Molecular model used to study the polarization effects on the vibrational energy transfer process. Dipole moments $\vec{\boldsymbol{\mu}}$ as well as intramolecular and intermolecular transition rate constants $\gamma$ are indicated in the figure. Notice that intermolecular transition constants involve simultaneous transitions in molecules A and B.}
\end{figure}

\begin{figure}[hb]
\includegraphics[clip]{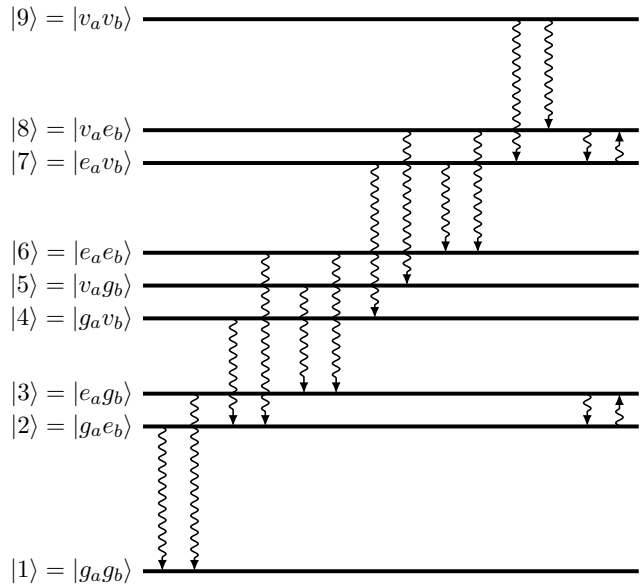}
\caption{The molecular model presented in the two-molecule combination state space. The wavy arrows are transitions due to intra and inter-molecular interactions, and they are represented by $\Gamma_{mmnn}$ type of transtion rate constants in the text.
}
\label{fig2}
\end{figure}

\begin{figure}[ht]
\includegraphics[clip]{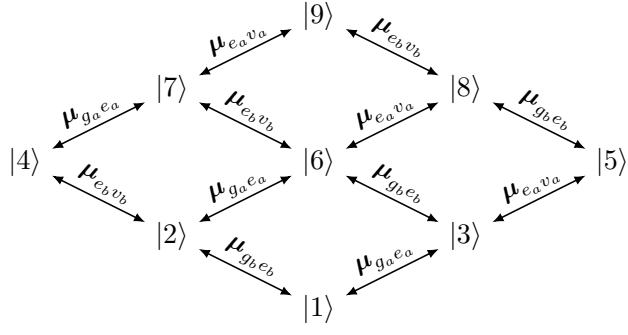}
\caption{
\label{fig3}
Diagram of the couplings between the states of the total molecular system. The coupling are induced by the laser-molecule interaction. For clarity, the corresponding transition dipole moments are labeled.}
\end{figure}

In the model used in the present work, the material system interacting with the nonlinear optical fields is made of two molecules undergoing vibrational energy transfer.
The vibrational states involved may be two anharmonic vibration modes of the donor and acceptor molecules, respectively. They may also be vibrational combination states of individual molecules. For simplicity, the first case will be considered. Thus, the energy transfer takes place between two anharmonic vibration modes $\boldsymbol{Q}_{\rm A}$ and $\boldsymbol{Q}_{\rm B}$, pertaining to molecules A and B, respectively. The corresponding vibrational Hamiltonian can be expressed as
\begin{multline}
\boldsymbol{H}=\sum_{i={\rm  A,B}}\hbar\Omega_i\left[\boldsymbol{B}^+_i\boldsymbol{B}_i+\tfrac{1}{2}\right]+W\left({\boldsymbol{B}^+_i}^3,\boldsymbol{B}^3_i,\ldots\right)+\sum_{\alpha=1}^N\hbar\omega_{\alpha}\left[\boldsymbol{b}^+_{\alpha}\boldsymbol{b}_{\alpha}+\tfrac{1}{2}\right] \\
+\sum_{i}\left[\Delta_{i,1\cdots N}\boldsymbol{B}^+_i\boldsymbol{b}_1\cdots\boldsymbol{b}_N+\Delta^{\star}_{i,1\cdots N}\boldsymbol{B}_i\boldsymbol{b}^+_1\cdots\boldsymbol{b}^+_N\right]
\label{3.1}
\end{multline}
where $\boldsymbol{B}^+_i$ and $\boldsymbol{B}_i$ stand for the creation and annihilation operators of the molecular vibrational modes and $\boldsymbol{b}^+_i$ and $\boldsymbol{b}_i$ for the corresponding operators of the bath modes with
$\Delta_{i,1\cdots N}$ and $\Delta^{\star}_{i,1\cdots N}$ their corresponding coupling constants. The energy level scheme of this reduced molecular space is shown in
Fig.~\ref{fig1} where all $\gamma_{\alpha\beta\gamma\delta}$ are the transition rate constants assuming a Markovian bath.
Subsequently, the set of levels required for the description of the two-molecule system are defined in Fig.~\ref{fig2}. The possible decays between these two-molecule combination states are also presented with the wavy arrows. Specifically, the couplings induced by the interaction Hamiltonian $\boldsymbol{V}\left(t\right)$ are represented by the diagram shown in Fig.~\ref{fig3}. It is worthy of noting that those coupling depicted in Fig.~\ref{fig3} correspond to all of the transitions drawn in Fig.~\ref{fig2} except that the transitions between states 2, 3 and between 7, 8 are not included, and that in Fig.~\ref{fig2} the couplings are bidrectional between initial and final states, while those decays are unidirectional.
Then, all the pathways participating in the evaluation of the 2DIR spectrum and satisfying the rotating wave approximation are determined for the rephasing
$\vec{k}_{\text{re}}=-\vec{k}_a+\vec{k}_b+\vec{k}_c$ and the non rephasing  $\vec{k}_{\text{nre}}=-\vec{k}_b+\vec{k}_a+\vec{k}_c$ directions. They are listed in Supplement A according to whether laser field $a$\/ or $b$\/ acts in the first, the second or the third interaction.
The contributions of individual pathways are calculated from the matrix elements of the evolution Liouvillians, interaction Liouvillians and the additional dipole propagators.
In the previous section we have already shown how these contributions depend on the chronological ordering of the pulse-molecule interactions.

In practice, the integrands $\mathcal{I}^{I\rightarrow IV}\left(t\right)$ of the heterodyne signal 
$\left\langle\Pi^{\rm I\rightarrow IV}\left(t\right)\right\rangle_{\text{or}}$ can be expressed in a general form:
\begin{multline}
\mathcal{I}^{I\rightarrow IV}\left(t\right)=\frac{i}{\hbar^3}Q^{\text{id}}\left(n,\alpha,r,q,p\right)e^{K^{\text{id}}\left(n,\alpha,r,q,p\right)t}e^{A^{\text{id}}\left(n,\alpha,r,q,p\right)\tau_3} e^{B^{\text{id}}\left(n,\alpha,r,q\right)\tau_2}e^{C^{\text{id}}\left(n,\alpha,r\right)\tau_1} \\
\times Q^{\text{or}}\left(n,\beta,r,q,p\right)e^{K^{\text{or}}\left(n,\beta,r,q,p\right)t}e^{A^{\text{or}}\left(n,\beta,r,q,p\right)\tau_3} e^{B^{\text{or}}\left(n,\beta,r,q\right)\tau_2}e^{C^{\text{or}}\left(n,\beta,r\right)\tau_1}
\label{3.2}
\end{multline}
Several comments would be valuable. First, the superscripts I through IV do not appear in all of the constants $A,B,C$ and $K,Q$. This is because a given pathway $n$ contributes to only one among the four heterodyne signals.
It suffices to specify the pathway $n$, without requiring any additional label.
The constants $Q^{\text{id}}\left(n,\alpha,r,q,p\right)$, $K^{\text{id}}\left(n,\alpha,r,q,p\right)$, $A^{\text{id}}\left(n,\alpha,r,q,p\right)$, $B^{\text{id}}\left(n,\alpha,r,q\right)$, and $C^{\text{id}}\left(n,\alpha,r\right)$
resulting from the internal dynamics are evaluated for the pathways listed in Supplement A and the results are shown in Supplement B.
The other set of constants
$Q^{\text{or}}_{n,\alpha,r,q,p}$, $K^{\text{or}}_{n,\alpha,r,q,p}$, $A^{\text{or}}_{n,\alpha,r,q,p}$, $B^{\text{or}}_{n,\alpha,r,q}$, and $C^{\text{or}}_{n,\alpha,r}$ are associated with the orientational average, and they are obtained with expressions \eqref{2.19}, \eqref{2.25}, \eqref{2.28} and \eqref{2.31}. The results are given Table~\ref{tab1}.
\begin{table}[ht]
\begin{tabular}{l@{\hspace{1em}}c@{\hspace{1em}}c@{\hspace{1em}}c@{\hspace{1em}}crrrr}
\hline
$\left\langle\Pi^{\rm I\rightarrow IV}\left(t\right)\right\rangle_{\text{or}}$ & $j$ & $k$ & $\beta$ & $Q^{\text{or}}_{n,\alpha,r,q,p}$ & $K^{\text{or}}_{n,\alpha,r,q,p}$
& $A^{\text{or}}_{n,\alpha,r,q,p}$ & $B^{\text{or}}_{n,\alpha,r,q}$ & $C^{\text{or}}_{n,\alpha,r}$\\
\hline
$\left\langle\Pi^{\rm I}\left(t\right)\right\rangle_{\text{or}}$ & A & $ $ & $ 1 $ & $E_{\text{lo}}\mu^{\left(\rm a\right)}/9 $ & $-2D^{\left(\rm a\right)}$ & $ 2D^{\left(\rm a\right)}$
& $-2D^{\left(\rm a\right)}$ & $ 2D^{\left(\rm a\right)}$\\
& & & 2 & $4E_{\text{lo}}\mu^{\left(\rm a\right)}/45$ & $-2D^{\left(\rm a\right)}$ & $-4D^{\left(\rm a\right)}$
& $ 4D^{\left(\rm a\right)}$ & $ 2D^{\left(\rm a\right)}$\\
& B & & 1 & $ E_{\text{lo}}\mu^{\left(\rm b\right)}/9 $ & $-2D^{\left(\rm b\right)}$ & $ 2D^{\left(\rm b\right)}$
& $-2D^{\left(\rm b\right)}$ & $ 2D^{\left(\rm b\right)}$\\
 & & & 2 & $4E_{\text{lo}}\mu^{\left(\rm b\right)}/45$ & $-2D^{\left(\rm b\right)}$ & $-4D^{\left(\rm b\right)}$
& $ 4D^{\left(\rm b\right)}$ & $2D^{\left(\rm b\right)}$\\
$\left\langle\Pi^{\rm II}\left(t\right)\right\rangle_{\text{or}}$ & A & B & 1 & $E_{\text{lo}}\mu^{\left(\rm b\right)}/9 $ & $-2D^{\left(\rm b\right)}$ & $ 2D^{\left(\rm b\right)}$
& $-2D^{\left(\rm a\right)}$ & $ 2D^{\left(\rm a\right)}$\\
 & B & A & 1 & $E_{\text{lo}}\mu^{\left(\rm a\right)}/9 $ & $-2D^{\left(\rm a\right)}$ & $ 2D^{\left(\rm a\right)}$
& $-2D^{\left(\rm b\right)}$ & $ 2D^{\left(\rm b\right)}$\\
$\left\langle\Pi^{\rm III}\left(t\right)\right\rangle_{\text{or}}$ & A & B & 1 & $E_{\text{lo}}\mu^{\left(\rm a\right)}/9 $ & $-2D^{\left(\rm a\right)}$ & $-2D^{\left(\rm b\right)}$
& $ 2D^{\left(\rm b\right)}$ & $ 2D^{\left(\rm a\right)}$\\
 & B & A & 1 & $E_{\text{lo}}\mu^{\left(\rm b\right)}/9$ & $-2D^{\left(\rm b\right)}$ & $-2D^{\left(\rm a\right)}$
& $2D^{\left(\rm a\right)}$ & $ 2D^{\left(\rm b\right)}$\\
$\left\langle\Pi^{\rm IV}\left(t\right)\right\rangle_{\text{or}}$ & A & B & 1 & $E_{\text{lo}}\mu^{\left(\rm b\right)}/9 $ & $-2D^{\left(\rm b\right)}$ & $-2D^{\left(\rm a\right)}$
& $ 2D^{\left(\rm b\right)}$ & $ 2D^{\left(\rm a\right)}$\\
 & B & A & 1 & $E_{\text{lo}}\mu^{\left(\rm a\right)}/9 $ & $-2D^{\left(\rm a\right)}$ & $-2D^{\left(\rm b\right)}$
& $ 2D^{\left(\rm a\right)}$ & $2D^{\left(\rm b\right)}$\\
\hline
\end{tabular}
\caption{\label{tab1}
Rotational constants participating in the dynamics, formally described by \eqref{3.2}.}
\end{table}
All these constants are listed in Supplement C according to the molecular sequences involved in the different pathways. Then, by introducing a new set of constants
\begin{equation}
\begin{aligned}[b]
&Q\left(n,\alpha,\beta,r,q,p\right)=Q^{\text{id}}\left(n,\alpha,r,q,p\right)Q^{\text{or}}\left(n,\beta,r,q,p\right)  \\
&K\left(n,\alpha,\beta,r,q,p\right)=K^{\text{id}}\left(n,\alpha,r,q,p\right)K^{\text{or}}\left(n,\beta,r,q,p\right)  \\
&A\left(n,\alpha,\beta,r,q,p\right)=A^{\text{id}}\left(n,\alpha,r,q,p\right)A^{\text{or}}\left(n,\beta,r,q,p\right)  \\
&B\left(n,\alpha,\beta,r,q\right)=B^{\text{id}}\left(n,\alpha,r,q\right)B^{\text{or}}\left(n,\beta,r,q\right)        \\
&C\left(n,\alpha,\beta,r\right)=C^{\text{id}}\left(n,\alpha,r\right)C^{\text{or}}\left(n,\beta,r\right)
\end{aligned}
\label{3.3}
\end{equation}
the formal expressions \eqref{3.2} can be rewritten as
\begin{equation}
\mathcal{I}^{\rm I\rightarrow IV}\left(t\right)=\frac{i}{\hbar^3}Q\left(n,\alpha,\beta,r,q,p\right)e^{K\left(n,\alpha,\beta,r,q,p\right)t}e^{A\left(n,\alpha,\beta,r,q,p\right)\tau_3}e^{B\left(n,\alpha,\beta,r,q\right)\tau_2}e^{C\left(n,\alpha,\beta,r\right)\tau_1}
\label{3.4}
\end{equation}
Besides, a complete evaluation of these constants requires the description of the free population evolutions satisfying the equation
$d\boldsymbol{\rho}\left(t\right)/dt=-\boldsymbol{\Gamma}\boldsymbol{\rho}\left(t\right)$. These populations are evaluated through diagonalizing the damping Liouvillian $\boldsymbol{\Gamma}$ associated with the model introduced in Fig.~\ref{fig1}. It consists of diagonal elements $\Gamma_{mmmm}$ except that for the ground state $\Gamma_{1111}=0$, together with the $\Gamma_{mmnn}$ terms presented in Fig.~\ref{fig2}.
\begin{equation}
\boldsymbol{\Gamma}=\left(
\begin{array}{*{9}{@{\extracolsep{1em}}c@{\extracolsep{1em}}}}
0&\Gamma_{1122}&\Gamma_{1133}&0&0&0&0&0&0 \\
0&\Gamma_{2222}&\Gamma_{2233}&\Gamma_{2244}&0&\Gamma_{2266}&0&0&0\\
0&\Gamma_{3322}&\Gamma_{3333}&0&\Gamma_{3355}&\Gamma_{3366}&0&0&0\\
0&0&0&\Gamma_{4444}&0&0&\Gamma_{4477}&0&0\\
0&0&0&0&\Gamma_{5555}&0&0&\Gamma_{5588}&0\\
0&0&0&0&0&\Gamma_{6666}&\Gamma_{6677}&\Gamma_{6688}&0\\
0&0&0&0&0&0&\Gamma_{7777}&\Gamma_{7788}&\Gamma_{7799}\\
0&0&0&0&0&0&\Gamma_{8877}&\Gamma_{8888}&\Gamma_{8899}\\
0&0&0&0&0&0&0&0&\Gamma_{9999}
\end{array}\right)
\label{3.3a}\end{equation}
From the sum rule $\boldsymbol{\Gamma}_{mmmm}=-\sum_{n\not=m}\boldsymbol{\Gamma}_{nnmm}$, the various total decay rates can be rewritten in terms of the individual molecular constants shown in Table \ref{tab2}.
\begin{table}[ht]
%\centering
\begin{tabular}{l@{\hspace{1em}}l@{\hspace{1em}}l@{\hspace{1em}}}
\hline
 \;$\Gamma_{2222}=\gamma_{e_be_be_be_b}$ &\; $\Gamma_{3333}=\gamma_{e_ae_ae_ae_a}$ &\; $\Gamma_{4444}=\gamma_{v_bv_bv_bv_b}$      \\
 \;$\Gamma_{5555}=\gamma_{v_av_av_av_a}$ &\; $\Gamma_{6666}=\gamma_{e_ae_ae_ae_a}+\gamma_{e_be_be_be_b}$ &\; $\Gamma_{7777}=\gamma_{e_ae_ae_ae_a}+\gamma_{v_bv_bv_bv_b}$      \\
 \;$\Gamma_{8888}=\gamma_{v_av_av_av_a}+\gamma_{e_be_be_be_b}$ &\; $\Gamma_{9999}=\Gamma_{v_av_av_av_a}+\gamma_{v_bv_bv_bv_b}$ &   \\
\hline
\end{tabular}
\caption{\label{tab2} Expressions of the decay rate constants of the total vibrational molecular system in terms of the individual molecular constants.}
\end{table}

We still need the expressions for the transition constants in terms of the individual molecular constants. They are shown in Table \ref{tab3}. Notice that the diagonal matrix elements in Eq.~\eqref{3.3a} are given in Table~\ref{tab2}. Besides, the vanishing matrix elements in Eq.~\eqref{3.3a} are not shown in Table~\ref{tab3}, for brevity.
\begin{table}[ht]
\begin{tabular}{l@{\hspace{2em}}l@{\hspace{2em}}l}
\hline
$\Gamma_{1122}=\gamma_{g_bg_be_be_b}$ &
$\Gamma_{1133}=\gamma_{g_ag_ae_ae_a}$ \\
$\Gamma_{2233}=\gamma_{g_ae_bg_ae_be_ag_be_ag_b}$ &
$\Gamma_{2244}=\gamma_{e_be_bv_bv_b}$ &
$\Gamma_{2266}=\gamma_{g_ag_ae_ae_a}$ \\
$\Gamma_{3322}=\gamma_{e_ag_be_ag_bg_ae_bg_ae_b}$  &
$\Gamma_{3355}=\gamma_{e_ae_av_av_a}$ &
$\Gamma_{3366}=\gamma_{g_bg_be_be_b}$ \\
$\Gamma_{4477}=\gamma_{g_ag_ae_ae_a}$ \\
$\Gamma_{5588}=\gamma_{g_bg_be_be_b}$ \\
$\Gamma_{6677}=\gamma_{e_be_bv_bv_b}$ &
$\Gamma_{6688}=\gamma_{e_ae_av_av_a}$ \\
$\Gamma_{7788}=\gamma_{e_av_be_av_bv_ae_bv_ae_b}$ &
$\Gamma_{7799}=\gamma_{e_ae_av_av_a}$ \\
$\Gamma_{8877}=\gamma_{v_ae_bv_ae_be_av_be_av_b}$ &
$\Gamma_{8899}=\gamma_{e_be_bv_bv_b}$ \\
\hline
\end{tabular}
\caption{\label{tab3} Relation between transition rate constants of the total system and transition rate constants of the individual molecules.}
\end{table}

Finally, the energy transfer constants satisfy detailed balance so that
\begin{equation}\begin{aligned}[b]
&\gamma_{mmnn}=\gamma_{nnmm}\exp\left[\left(E_{n}-E_{m}\right)/kT\right] & \text{if} & & m&=e_ag_b,& n&=g_ae_b \\
&\gamma_{ppqq}=\gamma_{qqpp}\exp\left[\left(E_{q}-E_{p}\right)/kT\right] & \text{if} & & p&=v_ae_b,& q&=e_av_b \;.
\end{aligned}\label{3.6}\end{equation}
and are related to the individual rates constants as expressed in Table \ref{tab4}.
\begin{table}[ht]
%\centering
\begin{tabular}{l@{\hspace{1em}}l}
\hline
 $\gamma_{v_av_av_av_a}=-\gamma_{e_av_be_av_bv_ae_bv_ae_b}-\gamma_{e_ae_av_av_a}$
& $\gamma_{v_bv_bv_bv_b}=-\gamma_{v_ae_bv_ae_be_av_be_av_b}-\gamma_{e_bv_bv_bv_b}$      \\
$\gamma_{e_ae_ae_ae_a}=-\gamma_{g_ae_bg_ae_be_ag_be_ag_b}-\gamma_{g_ag_ae_ae_a}$
& $\gamma_{e_be_be_be_b}=-\gamma_{e_ag_be_ag_bg_ae_bg_ae_b}-\gamma_{g_bg_be_be_b}$      \\
$\gamma_{g_ag_ag_ag_a}=\gamma_{g_bg_bg_bg_b}=0$
&  \\
\hline
\end{tabular}
\caption{\label{tab4}Relation between energy tranfer rate constants and transition rate constants for the individual molecules.}
\end{table}
With the parameters given in Table~\ref{tab1} through \ref{tab4}, we have all that we need for performing the numerical simulations to illustrate the influence of the polarization on the 2DIR spectra, starting from our analytical results.

\section{Numerical simulations and quantitative analysis of the polarization effects}

In this section, we present numerical simulations that illustrate the influence of the polarization on the 2DIR molecular spectra of a system undergoing vibrational
energy transfer.
To this end, the properties of the local oscillator (LO) field involve only in heterodyne detection but not in the molecular and nonlinear optical processes should be kept constant and made as simple as possible. The phase $\Psi$ of the LO field is chosen to be zero. Besides, other environmental factores and the properties of the other laser fields have definite influence on the simulated experimental results, but not on the nature of the effects we are investigating. Therefore, they are chosen to be rather realistic while suitable for demonstrating the effect under discussion. All laser field amplitudes are normalized to 1 and their pulse durations are equal to 55~fs.
The temperature of the system is assumed to be
$T=25\;\rm^{\circ}C$.

The parameters of the molecular system are more substantial but most of them should also be fixed in order to simplify the discussion. The energy levels of the molecules are $\omega_{g_a}=\omega_{g_b}=0$, $\omega_{e_a}=810\;\text{cm}^{-1}$, $\omega_{e_b}=730\;\text{cm}^{-1}$, $\omega_{v_a}=1480\;\text{cm}^{-1}$ and $\omega_{v_b}=1380\;\text{cm}^{-1}$. (Figure~\ref{fig1} is plotted to scale.)
The values of the relaxation and dephasing constants are also fixed. For the total decay rates we have chosen $\gamma_{e_ae_ae_ae_a}=4\;\text{cm}^{-1}$,
$\gamma_{v_av_av_av_a}=6\;\text{cm}^{-1}$, $\gamma_{e_be_be_be_b}=3\;\text{cm}^{-1}$ and $\gamma_{v_bv_bv_bv_b}=5\;\text{cm}^{-1}$. The transition constants for vibrational energy transfer are chosen to be $\gamma_{e_av_be_av_bv_ae_bv_ae_b}=2.5\;\text{cm}^{-1}$, $\gamma_{g_ae_bg_ae_be_ag_be_ag_b}=2.5\;\text{cm}^{-1}$ and their reverse constants are deduced from Eqs.~\eqref{3.6}. All the other transition constants are straightforwardly obtained from the sum rules and can be evaluated by using Tables \ref{tab2}, \ref{tab3} and \ref{tab4}. All of the dephasing constants are set to $\gamma_{pd}=3\;\text{cm}^{-1}$. Finally, in Table \ref{tab5}, we summarized the expressions of the dipole moments of the total molecular system in terms of the dipole moments of the individual molecules. Notice that all the other matrix elements not indicated in Table \ref{tab5} cancel.

\begin{table}[ht]
%\centering
\begin{tabular}{l@{\hspace{2em}}l}
\hline
$\left\langle 1\left\vert\boldsymbol{\mu}\right\vert 2\right\rangle=\left\langle g_b\left\vert\boldsymbol{\mu}\right\vert e_b\right\rangle$
&
$\left\langle 1\left\vert\boldsymbol{\mu}\right\vert 3\right\rangle=\left\langle g_a\left\vert\boldsymbol{\mu}\right\vert e_a\right\rangle$
\\
$\left\langle 1\left\vert\boldsymbol{\mu}\right\vert 6\right\rangle=\left\langle g_a\left\vert\boldsymbol{\mu}\right\vert e_a\right\rangle\left\langle g_b\vert e_b\right\rangle+\left\langle g_a\vert e_a\right\rangle\left\langle g_b\left\vert\boldsymbol{\mu}\right\vert e_b\right\rangle$
\\
$\left\langle 2\left\vert\boldsymbol{\mu}\right\vert 3\right\rangle=\left\langle g_a\left\vert\boldsymbol{\mu}\right\vert e_a\right\rangle\left\langle e_b\vert g_b\right\rangle+\left\langle g_a\vert e_a\right\rangle\left\langle e_b\left\vert\boldsymbol{\mu}\right\vert g_b\right\rangle$
&
$\left\langle 2\left\vert\boldsymbol{\mu}\right\vert 4\right\rangle=\left\langle e_b\left\vert\boldsymbol{\mu}\right\vert v_b\right\rangle$
\\
$\left\langle 2\left\vert\boldsymbol{\mu}\right\vert 7\right\rangle=\left\langle g_a\left\vert\boldsymbol{\mu}\right\vert e_a\right\rangle\left\langle e_b\vert v_b\right\rangle+\left\langle g_a\vert e_a\right\rangle\left\langle e_b\left\vert\boldsymbol{\mu}\right\vert v_b\right\rangle$
&
$\left\langle 2\left\vert\boldsymbol{\mu}\right\vert 6\right\rangle=\left\langle g_a\left\vert\boldsymbol{\mu}\right\vert e_a\right\rangle$
\\
$\left\langle 3\left\vert\boldsymbol{\mu}\right\vert 5\right\rangle=\left\langle e_a\left\vert\boldsymbol{\mu}\right\vert v_a\right\rangle$
&
$\left\langle 3\left\vert\boldsymbol{\mu}\right\vert 6\right\rangle=\left\langle g_b\left\vert\boldsymbol{\mu}\right\vert e_b\right\rangle$
\\
$\left\langle 3\left\vert\boldsymbol{\mu}\right\vert 8\right\rangle=\left\langle e_a\left\vert\boldsymbol{\mu}\right\vert v_a\right\rangle\left\langle g_b\vert e_b\right\rangle+\left\langle e_a\vert v_a\right\rangle\left\langle g_b\left\vert\boldsymbol{\mu}\right\vert e_b\right\rangle$
\\
$\left\langle 4\left\vert\boldsymbol{\mu}\right\vert 6\right\rangle=\left\langle g_a\left\vert\boldsymbol{\mu}\right\vert e_a\right\rangle\left\langle v_b\vert e_b\right\rangle+\left\langle g_a\vert e_a\right\rangle\left\langle v_b\left\vert\boldsymbol{\mu}\right\vert e_b\right\rangle$
&
$\left\langle 4\left\vert\boldsymbol{\mu}\right\vert 7\right\rangle=\left\langle g_a\left\vert\boldsymbol{\mu}\right\vert e_a\right\rangle$
\\
$\left\langle 5\left\vert\boldsymbol{\mu}\right\vert 6\right\rangle=\left\langle v_a\left\vert\boldsymbol{\mu}\right\vert e_a\right\rangle\left\langle g_b\vert e_b\right\rangle+\left\langle v_a\vert e_a\right\rangle\left\langle g_b\left\vert\boldsymbol{\mu}\right\vert e_b\right\rangle$
&
$\left\langle 5\left\vert\boldsymbol{\mu}\right\vert 8\right\rangle=\left\langle g_b\left\vert\boldsymbol{\mu}\right\vert e_b\right\rangle$
\\
$\left\langle 6\left\vert\boldsymbol{\mu}\right\vert 7\right\rangle=\left\langle e_b\left\vert\boldsymbol{\mu}\right\vert v_b\right\rangle$
&
$\left\langle 6\left\vert\boldsymbol{\mu}\right\vert 8\right\rangle=\left\langle e_a\left\vert\boldsymbol{\mu}\right\vert v_a\right\rangle$
\\
$\left\langle 6\left\vert\boldsymbol{\mu}\right\vert 9\right\rangle=\left\langle e_a\left\vert\boldsymbol{\mu}\right\vert v_a\right\rangle\left\langle e_b\vert v_b\right\rangle+\left\langle e_a\vert v_a\right\rangle\left\langle e_b\left\vert\boldsymbol{\mu}\right\vert v_b\right\rangle$
\\
$\left\langle 7\left\vert\boldsymbol{\mu}\right\vert 8\right\rangle=\left\langle e_a\left\vert\boldsymbol{\mu}\right\vert v_a\right\rangle\left\langle v_b\vert e_b\right\rangle+\left\langle e_a\vert v_a\right\rangle\left\langle v_b\left\vert\boldsymbol{\mu}\right\vert e_b\right\rangle$
&
$\left\langle 7\left\vert\boldsymbol{\mu}\right\vert 9\right\rangle=\left\langle e_a\left\vert\boldsymbol{\mu}\right\vert v_a\right\rangle$
\\
$\left\langle 8\left\vert\boldsymbol{\mu}\right\vert 9\right\rangle=\left\langle e_b\left\vert\boldsymbol{\mu}\right\vert v_b\right\rangle$
\\
\hline
\end{tabular}
\caption{\label{tab5}  Relation between dipole moments of the total system and individual dipole moments.}
\end{table}

The simulations presented below enable us to discuss the polarization effects, resulting from the rotational diffusion of donor and acceptor of an energy transfer process, on the 2DIR spectra.
We mainly focus our simulations on the frequency range related to the vibrational energy transfer between the two lowest vibrational excited states.
\begin{figure}[ht]
\includegraphics[clip]{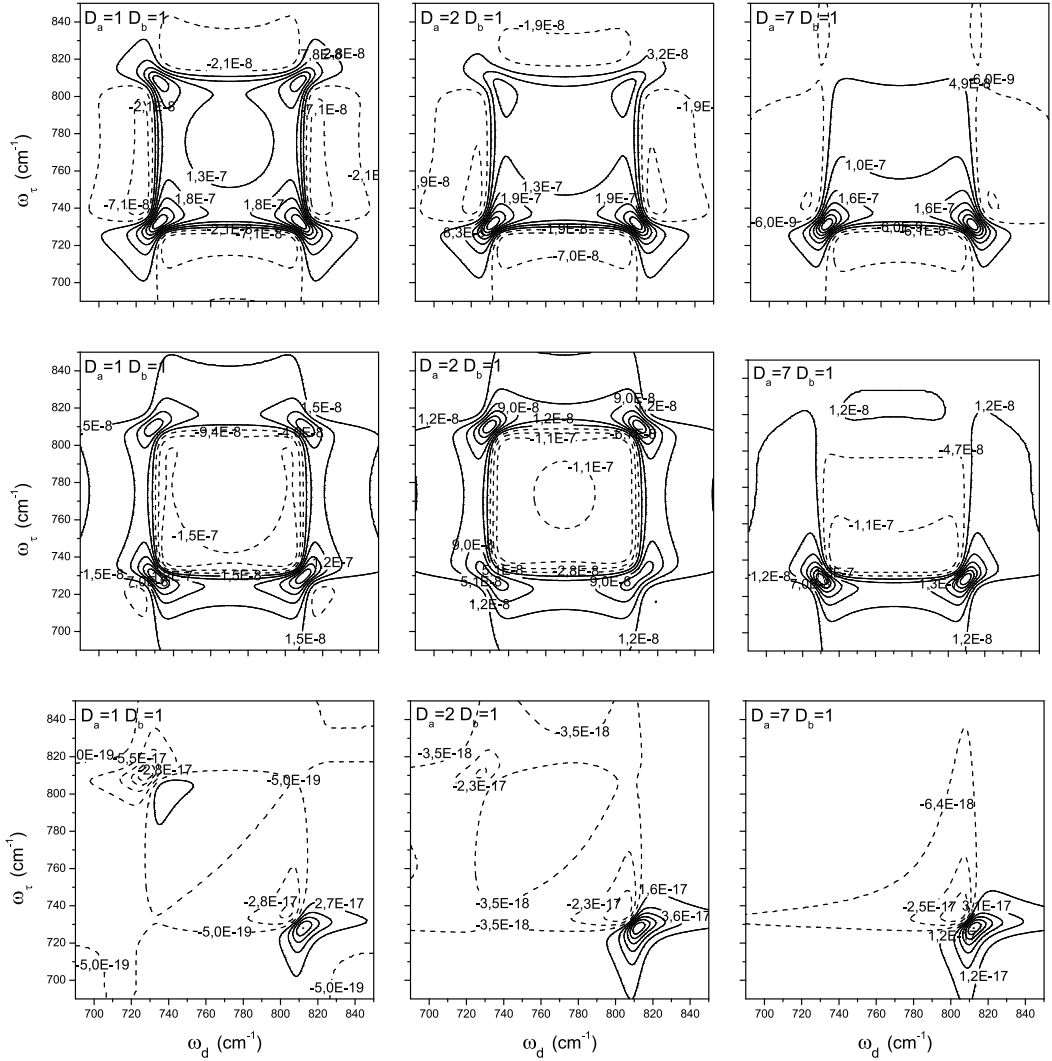}
\caption{The simulated 2DIR spectra when the laser field $a$ (rephasing term) or $b$ (non-rephasing term) acts as the first, the second or the third interaction. They correspond to the first, the second or the third row of panels, respectively. The waiting time between the two last interacting laser pulses, defined after Eq.~\eqref{2.5}, corresponds to $T=10\;{\rm ps}$. The frequencies of the exciting laser pulses are identical and correspond to $\omega_a=\omega_b=\omega_c=770\;\text{cm}^{-1}$ .}
\label{fig4}
\end{figure}

The first set of simulations are shown in Fig.~\ref{fig4}.
The three panels in the first row correspond to the cases where the laser field $a$ for the rephasing signal or $b$ for the non-rephasing signal act as the first interaction. The second and third rows are dedicated to cases where they act as the second or the third interactions, respectively. As indicated inside the panels, the three panels in the same row correspond to increasing values of the rotational diffusion constant of the molecule A, from left to right. Comparing the panels in the same row, a gradual smoothing of the peak structure can be observed, and this trend is more pronounced for the upper diagonal peak and cross peak. Since the effect is due to the increase of the rotational diffusion constant of molecule A, the upper peaks are more strongly influenced because the internal dynamics of molecule A plays a more relevant role in the building up of these peaks.
This observation is also true for the other cases (row 2 and 3). In addition, in the third row, there is no diagonal peaks in all panels because only coherences contribute in this case when fields $a$ or $b$ act as the third interaction. This can be seen in Supplement A and B.
Notice that the main axes of the contour lines of all the peaks in the panels in the upper row are rotated by $\pi/2$ with respect to those corresponding main axes in the second and third rows.
\begin{figure}[ht]
\includegraphics[clip]{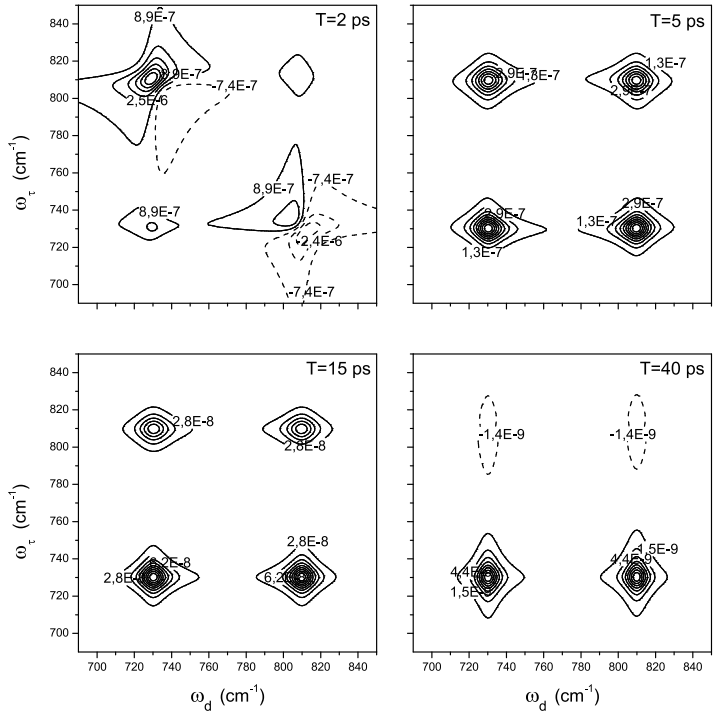}
\caption{2DIR spectra simulated with different waiting time T as marked inside each panel. The frequencies of the exciting laser pulses are fixed at
$\omega_a=\omega_b=\omega_c=770\;\text{cm}^{-1}$ to couple both transitions $\left\vert g_a\right\rangle \rightarrow \left\vert e_a \right\rangle$ and $\left\vert g_b\right\rangle \rightarrow \left\vert e_b \right\rangle$. The rotational diffusion constants are equal to $D_a=D_b=1\;\text{rad}^2/\text{ps}$. }
\label{fig5}
\end{figure}

Next, in Fig.5, we present the 2DIR spectra for increasing waiting times, ranging from 2 to 40~picoseconds.
A global decrease of the diagonal and cross peaks with increasing waiting time is observed, as expected. In these four simulation, the rotational diffusion constants are fixed. Therefore, the stronger damping observed in the upper diagonal and cross peaks can only be attributed to the vibrational energy transfer which is more efficient for transition from molecule A to B than that from B to A.

The next thing to further investigate is the dependence of the 2DIR spectra on the rotational diffusion constants of molecules A and B.
\begin{figure}[ht]
\includegraphics[clip]{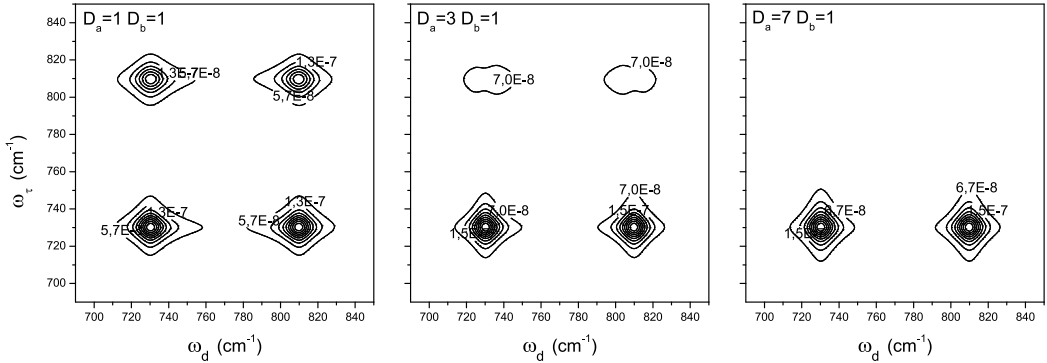}
\caption{Two-dimensional spectra obtained for different rotational diffusion constants of molecule A. The waiting time corresponds to T=10$\;$ps and the frequencies of the various laser pulses are chosen as $\omega_a=\omega_b=\omega_c=770\;\text{cm}^{-1}$.}
\label{fig6}
\end{figure}
In Fig.~\ref{fig6}, we present the simulated spectra with different $D_a$ values but with constant $D_b$ value.
The values are marked inside each panel. When $D_a=1$, both diagonal peaks and both cross peaks are clearly observed. When $D_a$ is increased to an intermediate value, say 3 as in the middle panel, a significant decrease of the upper cross and diagonal peaks can be noticed. For even faster diffusion, say $D_a=7$ as in the right panel, these peaks are completely washed-out.
\begin{figure}[ht]
\includegraphics[clip]{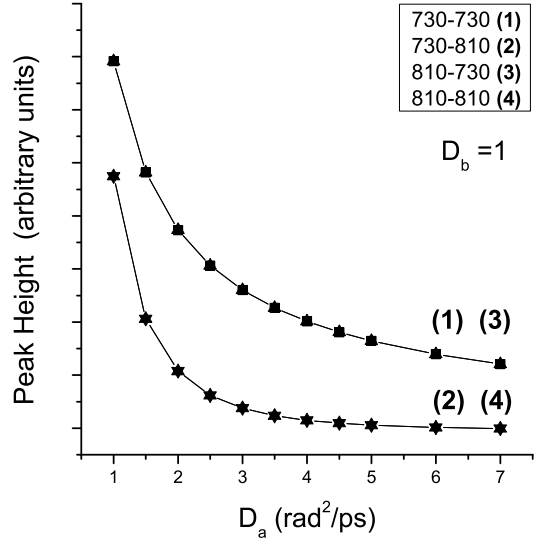}
\caption{Dependence of the peak heights with the rotational diffusion constant $D_a$. Peaks (1) and (3) exhibit similar variations and the same is true for peaks (2) and (4).}
\label{fig7}
\end{figure}
To better visualize these variations, we plot the peak heights as a function of $D_a$. From Fig.~\ref{fig7}, we see that lower peaks (1) and (3) have similar dependence, while upper peaks (2) and (4) share another trend. The function forms of thetwo trends are not much different, although the upper peaks diminish with increasing $D_a$ more rapidly than do the lower peaks.  

\begin{figure}[ht]
\includegraphics[clip]{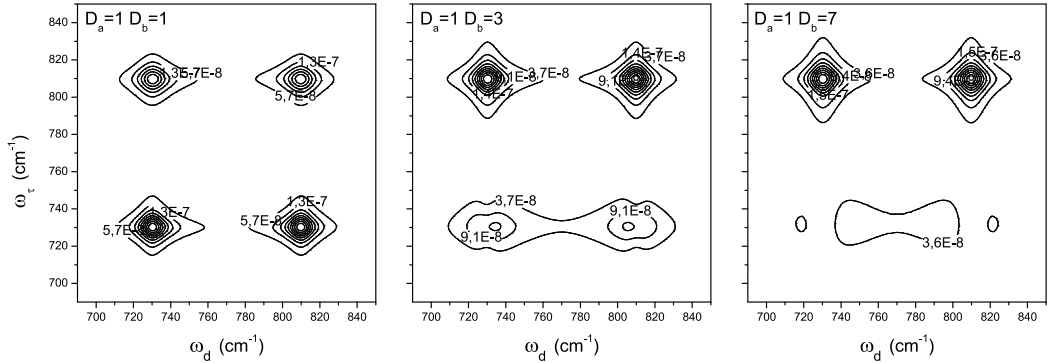}
\caption{Simulated 2DIR spectra with different values of $D_b$ and a fixed value of $D_a$. Other parameters used are identical to the ones used in the previous simulations.}
\label{fig8}
\end{figure}

Finally, in Fig.~\ref{fig8}, analogous simulations with different values of $D_b$ and fixed value of $D_a$ are presented. Features similar to those of the previous case are observed, but the trends of the upper and lower peaks are reversed. The changes in the heights of the upper diagonal and cross peaks are much less pronounced among these cases, while the heights of the lower peaks are now significantly reduced with increasing $D_b$. Obviously, this is because the internal dynamics of molecule B plays a more important role in the building up of the lower peaks and it has much weaker effects on the upper peaks.
\begin{figure}[ht]
\includegraphics[clip]{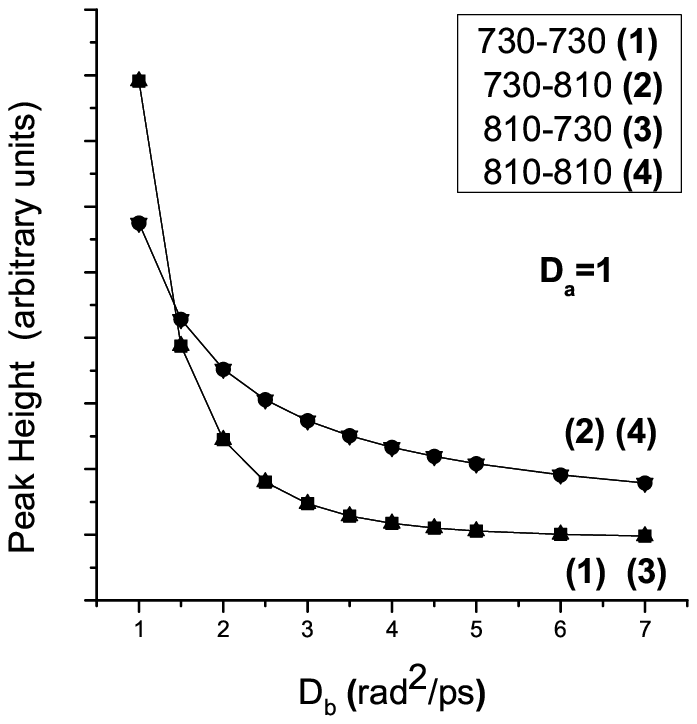}
\caption{Dependence of the peak heights on the rotational diffusion constant $D_b$.}
\label{fig9}
\end{figure}
We also plot the peak heights as functions of $D_b$ in Fig.~\ref{fig9}. This time, the heights of the lower peaks in Fig.~\ref{fig8} decrease much more rapidly with increasing value of $D_b$ than the heights of the upper peaks do. A crossing over of the relative heights can be seen.

\section{Conclusion}

In this work, we have attempted to tease out the influence of the polarization effects on intermolecular vibrational energy transfer. In the case of intermolecular processes monitored with 2DIR spectroscopy, the donor and acceptor molecules usually have different rotational diffusion constants. Then, it is important to understand how the relative values of their rotational diffusion constants affect the resulting 2DIR spectra which supposed to help identify and characterize the intermolecular vibrational energy transfer. Variations of the peak heights reflect the interplay between the donor and acceptor molecules participating in the process.

\bibliography{biblio}

\setcounter{table}{0}
\renewcommand{\appendixname}{Supplement}
\renewcommand{\thetable}{S.\Roman{table}}

\newpage
\appendix
\section{Pathways contributing to the 2DIR spectrum}

In this supplement section, we list the pathways contributing to the 2DIR spectra of the molecules undergoing intermolecular vibrational energy transfer with heterodyne detection in the directions   $\vec{k}_{\rm s}=-\vec{k}_a+\vec{k}_b+\vec{k}_c$ and $\vec{k}_{\rm s}=-\vec{k}_b+\vec{k}_a+\vec{k}_c$.
Only pathways satisfying the rotating wave approximation are retained.
There are totally 66 of them.
The superscripts $\left(\pm\right)$ indicate the directions $\left(\pm\vec{k}_j\right)$ with $j=a,b,c$.
In the following tables, the 66 pathways were separated into three tables, according to whether the field $a$ or $b$ acts in the first, the second and the third interaction terms.
Both rephasing and non-rephasing geometries were considered.
Since each pathway contributes to a specific term among those heterodyne signals
$\left\langle\Pi^{\rm X,IJK}\left(t\right)\right\rangle_{\text{or}}$, in the last column, the parameters X and IJK corresponding to that pathway are shown.
Notice that X runs from I to IV while the molecular interaction sequences IJK belongs to the set $\left\{\rm AAA, AAB, ABA,\cdots,BBB, BBA, BAB,\cdots\right\}$.
In all three tables, the second column is the index of the density matrix element of the final sate of the specific pathway. The ninth column is the index of the density matrix element of the initial state. In all of the pathways studied, the initial states are the same state. In column 8, 6 and 4, system-laser interactions probably change the population or the coherence of the system. In column 8 of row 1, for example, the index 1211 indicates that the photon interrupt the initial population in ground state to form a coherence between the ground state $\left(\left|1\right\rangle\right)$ and one of the excited state $\left(\left|2\right\rangle\right)$.
The processes in column 3, 5 and 7 are propagations of the system itself. In the pathways studied in this work, these are either the sustain of the coherence (say the coherence 21, in column 7 of row 1 of Table.~\ref{tab_6}), or the decay of population (say the decay from population 22 to population 33 in column 5 of row 4 of Table.~\ref{tab_6}).
Thus, in each pathway (row), the state (population or coherence) of the system propagates in time from column 9 to column 8, then to column 7, and so on, until column 2, following the change of the index of the density matrix element.
Some of the cells are left empty, if the content of it is the same as the cell above it.
\newpage
\setlength{\LTcapwidth}{0.9\linewidth}
\begin{center}
\begin{longtable*}{ccccccccccr}
\caption[Field $a$ or $b$ acting first]{Pathways participating in the 2DIR spectrum of the intermolecular vibrational energy transfer with the field $a$ or $b$ acting as the first interaction term. In columns 5 through 9, an empty cell means that its value is the same as the last non-empty cell above it in the same column.}
\label{tab_6}\\
\hline  \vspace{0.2em}\\
\scriptsize Path &
\scriptsize $\boldsymbol{\rho}\left(t\right)$ &
\scriptsize $\boldsymbol{G}\left(t-\tau_3\right)$ &
\scriptsize $\boldsymbol{L}_{v[p]}\left(\tau_3\right)$ &
\scriptsize $\boldsymbol{G}\left(\tau_3-\tau_2\right)$ &
\scriptsize $\boldsymbol{L}_{v[q]}\left(\tau_2\right)$ &
\scriptsize $\boldsymbol{G}\left(\tau_2-\tau_1\right)$ &
\scriptsize $\boldsymbol{L}_{v[a,b]}\left(\tau_1\right)$ &
\scriptsize $\boldsymbol{\rho}\left(t_0\right)$ &
Mol. dipoles &
X,IJK \\ \vspace{0.2em}\\
\hline
\endfirsthead
\multicolumn{11}{c}{{\bfseries \tablename\ \thetable{} -- continued from previous page}} \\
\vspace{0.2em} \\
\hline \scriptsize
Path & $\boldsymbol{\rho}\left(t\right)$ &
\scriptsize $\boldsymbol{G}\left(t-\tau_3\right)$ &
\scriptsize $\boldsymbol{L}_{v[p]}\left(\tau_3\right)$ &
\scriptsize $\boldsymbol{G}\left(\tau_3-\tau_2\right)$ &
\scriptsize $\boldsymbol{L}_{v[q]}\left(\tau_2\right)$ &
\scriptsize $\boldsymbol{G}\left(\tau_2-\tau_1\right)$ &
\scriptsize $\boldsymbol{L}_{v[a,b]}\left(\tau_1\right)$ &
\scriptsize $\boldsymbol{\rho}\left(t_0\right)$ &
Mol. dipoles &
X,IJK \\ \vspace{0.2em} \\
\hline
\endhead
\hline
\multicolumn{11}{|r|}{{Continued on next page}} \\
\hline
\endfoot
\vspace{0.2em} \\
\hline
\endlastfoot
\vspace{0.2em}\\$1$ & $21$ & $2121$ & $2122^{\left(+\right)}$ & $2222$ & $2212^{\left(+\right)}$ & $1212$ & $1211^{\left(-\right)}$ & $11$ & $\vec{\mu}_{e_bg_b}\vec{\mu}_{e_bg_b}\vec{\mu}_{g_be_b}$ & I,BBB\\
$2$ & $42$ & $4242$ & $4222^{\left(+\right)}$ & & & & & & $\vec{\mu}_{v_be_b}\vec{\mu}_{e_bg_b}\vec{\mu}_{g_be_b}$ & I,BBB\\
$3$ & $62$ & $6262$ & $6222^{\left(+\right)}$ & & & & & & $\vec{\mu}_{e_ag_a}\vec{\mu}_{e_bg_b}\vec{\mu}_{g_be_b}$ & II,ABB\\
$4$ & $31$ & $3131$ & $3133^{\left(+\right)}$ & $3322$ & & & & & $\vec{\mu}_{e_ag_a}\vec{\mu}_{e_bg_b}\vec{\mu}_{g_be_b}$ & II,ABB\\
$5$ & $63$ & $6363$ & $6333^{\left(+\right)}$ & & & & & & $\vec{\mu}_{e_bg_b}\vec{\mu}_{e_bg_b}\vec{\mu}_{g_be_b}$ & I,BBB\\
$6$ & $53$ & $5353$ & $5333^{\left(+\right)}$ & & & & & & $\vec{\mu}_{v_ae_a}\vec{\mu}_{e_bg_b}\vec{\mu}_{g_be_b}$ & II,ABB\\
$7$ & $21$ & $2121$ & $2111^{\left(+\right)}$ & $1122$ & & & & & $\vec{\mu}_{e_bg_b}\vec{\mu}_{e_bg_b}\vec{\mu}_{g_be_b}$ & I,BBB\\
$8$ & $31$ & $3131$ & $3111^{\left(+\right)}$ & & & & & & $\vec{\mu}_{e_ag_a}\vec{\mu}_{e_bg_b}\vec{\mu}_{g_be_b}$ & II,ABB\\
$9$ & $31$ & $3131$ & $3132^{\left(+\right)}$ & $3232$ & $3212^{\left(+\right)}$ & & & & $\vec{\mu}_{e_bg_b}\vec{\mu}_{e_ag_a}\vec{\mu}_{g_be_b}$ & IV,BAB\\
$10$ & $62$ & $6262$ & $6232^{\left(+\right)}$ & & & & & & $\vec{\mu}_{e_bg_b}\vec{\mu}_{e_ag_a}\vec{\mu}_{g_be_b}$ & IV,BAB\\
$11$ & $52$ & $5252$ & $5232^{\left(+\right)}$ & & & & & & $\vec{\mu}_{v_ae_a}\vec{\mu}_{e_ag_a}\vec{\mu}_{g_be_b}$ & III,AAB\\
$12$ & $21$ & $2121$ & $2111^{\left(+\right)}$ & $1111$ & $1112^{\left(+\right)}$ & & & & $\vec{\mu}_{e_bg_b}\vec{\mu}_{e_bg_b}\vec{\mu}_{g_be_b}$ & I,BBB\\
$13$ & $31$ & $3131$ & $3111^{\left(+\right)}$ & & & & & & $\vec{\mu}_{e_ag_a}\vec{\mu}_{e_bg_b}\vec{\mu}_{g_be_b}$ & II,ABB\\
$14$ & $31$ & $3131$ & $3133^{\left(+\right)}$ & $3333$ & $3313^{\left(+\right)}$ & $1313$ & $1311^{\left(-\right)}$ & & $\vec{\mu}_{e_ag_a}\vec{\mu}_{e_ag_a}\vec{\mu}_{g_ae_a}$ & I,AAA\\
$15$ & $63$ & $6363$ & $6333^{\left(+\right)}$ & & & & & & $\vec{\mu}_{e_bg_b}\vec{\mu}_{e_ag_a}\vec{\mu}_{g_ae_a}$ & II,BAA\\
$16$ & $53$ & $5353$ & $5333^{\left(+\right)}$ & & & & & & $\vec{\mu}_{v_ae_a}\vec{\mu}_{e_ag_a}\vec{\mu}_{g_ae_a}$ & I,AAA\\
$17$ & $21$ & $2121$ & $2122^{\left(+\right)}$ & $2233$ & & & & & $\vec{\mu}_{e_bg_b}\vec{\mu}_{e_ag_a}\vec{\mu}_{g_ae_a}$ & II,BAA\\
$18$ & $42$ & $4242$ & $4222^{\left(+\right)}$ & & & & & & $\vec{\mu}_{v_be_b}\vec{\mu}_{e_ag_a}\vec{\mu}_{g_ae_a}$ & II,BAA\\
$19$ & $62$ & $6262$ & $6222^{\left(+\right)}$ & & & & & & $\vec{\mu}_{e_ag_a}\vec{\mu}_{e_ag_a}\vec{\mu}_{g_ae_a}$ & I,AAA\\
$20$ & $21$ & $2121$ & $2111^{\left(+\right)}$ & $1133$ & & & & & $\vec{\mu}_{e_bg_b}\vec{\mu}_{e_ag_a}\vec{\mu}_{g_ae_a}$ & II,BAA\\
$21$ & $31$ & $3131$ & $3111^{\left(+\right)}$ & & & & & & $\vec{\mu}_{e_ag_a}\vec{\mu}_{e_ag_a}\vec{\mu}_{g_ae_a}$ & I,AAA\\
$22$ & $21$ & $2121$ & $2111^{\left(+\right)}$ & $1111$ & $1113^{\left(+\right)}$ & & & & $\vec{\mu}_{e_bg_b}\vec{\mu}_{e_ag_a}\vec{\mu}_{g_ae_a}$ & II,BAA\\
$23$ & $31$ & $3131$ & $3111^{\left(+\right)}$ & & & & & & $\vec{\mu}_{e_ag_a}\vec{\mu}_{e_ag_a}
\vec{\mu}_{g_ae_a}$ & I,AAA\\
$24$ & $21$ & $2121$ & $2123^{\left(+\right)}$ & $2323$ & $2313^{\left(+\right)}$ & & & & $\vec{\mu}_{e_ag_a}\vec{\mu}_{e_bg_b}
\vec{\mu}_{g_ae_a}$ & IV,ABA\\
$25$ & $43$ & $4343$ & $4323^{\left(+\right)}$ & & & & & & $\vec{\mu}_{v_be_b}\vec{\mu}_{e_bg_b}
\vec{\mu}_{g_ae_a}$ & III,BBA\\
$26$ & $63$ & $6363$ & $6323^{\left(+\right)}$ & & & & & & $\vec{\mu}_{e_ag_a}\vec{\mu}_{e_bg_b}
\vec{\mu}_{g_ae_a}$ & IV,ABA\\
\end{longtable*}
\end{center}

\newpage

\begin{center}
\begin{longtable*}{ccccccccccr}
\caption[Field $a$ or $b$ acting second]{Pathways participating in the 2DIR spectrum of the intermolecular energy transfer process with the field $a$ or $b$ acting in the second interaction term. Other quantities are the same like in previous table.\label{tab_7}}\\
\hline \vspace{0.2em} \\
\scriptsize Path &
\scriptsize $\boldsymbol{\rho}\left(t\right)$ &
\scriptsize $\boldsymbol{G}\left(t-\tau_3\right)$ &
\scriptsize $\boldsymbol{L}_{v[p]}\left(\tau_3\right)$ &
\scriptsize $\boldsymbol{G}\left(\tau_3-\tau_2\right)$ &
\scriptsize $\boldsymbol{L}_{v[a,b]}\left(\tau_2\right)$ &
\scriptsize $\boldsymbol{G}\left(\tau_2-\tau_1\right)$ &
\scriptsize $\boldsymbol{L}_{v[r]}\left(\tau_1\right)$ &
\scriptsize $\boldsymbol{\rho}\left(t_0\right)$ &
Mol. dipoles &
X,IJK \\ \vspace{0.2em} \\
\hline
\endfirsthead
\multicolumn{11}{c}{{\bfseries \tablename\ \thetable{} -- continued from previous page}} \\
\hline \vspace{0.2em} \\
\scriptsize Path &
\scriptsize $\boldsymbol{\rho}\left(t\right)$ &
\scriptsize $\boldsymbol{G}\left(t-\tau_3\right)$ &
\scriptsize $\boldsymbol{L}_{v[p]}\left(\tau_3\right)$ &
\scriptsize $\boldsymbol{G}\left(\tau_3-\tau_2\right)$ &
\scriptsize $\boldsymbol{L}_{v[a,b]}\left(\tau_2\right)$ &
\scriptsize $\boldsymbol{G}\left(\tau_2-\tau_1\right)$ &
\scriptsize $\boldsymbol{L}_{v[r]}\left(\tau_1\right)$ &
\scriptsize $\boldsymbol{\rho}\left(t_0\right)$ &
Mol. dipoles &
X,IJK \\ \vspace{0.2em} \\
\hline
\endhead
\hline
\multicolumn{11}{|r|}{{Continued on next page}} \\
\hline
\endfoot \vspace{0.2em} \\
\hline
\endlastfoot
\vspace{0.2em}\\$27$ & $21$ & $2121$ & $2123^{\left(+\right)}$ & $2323$ & $2321^{\left(-\right)}$ & $2121$ & $2111^{\left(+\right)}$ & $11$ & $\vec{\mu}_{e_ag_a}\vec{\mu}_{g_ae_a}
\vec{\mu}_{e_bg_b}$ & III,AAB \\
$28$ & $43$ & $4343$ & $4323^{\left(+\right)}$ & & & & & & $\vec{\mu}_{v_be_b}\vec{\mu}_{g_ae_a}
\vec{\mu}_{e_bg_b}$ & IV,BAB \\
$29$ & $63$ & $6363$ & $6323^{\left(+\right)}$ & & & & & & $\vec{\mu}_{e_ag_a}\vec{\mu}_{g_ae_a}
\vec{\mu}_{e_bg_b}$ & III,AAB \\
$30$ & $21$ & $2121$ & $2122^{\left(+\right)}$ & $2222$ & $2221^{\left(-\right)}$ & &  & & $\vec{\mu}_{e_bg_b}\vec{\mu}_{g_be_b} \vec{\mu}_{e_bg_b}$ & I,BBB \\
$31$ & $42$ & $4242$ & $4222^{\left(+\right)}$ & & & & & & $\vec{\mu}_{v_be_b}\vec{\mu}_{g_be_b}
\vec{\mu}_{e_bg_b}$ & I,BBB \\
$32$ & $62$ & $6262$ & $6222^{\left(+\right)}$ & & & & & & $\vec{\mu}_{e_ag_a}\vec{\mu}_{g_be_b}
\vec{\mu}_{e_bg_b}$ & II,ABB \\
$33$ & $31$ & $3131$ & $3133^{\left(+\right)}$ & $3322$ & & & & & $\vec{\mu}_{e_ag_a}\vec{\mu}_{g_be_b}
\vec{\mu}_{e_bg_b}$ & II,ABB \\
$34$ & $63$ & $6363$ & $6333^{\left(+\right)}$ & & & & & & $\vec{\mu}_{e_bg_b}\vec{\mu}_{g_be_b}
\vec{\mu}_{e_bg_b}$ & I,BBB \\
$35$ & $53$ & $5353$ & $5333^{\left(+\right)}$ & & & & & & $\vec{\mu}_{v_ae_a}\vec{\mu}_{g_be_b}
\vec{\mu}_{e_bg_b}$ & II,ABB \\
$36$ & $21$ & $2121$ & $2111^{\left(+\right)}$ & $1122$ & & & & & $\vec{\mu}_{e_bg_b}\vec{\mu}_{g_be_b}
\vec{\mu}_{e_bg_b}$ & I,BBB \\
$37$ & $31$ & $3131$ & $3111^{\left(+\right)}$ & & & & & & $\vec{\mu}_{e_ag_a}\vec{\mu}_{g_be_b}
\vec{\mu}_{e_bg_b}$ & II,ABB \\
$38$ & $21$ & $2121$ & $2111^{\left(+\right)}$ & $1111$ & $1121^{\left(-\right)}$ & & & & $\vec{\mu}_{e_bg_b} \vec{\mu}_{g_be_b}\vec{\mu}_{e_bg_b}$ & I,BBB \\
$39$ & $31$ & $3131$ & $3111^{\left(+\right)}$ & & & & & & $\vec{\mu}_{e_ag_a}\vec{\mu}_{g_be_b}
\vec{\mu}_{e_bg_b}$ & II,ABB \\
$40$ & $31$ & $3131$ & $3133^{\left(+\right)}$ & $3333$ & $3331^{\left(-\right)}$ & $3131$ & $3111^{\left(+\right)}$ & & $\vec{\mu}_{e_ag_a} \vec{\mu}_{g_ae_a}\vec{\mu}_{e_ag_a}$ & I,AAA \\
$41$ & $63$ & $6363$ & $6333^{\left(+\right)}$ & & & & & & $\vec{\mu}_{e_bg_b}\vec{\mu}_{g_ae_a}
\vec{\mu}_{e_ag_a}$ & II,BAA \\
$42$ & $53$ & $5353$ & $5333^{\left(+\right)}$ & & & & & & $\vec{\mu}_{v_ae_a}\vec{\mu}_{g_ae_a}
\vec{\mu}_{e_ag_a}$ & I,AAA \\
$43$ & $21$ & $2121$ & $2122^{\left(+\right)}$ & $2233$ & & & & & $\vec{\mu}_{e_bg_b}\vec{\mu}_{g_ae_a}
\vec{\mu}_{e_ag_a}$ & II,BAA \\
$44$ & $42$ & $4242$ & $4222^{\left(+\right)}$ & & & & & & $\vec{\mu}_{v_be_b}\vec{\mu}_{g_ae_a}
\vec{\mu}_{e_ag_a}$ & II,BAA \\
$45$ & $62$ & $6262$ & $6222^{\left(+\right)}$ & & & & & & $\vec{\mu}_{e_ag_a}\vec{\mu}_{g_ae_a}
\vec{\mu}_{e_ag_a}$ & I,AAA \\
$46$ & $21$ & $2121$ & $2111^{\left(+\right)}$ & $1133$ & & & & & $\vec{\mu}_{e_bg_b}\vec{\mu}_{g_ae_a}
\vec{\mu}_{e_ag_a}$ & II,BAA \\
$47$ & $31$ & $3131$ & $3111^{\left(+\right)}$ & & & & & & $\vec{\mu}_{e_ag_a}\vec{\mu}_{g_ae_a}
\vec{\mu}_{e_ag_a}$ & I,AAA \\
$48$ & $31$ & $3131$ & $3132^{\left(+\right)}$ & $3232$ & $3231^{\left(-\right)}$ & & & & $\vec{\mu}_{e_bg_b} \vec{\mu}_{g_be_b} \vec{\mu}_{e_ag_a}$ & III,BBA \\
$49$ & $62$ & $6262$ & $6232^{\left(+\right)}$ & & & & & & $\vec{\mu}_{e_bg_b}\vec{\mu}_{g_be_b}
\vec{\mu}_{e_ag_a}$ & III,BBA \\
$50$ & $52$ & $5252$ & $5232^{\left(+\right)}$ & & & & & & $\vec{\mu}_{v_ae_a}\vec{\mu}_{g_be_b}
\vec{\mu}_{e_ag_a}$ & IV,ABA \\
$51$ & $21$ & $2121$ & $2111^{\left(+\right)}$ & $1111$ & $1131^{\left(-\right)}$ & & & & $\vec{\mu}_{e_bg_b} \vec{\mu}_{g_ae_a} \vec{\mu}_{e_ag_a}$ & II,BAA \\
$52$ & $31$ & $3131$ & $3111^{\left(+\right)}$ & & & & & & $\vec{\mu}_{e_ag_a}\vec{\mu}_{g_ae_a}
\vec{\mu}_{e_ag_a}$ & I,AAA \\
\end{longtable*}
\end{center}

\newpage
\begin{center}
\begin{longtable*}{ccccccccccr}
\caption[Field $a$ or $b$ acting third]{Pathways participating in the 2DIR spectrum of the intermolecular energy transfer process with the field $a$ or $b$ acting in the third interaction term. Other quantities are
the same like in previous tables.\label{tab_8}}\\
\hline \vspace{0.2em}\\
\scriptsize Path &
\scriptsize $\boldsymbol{\rho}\left(t\right)$ &
\scriptsize $\boldsymbol{G}\left(t-\tau_3\right)$ &
\scriptsize $\boldsymbol{L}_{v[a,b]}\left(\tau_3\right)$ &
\scriptsize $\boldsymbol{G}\left(\tau_3-\tau_2\right)$ &
\scriptsize $\boldsymbol{L}_{v[q]}\left(\tau_2\right)$ &
\scriptsize $\boldsymbol{G}\left(\tau_2-\tau_1\right)$ &
\scriptsize $\boldsymbol{L}_{v[r]}\left(\tau_1\right)$ &
\scriptsize $\boldsymbol{\rho}\left(t_0\right)$ &
Mol. dipoles &
X,IJK \\ \vspace{0.2em}\\
\hline
\endfirsthead
\multicolumn{11}{c}{{\bfseries \tablename\ \thetable{} -- continued from previous page}} \\
\hline \vspace{0.2em}\\
\scriptsize Path &
\scriptsize $\boldsymbol{\rho}\left(t\right)$ &
\scriptsize $\boldsymbol{G}\left(t-\tau_3\right)$ &
\scriptsize $\boldsymbol{L}_{v[a,b]}\left(\tau_3\right)$ &
\scriptsize $\boldsymbol{G}\left(\tau_3-\tau_2\right)$ &
\scriptsize $\boldsymbol{L}_{v[q]}\left(\tau_2\right)$ &
\scriptsize $\boldsymbol{G}\left(\tau_2-\tau_1\right)$ &
\scriptsize $\boldsymbol{L}_{v[r]}\left(\tau_1\right)$ &
\scriptsize $\boldsymbol{\rho}\left(t_0\right)$ &
Mol. dipoles &
X,IJK \\ \vspace{0.2em}\\
\hline
\endhead
\hline
\multicolumn{11}{|r|}{{Continued on next page}} \\
\hline
\endfoot \vspace{0.2em} \\
\hline
\endlastfoot
\vspace{0.2em}\\${53}$ & ${21}$ & ${2121}$ & ${2141^{\left(-\right)}}$ & ${4141}$ & ${4121^{\left(+\right)}}$ & ${2121}$ & ${2111^{\left(+\right)}}$ & ${11}$ & ${\vec{\mu}_{e_bv_b}\vec{\mu}_{v_be_b}
\vec{\mu}_{e_bg_b}}$ & I,BBB\\
${54}$ & ${42}$ & ${4242}$ & ${4241^{\left(-\right)}}$ & & & & & & ${\vec{\mu}_{g_be_b}\vec{\mu}_{v_be_b}
\vec{\mu}_{e_bg_b}}$ & I,BBB\\
${55}$ & ${43}$ & ${4343}$ & ${4341^{\left(-\right)}}$ & & & & & & ${\vec{\mu}_{g_ae_a}\vec{\mu}_{v_be_b}
\vec{\mu}_{e_bg_b}}$ & II,ABB \\
${56}$ & ${21}$ & ${2121}$ & ${2161^{\left(-\right)}}$ & ${6161}$ & ${6121^{\left(+\right)}}$ & & & & ${\vec{\mu}_{g_ae_a}\vec{\mu}_{e_ag_a} \vec{\mu}_{e_bg_b}}$ & III,AAB \\
${57}$ & ${31}$ & ${3131}$ & ${3161^{\left(-\right)}}$ & & & & & & ${\vec{\mu}_{g_be_b}\vec{\mu}_{e_ag_a}
\vec{\mu}_{e_bg_b}}$ & IV,BAB \\
${58}$ & ${62}$ & ${6262}$ & ${6251^{\left(-\right)}}$ & & & & & & ${\vec{\mu}_{g_be_b}\vec{\mu}_{e_ag_a}
\vec{\mu}_{e_bg_b}}$ & IV,BAB \\
${59}$ & ${63}$ & ${6363}$ & ${6361^{\left(-\right)}}$ & & & & & & ${\vec{\mu}_{g_ae_a}\vec{\mu}_{e_ag_a}
\vec{\mu}_{e_bg_b}}$ & III,AAB \\
${60}$ & ${31}$ & ${3131}$ & ${3151^{\left(-\right)}}$ & ${5151}$ & ${5131^{\left(+\right)}}$ & ${3131}$ & ${3111^{\left(+\right)}}$ & & ${\vec{\mu}_{e_av_a}\vec{\mu}_{v_ae_a} \vec{\mu}_{e_ag_a}}$ & I,AAA \\
${61}$ & ${52}$ & ${5252}$ & ${5251^{\left(-\right)}}$ & & & & & & ${\vec{\mu}_{g_be_b}\vec{\mu}_{v_ae_a}
\vec{\mu}_{e_ag_a}}$ & II,BAA \\
${62}$ & ${53}$ & ${5353}$ & ${5351^{\left(-\right)}}$ & & & & & & ${\vec{\mu}_{g_ae_a}\vec{\mu}_{v_ae_a}
\vec{\mu}_{e_ag_a}}$ & I,AAA\\
${63}$ & ${21}$ & ${2121}$ & ${2161^{\left(-\right)}}$ & ${6161}$ & ${6131^{\left(+\right)}}$ & & & & ${\vec{\mu}_{g_ae_a} \vec{\mu}_{e_bg_b}
\vec{\mu}_{e_ag_a}}$ & IV,ABA \\
${64}$ & ${31}$ & ${3131}$ & ${3161^{\left(-\right)}}$ & & & & & & ${\vec{\mu}_{g_be_b}\vec{\mu}_{e_bg_b}
\vec{\mu}_{e_ag_a}}$ & III,BBA \\
${65}$ & ${62}$ & ${6262}$ & ${6261^{\left(-\right)}}$ & & & & & & ${\vec{\mu}_{g_be_b}\vec{\mu}_{e_bg_b}
\vec{\mu}_{e_ag_a}}$ & III,BBA \\
${66}$ & ${63}$ & ${6363}$ & ${6361^{\left(-\right)}}$ & & & & & & ${\vec{\mu}_{g_ae_a}\vec{\mu}_{e_bg_b}
\vec{\mu}_{e_ag_a}}$ & IV,ABA \\
\end{longtable*}
\end{center}

\newpage

\section{Constants associated with the internal dynamics}

In Equation~\eqref{3.2}, the integrands of the Factors contributing to the individual pathways presented in Supplement~A, and participating in the 2DIR spectra of the molecules undergoing the vibrational energy transfer, are expressed in terms of some constants. In this supplement, the constants accounting for the internal dynamical associated with the rephasing contribution are presented. The corresponding constants associated with the non-rephasing terms can be deduced straightforwardly.
Notice that we have introduced the simplifying notation $\Xi^{pqr}_{ij,kl,mn}=E_pE_qE_r\mu_{ij}\mu_{kl}\mu_{mn}$.
For simplicity we have introduced the formal decomposition $\boldsymbol{G}(\tau_3-\tau_2)=\sum_{\alpha}W_{\alpha}^{iijj}\exp\left\{r_{\alpha}\left(\tau_3-\tau_2\right)\right\}$. In the tables of this supplement, $W_{\alpha}^{iijj}$ and $r_{\alpha}$ are used in the expressions.

The expressions $Q^{\text{id}}$, $K^{\text{id}}$, $A^{\text{id}}$, $B^{\text{id}}$ and $C^{\text{id}}$ depend on all or some of the parameters $n$, $\alpha$, $r$, $p$ and $q$, defined in the text. Except for $Q^{\text{id}}$, these quantities are rather simply expressed in terms of molecular parameters.
$C^{\text{id}}$, for example, has very simple dependence on its parameters. Among the 66 pathways considered, there are only 6 different values of $C^{\text{id}}$
For brevity, these quantities are arranged into six separate tables.
The first and second table contain pathways $n=1$ through $n=26$.
They correspond to the pathways in Table~\ref{tab_6}.
Further, The first table contains pathways $n=1$ through $n=13$,
and the second table contains pathways $n=14$ through $n=26$.
The $C^{\text{id}}$ values of all 13 pathways in the Table~\ref{tab_91} equal
$i\omega_{12}+\Gamma_{1212}+i\omega_{a}$, while in Table~\ref{tab_92} all of the pathways have $C^{\text{id}}=i\omega_{13}+\Gamma_{1313}+i\omega_{a}$.

By separating the pathways this way, it can be seen that $A^{\text{id}}$ and $B^{\text{id}}$ values of the pathways in the same table also share common terms. In the tables, we explicitly take out the common terms among these values to make the table more concise. These can be easily seen in the tables themselves.

The third through sixth tables of this section are compiled in the same manner, which will be apparent by comparing them to the explanation given to Tables~\ref{tab_91} and \ref{tab_92} in the previous paragraph. Table~\ref{tab_93} and \ref{tab_94} are related to pathways in Table~\ref{tab_7}, while Table~\ref{tab_95} and Table~\ref{tab_96} correspond to pathways in Table~\ref{tab_8}.

\newpage
\begin{center}
\begin{longtable*}{ccccc}
\caption[ ]{The constants contributing to the calculation of the integrands of pathways, as explained at the beginning of this section. In this table, all of the pathways have $C^{\text{id}}\left(n,\alpha,r\right)=i\omega_{12}+\Gamma_{1212}+i\omega_{a}$.
All of the ${B^{\text{id}}\left(n,\alpha,r,q\right)}$ constants have common terms $-i\omega_q-\Gamma_{1212}$, and all of the ${A^{\text{id}}\left(n,\alpha,r,q,p\right)}$ constant have a common term $-i\omega_p$.}
\label{tab_91}\\
\hline \vspace{0.2em}\\
${n}$ &
${{Q^{\text{id}}\left(n,\alpha,r,q,p\right)}}$ &
${K^{\text{id}}\left(n,\alpha,r,q,p\right)}$ &
${A^{\text{id}}\left(n,\alpha,r,q,p\right)}+i\omega_p$ &
${B^{\text{id}}\left(n,\alpha,r,q\right)}+i\omega_q+\Gamma_{1212}$ \\
\vspace{0.2em}\\
\hline
\endfirsthead \vspace{0.2em}\\
\multicolumn{5}{c}{{\bfseries \tablename\ \thetable{} -- continued from previous page}} \\
\hline \vspace{0.2em}\\
${n}$ &
${{Q^{\text{id}}\left(n,\alpha,r,q,p\right)}}$ &
${K^{\text{id}}\left(n,\alpha,r,q,p\right)}$ &
${A^{\text{id}}\left(n,\alpha,r,q,p\right)}+i\omega_p$ &
${B^{\text{id}}\left(n,\alpha,r,q\right)}+i\omega_q+\Gamma_{1212}$ \\
\vspace{0.2em}\\
\hline
\endhead \vspace{0.2em}\\
\hline
\multicolumn{5}{|r|}{{Continued on next page}} \\
\hline
\endfoot \vspace{0.2em}\\
\hline
\endlastfoot
\vspace{0.2em}\\
${1}$
& ${-W_{\alpha}^{2222}\Xi^{pqa}_{{e_bg_b},{e_bg_b},{g_be_b}}}$
& ${-i\omega_{21}-\Gamma_{2121}}$
& ${i\omega_{21}+\Gamma_{2121}+r_{\alpha}}$
& ${-r_{\alpha}-i\omega_{12}}$
\\ \vspace{0.2em}\\
${2}$
& ${W_{\alpha}^{2222}\Xi^{pqa}_{{v_be_b},{e_bg_b},{g_be_b}}}$
& ${-i\omega_{42}-\Gamma_{4242}}$
& ${i\omega_{42}+\Gamma_{4242}+r_{\alpha}}$
& ${-r_{\alpha}-i\omega_{12}}$
\\ \vspace{0.2em}\\
${3}$
& ${W_{\alpha}^{2222}\Xi^{pqa}_{{e_ag_a},{e_bg_b},{g_be_b}}}$
& ${-i\omega_{62}-\Gamma_{6262}}$
& ${i\omega_{62}+\Gamma_{6262}+r_{\alpha}}$
& ${-r_{\alpha}-i\omega_{12}}         $
\\ \vspace{0.2em}\\
${4}$
& ${-W_{\alpha}^{3322}\Xi^{pqa}_{{e_ag_a},{e_bg_b},{g_be_b}}}$
& ${-i\omega_{31}-\Gamma_{3131}}$
& ${i\omega_{31}+\Gamma_{3131}+r_{\alpha}}$
& ${-r_{\alpha}-i\omega_{12}}$
\\ \vspace{0.2em}\\
${5}$
& ${W_{\alpha}^{3322}\Xi^{pqa}_{{e_bg_b},{e_bg_b},{g_be_b}}}$
& ${-i\omega_{63}-\Gamma_{6363}}$
& ${i\omega_{63}+\Gamma_{6363}+r_{\alpha}}$
& ${-r_{\alpha}-i\omega_{12}}$
\\ \vspace{0.2em}\\
${6}$
& ${W_{\alpha}^{3322}\Xi^{pqa}_{{v_ae_a},{e_bg_b},{g_be_b}}}$
& ${-i\omega_{53}-\Gamma_{5353}}$
& ${i\omega_{53}+\Gamma_{5353}+r_{\alpha}}$
& ${-r_{\alpha}-i\omega_{12}}$
\\ \vspace{0.2em}\\
${7}$
& ${W_{\alpha}^{1122}\Xi^{pqa}_{{e_bg_b},{e_bg_b},{g_be_b}}}$
& ${-i\omega_{21}-\Gamma_{2121}}$
& ${i\omega_{21}+\Gamma_{2121}+r_{\alpha}}$
& ${-r_{\alpha}-i\omega_{12}}$
\\ \vspace{0.2em}\\
${8}$
& ${W_{\alpha}^{1122}\Xi^{pqa}_{{e_ag_a},{e_bg_b},{g_be_b}}}$
& ${-i\omega_{31}-\Gamma_{3131}}$
& ${i\omega_{31}+\Gamma_{3131}+r_{\alpha}}$
& ${-r_{\alpha}-i\omega_{12}}$
\\ \vspace{0.2em}\\
${9}$
& ${-\Xi^{pqa}_{{e_bg_b},{e_ag_a},{g_be_b}}}$
& ${-i\omega_{31}-\Gamma_{3131}}$
& ${i\omega_{21}+\Gamma_{3131}-\Gamma_{3232}}$
& ${i\omega_{31}+\Gamma_{3232}}$
\\ \vspace{0.2em}\\
${10}$
& ${\Xi^{pqa}_{{e_bg_b},{e_ag_a},{g_be_b}}}$
& ${-i\omega_{62}-\Gamma_{6262}}$
& ${i\omega_{63}+\Gamma_{6262}-\Gamma_{3232}}$
& ${i\omega_{31}+\Gamma_{3232}}$
\\ \vspace{0.2em}\\
${11}$
& ${\Xi^{pqa}_{{v_ae_a},{e_ag_a},{g_be_b}}}$
& ${-i\omega_{52}-\Gamma_{5252}}$
& ${i\omega_{53}+\Gamma_{5252}-\Gamma_{3232}}$
& ${i\omega_{31}+\Gamma_{3232}}$
\\ \vspace{0.2em}\\
${12}$
& ${-\Xi^{pqa}_{{e_bg_b},{e_bg_b},{g_be_b}}}$
& ${-i\omega_{21}-\Gamma_{2121}}$
& ${i\omega_{21}+\Gamma_{2121}}$
& ${-i\omega_{12}}$
\\ \vspace{0.2em}\\
${13}$
& ${-\Xi^{pqa}_{{e_ag_a}{e_bg_b}{g_be_b}}}$
& ${-i\omega_{31}-\Gamma_{3131}}$
& ${i\omega_{31}+\Gamma_{3131}}$
& ${-i\omega_{12}}$
\\
\end{longtable*}
\end{center}

\begin{center}
\begin{longtable*}{ccccc}
\caption[ ]{The constants contributing to the calculation of the integrands of pathways, as explained at the beginning of this section. In this table, all of the pathways have $C^{\text{id}}\left(n,\alpha,r\right)=i\omega_{13}+\Gamma_{1313}+i\omega_{a}$.
All of the ${B^{\text{id}}\left(n,\alpha,r,q\right)}$ constants have common terms $-i\omega_q-\Gamma_{1313}$, and all of the ${A^{\text{id}}\left(n,\alpha,r,q,p\right)}$ constant have a common term $-i\omega_p$.}
\label{tab_92}\\
\hline \vspace{0.2em}\\
${n}$ &
${{Q^{\text{id}}\left(n,\alpha,r,q,p\right)}}$ & ${K^{\text{id}}\left(n,\alpha,r,q,p\right)}$ & ${A^{\text{id}}\left(n,\alpha,r,q,p\right)}+i\omega_p$ & ${B^{\text{id}}\left(n,\alpha,r,q\right)}+i\omega_q+\Gamma_{1313}$ \\
\vspace{0.2em}\\
\hline
\endfirsthead
\multicolumn{5}{c}{{\bfseries \tablename\ \thetable{} -- continued from previous page}} \\
\hline \vspace{0.2em}\\
${n}$ & ${Q^{\text{id}}\left(n,\alpha,r,q,p\right)}$ & ${K^{\text{id}}\left(n,\alpha,r,q,p\right)}$ & ${A^{\text{id}}\left(n,\alpha,r,q,p\right)}+i\omega_p$ & ${B^{\text{id}}\left(n,\alpha,r,q\right)}+i\omega_q+\Gamma_{1313}$ \\
\vspace{0.2em}\\ \hline
\endhead
\hline
\multicolumn{5}{|r|}{{Continued on next page}} \\
\hline
\endfoot \vspace{0.2em}\\
\hline
\endlastfoot
\vspace{0.2em}\\ ${14}$
& ${-W_{\alpha}^{3333}\Xi^{pqa}_{{e_ag_a},{e_ag_a},{g_ae_a}}}$
& ${-i\omega_{31}-\Gamma_{3131}}$
& ${i\omega_{31}+\Gamma_{3131}+r_{\alpha}}$
& ${-r_{\alpha}-i\omega_{13}}$
\\ \vspace{0.2em}\\
${15}$
& ${W_{\alpha}^{3333}\Xi^{pqa}_{{e_bg_b},{e_ag_a},{g_ae_a}}}$
& ${-i\omega_{63}-\Gamma_{6363}}$
& ${i\omega_{63}+\Gamma_{6363}+r_{\alpha}}$
& ${-r_{\alpha}-i\omega_{13}}$
\\ \vspace{0.2em}\\
${16}$
& ${W_{\alpha}^{3333}\Xi^{pqa}_{{v_ae_a},{e_ag_a},{g_ae_a}}}$
& ${-i\omega_{53}-\Gamma_{5353}}$
& ${i\omega_{53}+\Gamma_{5353}+r_{\alpha}}$
& ${-r_{\alpha}-i\omega_{13}}$
\\ \vspace{0.2em}\\
${17}$
& ${-W_{\alpha}^{2233}\Xi^{pqa}_{{e_bg_b},{e_ag_a},{g_ae_a}}}$
& ${-i\omega_{21}-\Gamma_{2121}}$
& ${i\omega_{21}+\Gamma_{2121}+r_{\alpha}}$
& ${-r_{\alpha}-i\omega_{13}}$
\\ \vspace{0.2em}\\
${18}$
& ${W_{\alpha}^{2233}\Xi^{pqa}_{{v_be_b},{e_ag_a},{g_ae_a}}}$
& ${-i\omega_{42}-\Gamma_{4242}}$
& ${i\omega_{42}+\Gamma_{4242}+r_{\alpha}}$
& ${-r_{\alpha}-i\omega_{13}}$
\\ \vspace{0.2em}\\
${19}$
& ${W_{\alpha}^{2233}\Xi^{pqa}_{{e_ag_a},{e_ag_a},{g_ae_a}}}$
& ${-i\omega_{62}-\Gamma_{6262}}$
& ${i\omega_{62}+\Gamma_{6262}+r_{\alpha}}$
& ${-r_{\alpha}-i\omega_{13}}$
\\ \vspace{0.2em}\\
${20}$
& ${W_{\alpha}^{1133}\Xi^{pqa}_{{e_bg_b},{e_ag_a},{g_ae_a}}}$
& ${-i\omega_{21}-\Gamma_{2121}}$
& ${i\omega_{21}+\Gamma_{2121}+r_{\alpha}}$
& ${-r_{\alpha}-i\omega_{13}}$
\\ \vspace{0.2em}\\
${21}$
& ${W_{\alpha}^{1133}\Xi^{pqa}_{{e_ag_a},{e_ag_a},{g_ae_a}}}$
& ${-i\omega_{31}-\Gamma_{3131}}$
& ${i\omega_{31}+\Gamma_{3131}+r_{\alpha}}$
& ${-r_{\alpha}-i\omega_{13}}$
\\ \vspace{0.2em}\\
${22}$
& ${-\Xi^{pqa}_{{e_bg_b},{e_ag_a},{g_ae_a}}}$
& ${-i\omega_{21}-\Gamma_{2121}}$
& ${i\omega_{21}+\Gamma_{2121}}$
& ${-i\omega_{13}}$
\\ \vspace{0.2em}\\
${23}$
& ${-\Xi^{pqa}_{{e_ag_a},{e_ag_a},{g_ae_a}}}$
& ${-i\omega_{31}-\Gamma_{3131}}$
& ${i\omega_{31}+\Gamma_{3131}}$
& ${-i\omega_{13}}$
\\ \vspace{0.2em}\\
${24}$
& ${-\Xi^{pqa}_{{e_ag_a},{e_bg_b},{g_ae_a}}}$
& ${-i\omega_{21}-\Gamma_{2121}}$
& ${i\omega_{31}+\Gamma_{2121}-\Gamma_{2323}}$
& ${i\omega_{21}+\Gamma_{2323}}$
\\ \vspace{0.2em}\\
${25}$
& ${\Xi^{pqa}_{{v_be_b},{e_bg_b},{g_ae_a}}}$
& ${-i\omega_{43}-\Gamma_{4343}}$
& ${i\omega_{42}+\Gamma_{4343}-\Gamma_{2323}}$
& ${i\omega_{21}+\Gamma_{2323}}$
\\ \vspace{0.2em}\\
${26}$
& ${\Xi^{pqa}_{{e_ag_a},{e_bg_b},{g_ae_a}}}$
& ${-i\omega_{63}-\Gamma_{6363}}$
& ${i\omega_{62}+\Gamma_{6363}-\Gamma_{2323}}$
& ${i\omega_{21}+\Gamma_{2323}}$
\\ 
\end{longtable*}
\end{center}

\newpage
\begin{center}
\begin{longtable*}{ccccc}
\caption[ ]{The constants contributing to the calculation of the integrands of pathways, as explained at the beginning of this section. In this table, all of the pathways have $C^{\text{id}}\left(n,\alpha,r\right)=i\omega_{21}+\Gamma_{2121}-i\omega_{r}$.
All of the ${B^{\text{id}}\left(n,\alpha,r,q\right)}$ constants have common terms $i\omega_a-\Gamma_{2121}$, and all of the ${A^{\text{id}}\left(n,\alpha,r,q,p\right)}$ constant have a common term $-i\omega_p$.}
\label{tab_93}\\
\hline \vspace{0.2em}\\
${n}$ &
${{Q^{\text{id}}\left(n,\alpha,r,q,p\right)}}$ & ${K^{\text{id}}\left(n,\alpha,r,q,p\right)}$ & ${A^{\text{id}}\left(n,\alpha,r,q,p\right)}+i\omega_p$ & ${B^{\text{id}}\left(n,\alpha,r,q\right)}-i\omega_a+\Gamma_{2121}$ \\
\vspace{0.2em}\\
\hline
\endfirsthead
\multicolumn{5}{c}{{\bfseries \tablename\ \thetable{} -- continued from previous page}} \\
\hline \vspace{0.2em}\\
${n}$ & ${Q^{\text{id}}\left(n,\alpha,r,q,p\right)}$ & ${K^{\text{id}}\left(n,\alpha,r,q,p\right)}$ & ${A^{\text{id}}\left(n,\alpha,r,q,p\right)}+i\omega_p$ & ${B^{\text{id}}\left(n,\alpha,r,q\right)}-i\omega_a+\Gamma_{2121}$ \\
\vspace{0.2em}\\ \hline
\endhead
\hline
\multicolumn{5}{|r|}{{Continued on next page}} \\
\hline
\endfoot \vspace{0.2em}\\
\hline
\endlastfoot
\vspace{0.2em}\\ ${27}$
& ${-\Xi^{par}_{{e_ag_a},{g_ae_a},{e_bg_b}}}$
& ${-i\omega_{21}-\Gamma_{2121}}$
& ${i\omega_{31}+\Gamma_{2121}-\Gamma_{2323}}$
& ${i\omega_{13}+\Gamma_{2323}}$
\\ \vspace{0.2em}\\
${28}$
& ${\Xi^{par}_{{v_be_b},{g_ae_a},{e_bg_b}}}$
& ${-i\omega_{43}-\Gamma_{4343}}$
& ${i\omega_{42}+\Gamma_{4343}-\Gamma_{2323}}$
& ${i\omega_{13}+\Gamma_{2323}}$
\\ \vspace{0.2em}\\
${29}$
& ${\Xi^{par}_{{e_ag_a},{g_ae_a},{e_bg_b}}}$
& ${-i\omega_{63}-\Gamma_{6363}}$
& ${i\omega_{62}+\Gamma_{6363}-\Gamma_{2323}}$
& ${i\omega_{13}+\Gamma_{2323}}$
\\ \vspace{0.2em}\\
${30}$
& ${-W_{\alpha}^{2222}\Xi^{par}_{{e_bg_b},{g_be_b},{e_bg_b}}}$
& ${-i\omega_{21}-\Gamma_{2121}}$
& ${i\omega_{21}+\Gamma_{2121}+r_{\alpha}}$
& ${-r_{\alpha}-i\omega_{21}}$
\\ \vspace{0.2em}\\
${31}$
& ${W_{\alpha}^{2222}\Xi^{par}_{{v_be_b},{g_be_b},{e_bg_b}}}$
& ${-i\omega_{42}-\Gamma_{4242}}$
& ${i\omega_{42}+\Gamma_{4242}+r_{\alpha}}$
& ${-r_{\alpha}-i\omega_{21}}$
\\ \vspace{0.2em}\\
${32}$
& ${W_{\alpha}^{2222}\Xi^{par}_{{e_ag_a},{g_be_b},{e_bg_b}}}$
& ${-i\omega_{62}-\Gamma_{6262}}$
& ${i\omega_{62}+\Gamma_{6262}+r_{\alpha}}$
& ${-r_{\alpha}-i\omega_{21}}$
\\ \vspace{0.2em}\\
${33}$
& ${-W_{\alpha}^{3322}\Xi^{par}_{{e_ag_a},{g_be_b},{e_bg_b}}}$
& ${-i\omega_{31}-\Gamma_{3131}}$
& ${i\omega_{31}+\Gamma_{3131}+r_{\alpha}}$
& ${-r_{\alpha}-i\omega_{21}}$
\\ \vspace{0.2em}\\
${34}$
& ${W_{\alpha}^{3322}\Xi^{par}_{{e_bg_b},{g_be_b},{e_bg_b}}}$
& ${-i\omega_{63}-\Gamma_{6363}}$
& ${i\omega_{63}+\Gamma_{6363}+r_{\alpha}}$
& ${-r_{\alpha}-i\omega_{21}}$
\\ \vspace{0.2em}\\
${35}$
& ${W_{\alpha}^{3322}\Xi^{par}_{{v_ae_a},{g_be_b},{e_bg_b}}}$
& ${-i\omega_{53}-\Gamma_{5353}}$
& ${i\omega_{53}+\Gamma_{5353}+r_{\alpha}}$
& ${-r_{\alpha}-i\omega_{21}}$
\\ \vspace{0.2em}\\
${36}$
& ${W_{\alpha}^{1122}\Xi^{par}_{{e_bg_b},{g_be_b},{e_bg_b}}}$
& ${-i\omega_{21}-\Gamma_{2121}}$
& ${i\omega_{21}+\Gamma_{2121}+r_{\alpha}}$
& ${-r_{\alpha}-i\omega_{21}}$
\\ \vspace{0.2em}\\
${37}$
& ${W_{\alpha}^{1122}\Xi^{par}_{{e_ag_a},{g_be_b},{e_bg_b}}}$
& ${-i\omega_{31}-\Gamma_{3131}}$
& ${i\omega_{31}+\Gamma_{3131}+r_{\alpha}}$
& ${-r_{\alpha}-i\omega_{21}}$
\\ \vspace{0.2em}\\
${38}$
& ${-\Xi^{par}_{{e_bg_b},{g_be_b},{e_bg_b}}}$
& ${-i\omega_{21}-\Gamma_{2121}}$
& ${i\omega_{21}+\Gamma_{2121}}$
& ${-i\omega_{21}}$
\\ \vspace{0.2em}\\
${39}$
& ${-\Xi^{par}_{{e_ag_a},{g_be_b},{e_bg_b}}}$
& ${-i\omega_{31}-\Gamma_{3131}}$
& ${i\omega_{31}+\Gamma_{3131}}$
& ${-i\omega_{21}}$
\\ \vspace{0.2em}\\
\end{longtable*}
\end{center}

\begin{center}
\begin{longtable*}{ccccc}
\caption[ ]{The constants contributing to the calculation of the integrands of pathways, as explained at the beginning of this section. In this table, all of the pathways have $C^{\text{id}}\left(n,\alpha,r\right)=i\omega_{31}+\Gamma_{3131}-i\omega_{r}$.
All of the ${B^{\text{id}}\left(n,\alpha,r,q\right)}$ constants have common terms $i\omega_a-\Gamma_{3131}$, and all of the ${A^{\text{id}}\left(n,\alpha,r,q,p\right)}$ constant have a common term $-i\omega_p$.}
\label{tab_94}\\
\hline \vspace{0.2em}\\
${n}$ &
${{Q^{\text{id}}\left(n,\alpha,r,q,p\right)}}$ & ${K^{\text{id}}\left(n,\alpha,r,q,p\right)}$ & ${A^{\text{id}}\left(n,\alpha,r,q,p\right)}+i\omega_p$ & ${B^{\text{id}}\left(n,\alpha,r,q\right)}+i\omega_q+\Gamma_{3131}$ \\
\vspace{0.2em}\\
\hline
\endfirsthead
\multicolumn{5}{c}{{\bfseries \tablename\ \thetable{} -- continued from previous page}} \\
\hline \vspace{0.2em}\\
${n}$ & ${Q^{\text{id}}\left(n,\alpha,r,q,p\right)}$ & ${K^{\text{id}}\left(n,\alpha,r,q,p\right)}$ & ${A^{\text{id}}\left(n,\alpha,r,q,p\right)}+i\omega_p$ & ${B^{\text{id}}\left(n,\alpha,r,q\right)}+i\omega_q+\Gamma_{1313}$ \\
\vspace{0.2em}\\ \hline
\endhead
\hline
\multicolumn{5}{|r|}{{Continued on next page}} \\
\hline
\endfoot \vspace{0.2em}\\
\hline
\endlastfoot
\vspace{0.2em}\\ ${40}$
& ${-W_{\alpha}^{3333}\Xi^{par}_{{e_ag_a},{g_ae_a},{e_ag_a}}}$
& ${-i\omega_{31}-\Gamma_{3131}}$
& ${i\omega_{31}+\Gamma_{3131}+r_{\alpha}}$
& ${-r_{\alpha}-i\omega_{31}}$
\\ \vspace{0.2em}\\
${41}$
& ${W_{\alpha}^{3333}\Xi^{par}_{{e_bg_b},{g_ae_a},{e_ag_a}}}$
& ${-i\omega_{63}-\Gamma_{6363}}$
& ${i\omega_{63}+\Gamma_{6363}+r_{\alpha}}$
& ${-r_{\alpha}-i\omega_{31}}$
\\ \vspace{0.2em}\\
${42}$
& ${W_{\alpha}^{3333}\Xi^{par}_{{v_ae_a},{g_ae_a},{e_ag_a}}}$
& ${-i\omega_{53}-\Gamma_{5353}}$
& ${i\omega_{53}+\Gamma_{5353}+r_{\alpha}}$
& ${-r_{\alpha}-i\omega_{31}}$
\\ \vspace{0.2em}\\
${43}$
& ${-W_{\alpha}^{2233}\Xi^{par}_{{e_bg_b},{g_ae_a},{e_ag_a}}}$
& ${-i\omega_{21}-\Gamma_{2121}}$
& ${i\omega_{21}+\Gamma_{2121}+r_{\alpha}}$
& ${-r_{\alpha}-i\omega_{31}}$
\\ \vspace{0.2em}\\
${44}$
& ${W_{\alpha}^{2233}\Xi^{par}_{{v_be_b},{g_ae_a},{e_ag_a}}}$
& ${-i\omega_{42}-\Gamma_{4242}}$
& ${i\omega_{42}+\Gamma_{4242}+r_{\alpha}}$
& ${-r_{\alpha}-i\omega_{31}}$
\\ \vspace{0.2em}\\
${45}$
& ${W_{\alpha}^{2233}\Xi^{par}_{{e_ag_a},{g_ae_a},{e_ag_a}}}$
& ${-i\omega_{62}-\Gamma_{6262}}$
& ${i\omega_{62}+\Gamma_{6262}+r_{\alpha}}$
& ${-r_{\alpha}-i\omega_{31}}$
\\ \vspace{0.2em}\\
${46}$
& ${W_{\alpha}^{1133}\Xi^{par}_{{e_bg_b},{g_ae_a},{e_ag_a}}}$
& ${-i\omega_{21}-\Gamma_{2121}}$
& ${i\omega_{21}+\Gamma_{2121}+r_{\alpha}}$
& ${-r_{\alpha}-i\omega_{31}}$
\\ \vspace{0.2em}\\
${47}$
& ${W_{\alpha}^{1133}\Xi^{par}_{{e_ag_a},{g_ae_a},{e_ag_a}}}$
& ${-i\omega_{31}-\Gamma_{3131}}$
& ${i\omega_{31}+\Gamma_{3131}+r_{\alpha}}$
& ${-r_{\alpha}-i\omega_{31}}$
\\ \vspace{0.2em}\\
${48}$
& ${-\Xi^{par}_{{e_bg_b},{g_be_b},{e_ag_a}}}$
& ${-i\omega_{31}-\Gamma_{3131}}$
& ${i\omega_{21}+\Gamma_{3131}-\Gamma_{3232}}$
& ${i\omega_{12}+\Gamma_{3232}}$
\\ \vspace{0.2em}\\
${49}$
& ${\Xi^{par}_{{e_bg_b},{g_be_b},{e_ag_a}}}$
& ${-i\omega_{62}-\Gamma_{6262}}$
& ${i\omega_{63}+\Gamma_{6262}-\Gamma_{3232}}$
& ${i\omega_{12}+\Gamma_{3232}}$
\\ \vspace{0.2em}\\
${50}$
& ${\Xi^{par}_{{v_ae_a},{g_be_b},{e_ag_a}}}$
& ${-i\omega_{52}-\Gamma_{5252}}$
& ${i\omega_{53}+\Gamma_{5252}-\Gamma_{3232}}$
& ${i\omega_{12}+\Gamma_{3232}}$
\\ \vspace{0.2em}\\
${51}$
& ${-\Xi^{par}_{{e_bg_b},{g_ae_a},{e_ag_a}}}$
& ${-i\omega_{21}-\Gamma_{2121}}$
& ${i\omega_{21}+\Gamma_{2121}}$
& ${-i\omega_{31}}$
\\ \vspace{0.2em}\\
${52}$
& ${-\Xi^{par}_{{e_ag_a},{g_ae_a},{e_ag_a}}}$
& ${-i\omega_{31}-\Gamma_{3131}}$
& ${i\omega_{31}+\Gamma_{3131}}$
& ${-i\omega_{31}}$
\\
\end{longtable*}
\end{center}

\begin{center}
\begin{longtable*}{ccccc}
\caption[ ]{The constants contributing to the calculation of the integrands of pathways, as explained at the beginning of this section. In this table, all of the pathways have $C^{\text{id}}\left(n,\alpha,r\right)=i\omega_{21}+\Gamma_{2121}-i\omega_{r}$.
All of the ${B^{\text{id}}\left(n,\alpha,r,q\right)}$ constants have common terms $-i\omega_q-\Gamma_{2121}$, and all of the ${A^{\text{id}}\left(n,\alpha,r,q,p\right)}$ constant have a common term $i\omega_a$.}
\label{tab_95}\\
\hline \vspace{0.2em}\\
${n}$ &
${{Q^{\text{id}}\left(n,\alpha,r,q,p\right)}}$ & ${K^{\text{id}}\left(n,\alpha,r,q,p\right)}$ & ${A^{\text{id}}\left(n,\alpha,r,q,p\right)}-i\omega_a$ & ${B^{\text{id}}\left(n,\alpha,r,q\right)}+i\omega_q+\Gamma_{2121}$ \\
\vspace{0.2em}\\
\hline
\endfirsthead
\multicolumn{5}{c}{{\bfseries \tablename\ \thetable{} -- continued from previous page}} \\
\hline \vspace{0.2em}\\
${n}$ & ${Q^{\text{id}}\left(n,\alpha,r,q,p\right)}$ & ${K^{\text{id}}\left(n,\alpha,r,q,p\right)}$ & ${A^{\text{id}}\left(n,\alpha,r,q,p\right)}-i\omega_a$ & ${B^{\text{id}}\left(n,\alpha,r,q\right)}+i\omega_q+\Gamma_{2121}$ \\
\vspace{0.2em}\\ \hline
\endhead
\hline
\multicolumn{5}{|r|}{{Continued on next page}} \\
\hline
\endfoot \vspace{0.2em}\\
\hline
\endlastfoot
\vspace{0.2em}\\ ${53}$
& ${-\Xi^{aqr}_{{e_bv_b},{v_be_b},{e_bg_b}}}$
& ${-i\omega_{21}-\Gamma_{2121}}$
& ${i\omega_{24}+\Gamma_{2121}-\Gamma_{4141}}$
& ${i\omega_{42}+\Gamma_{4141}}$
\\ \vspace{0.2em}\\
${54}$
& ${\Xi^{aqr}_{{g_be_b},{v_be_b},{e_bg_b}}}$
& ${-i\omega_{42}-\Gamma_{4242}}$
& ${i\omega_{12}+\Gamma_{4242}-\Gamma_{4141}}$
& ${i\omega_{42}+\Gamma_{4141}}$
\\ \vspace{0.2em}\\
${55}$
& ${\Xi^{aqr}_{{g_ae_a},{v_be_b},{e_bg_b}}}$
& ${-i\omega_{43}-\Gamma_{4343}}$
& ${i\omega_{13}+\Gamma_{4343}-\Gamma_{4141}}$
& ${i\omega_{42}+\Gamma_{4141}}$
\\ \vspace{0.2em}\\
${56}$
& ${-\Xi^{aqr}_{{g_ae_a},{e_ag_a},{e_bg_b}}}$
& ${-i\omega_{21}-\Gamma_{2121}}$
& ${i\omega_{26}+\Gamma_{2121}-\Gamma_{6161}}$
& ${i\omega_{62}+\Gamma_{6161}}$
\\ \vspace{0.2em}\\
${57}$
& ${-\Xi^{aqr}_{{g_be_b},{e_ag_a},{e_bg_b}}}$
& ${-i\omega_{31}-\Gamma_{3131}}$
& ${i\omega_{36}+\Gamma_{3131}-\Gamma_{6161}}$
& ${i\omega_{62}+\Gamma_{6161}}$
\\ \vspace{0.2em}\\
${58}$
& ${\Xi^{aqr}_{{g_be_b},{e_ag_a},{e_bg_b}}}$
& ${-i\omega_{62}-\Gamma_{6262}}$
& ${i\omega_{12}+\Gamma_{6262}-\Gamma_{6161}}$
& ${i\omega_{62}+\Gamma_{6161}}$
\\ \vspace{0.2em}\\
${59}$
& ${\Xi^{aqr}_{{g_ae_a},{e_ag_a},{e_bg_b}}}$
& ${-i\omega_{63}-\Gamma_{6363}}$
& ${i\omega_{13}+\Gamma_{6363}-\Gamma_{6161}}$
& ${i\omega_{62}+\Gamma_{6161}}$
\\
\end{longtable*}
\end{center}

\begin{center}
\begin{longtable*}{ccccc}
\caption[ ]{The constants contributing to the calculation of the integrands of pathways, as explained at the beginning of this section. In this table, all of the pathways have $C^{\text{id}}\left(n,\alpha,r\right)=i\omega_{31}+\Gamma_{3131}-i\omega_{r}$.
All of the ${B^{\text{id}}\left(n,\alpha,r,q\right)}$ constants have common terms $-i\omega_q-\Gamma_{3131}$, and all of the ${A^{\text{id}}\left(n,\alpha,r,q,p\right)}$ constant have a common term $i\omega_a$.}
\label{tab_96}\\
\hline \vspace{0.2em}\\
${n}$ &
${{Q^{\text{id}}\left(n,\alpha,r,q,p\right)}}$ & ${K^{\text{id}}\left(n,\alpha,r,q,p\right)}$ & ${A^{\text{id}}\left(n,\alpha,r,q,p\right)}-i\omega_a$ & ${B^{\text{id}}\left(n,\alpha,r,q\right)}+i\omega_q+\Gamma_{3131}$ \\
\vspace{0.2em}\\
\hline
\endfirsthead
\multicolumn{5}{c}{{\bfseries \tablename\ \thetable{} -- continued from previous page}} \\
\hline \vspace{0.2em}\\
${n}$ & ${Q^{\text{id}}\left(n,\alpha,r,q,p\right)}$ & ${K^{\text{id}}\left(n,\alpha,r,q,p\right)}$ & ${A^{\text{id}}\left(n,\alpha,r,q,p\right)}-i\omega_a$ & ${B^{\text{id}}\left(n,\alpha,r,q\right)}+i\omega_q+\Gamma_{3131}$ \\
\vspace{0.2em}\\ \hline
\endhead
\hline
\multicolumn{5}{|r|}{{Continued on next page}} \\
\hline
\endfoot \vspace{0.2em}\\
\hline
\endlastfoot
\vspace{0.2em}\\ ${60}$
& ${-\Xi^{aqr}_{{e_av_a},{v_ae_a},{e_ag_a}}}$
& ${-i\omega_{31}-\Gamma_{3131}}$
& ${i\omega_{35}+\Gamma_{3131}-\Gamma_{5151}}$
& ${i\omega_{53}+\Gamma_{5151}}$
\\ \vspace{0.2em}\\
${61}$
& ${\Xi^{aqr}_{{g_be_b},{v_ae_a},{e_ag_a}}}$
& ${-i\omega_{52}-\Gamma_{5252}}$
& ${i\omega_{12}+\Gamma_{5252}-\Gamma_{5151}}$
& ${i\omega_{53}+\Gamma_{5151}}$
\\ \vspace{0.2em}\\
${62}$
& ${\Xi^{aqr}_{{g_ae_a},{v_ae_a},{e_ag_a}}}$
& ${-i\omega_{53}-\Gamma_{5353}}$
& ${i\omega_{13}+\Gamma_{5353}-\Gamma_{5151}}$
& ${i\omega_{53}+\Gamma_{5151}}$
\\ \vspace{0.2em}\\
${63}$
& ${-\Xi^{aqr}_{{g_ae_a},{e_bg_b},{e_ag_a}}}$
& ${-i\omega_{21}-\Gamma_{2121}}$
& ${i\omega_{26}+\Gamma_{2121}-\Gamma_{6161}}$
& ${i\omega_{63}+\Gamma_{6161}}$
\\ \vspace{0.2em}\\
${64}$
& ${-\Xi^{ }_{{g_be_b},{e_bg_b},{e_ag_a}}}$
& ${-i\omega_{31}-\Gamma_{3131}}$
& ${i\omega_{36}+\Gamma_{3131}-\Gamma_{6161}}$
& ${i\omega_{63}+\Gamma_{6161}}$
\\ \vspace{0.2em}\\
${65}$
& ${\Xi^{aqr}_{{g_be_b},{e_bg_b},{e_ag_a}}}$
& ${-i\omega_{62}-\Gamma_{6262}}$
& ${i\omega_{12}+\Gamma_{6262}-\Gamma_{6161}}$
& ${i\omega_{63}+\Gamma_{6161}}$
\\ \vspace{0.2em}\\
${66}$
& ${\Xi^{aqr}_{{g_ae_a},{e_bg_b},{e_ag_a}}}$
& ${-i\omega_{63}-\Gamma_{6363}}$
& ${i\omega_{13}+\Gamma_{6363}-\Gamma_{6161}}$
& ${i\omega_{63}+\Gamma_{6161}}$
\\
\end{longtable*}
\end{center}

\newpage

\section{Constants associated with the orientational average}
In Equation~\eqref{3.2}, there are constants accounting for the orientational motion of the molecules. In this supplement, these constants are presented.
These constants are arranged into three tables. The grouping is exactly the same as in the first supplement. In other words, they are separated into the three tables according to whether the fields $a$ or $b$ act in the first, the second, or the third interaction term. However, in contrast with the constants presented in the previous section, these orientational constants depend on another parameter $\beta$, instead of $\alpha$, and $\beta$ may be 1 or 2 for certain pathways $n$, as explained in the text.

\newpage
\begin{center}
\begin{longtable*}{cccccc}
\caption{Orientational molecular constants for pathways in Table~\ref{tab_6}.}
\label{tab_101}\\
\hline\vspace{0.2em}\\
${n,\beta}$ & ${Q^{\text{or}}_{n,\beta,r,q,p}}$ & ${K^{\text{or}}_{n,\beta,r,q,p}}$ & ${A^{\text{or}}_{n,\beta,r,q,p}}$ & ${B^{\text{or}}_{n,\beta,r,q}}$ & ${C^{\text{or}}_{n,\beta,r}}$\\
\vspace{0.2em}\\ \hline
\endfirsthead
\multicolumn{6}{c}{{\bfseries \tablename\ \thetable{} -- continued from previous page}} \\
\hline\vspace{0.2em}\\
${n,\beta}$ & ${Q^{\text{or}}_{n,\beta,r,q,p}}$ & ${K^{\text{or}}_{n,\beta,r,q,p}}$ & ${A^{\text{or}}_{n,\beta,r,q,p}}$ & ${B^{\text{or}}_{n,\beta,r,q}}$ & ${C^{\text{or}}_{n,\beta,r}}$\\
\vspace{0.2em}\\ \hline
\endhead
\hline
\multicolumn{6}{|r|}{{Continued on next page}} \\
\hline
\endfoot
\vspace{0.2em}\\ \hline
\endlastfoot
\vspace{0.2em}\\${1,1}$
& ${\frac{1}{9}E_{\text{lo}}\mu^{\left(b\right)}}$
& ${-2D^{\left(b\right)}}$
& ${2D^{\left(b\right)}}$
& ${-2D^{\left(b\right)}}$
& ${2D^{\left(b\right)}}$ \\ \vspace{0.2em}\\
${1,2}$
& ${\frac{4}{45}E_{\text{lo}}\mu^{\left(b\right)}}$
& ${-2D^{\left(b\right)}}$
& ${-4D^{\left(b\right)}}$
& ${4D^{\left(b\right)}}$
& ${2D^{\left(b\right)}}$\\ \vspace{0.2em}\\
${2,1}$
& ${\frac{1}{9}E_{\text{lo}}\mu^{\left(b\right)}}$
& ${-2D^{\left(b\right)}}$
& ${2D^{\left(b\right)}}$
& ${-2D^{\left(b\right)}}$
& ${2D^{\left(b\right)}}$ \\ \vspace{0.2em}\\
${2,2}$
& ${\frac{4}{45}E_{\text{lo}}\mu^{\left(b\right)}}$
& ${-2D^{\left(b\right)}}$
& ${-4D^{\left(b\right)}}$
& ${4D^{\left(b\right)}}$
& ${2D^{\left(b\right)}}$\\ \vspace{0.2em}\\
${3,1}$
& ${\frac{1}{9}E_{\text{lo}}\mu^{\left(a\right)}}$
& ${-2D^{\left(a\right)}}$
& ${2D^{\left(a\right)}}$
& ${-2D^{\left(b\right)}}$
& ${2D^{\left(b\right)}}$ \\ \vspace{0.2em}\\
${4,1}$
& ${\frac{1}{9}E_{\text{lo}}\mu^{\left(a\right)}}$
& ${-2D^{\left(a\right)}}$
& ${2D^{\left(a\right)}}$
& ${-2D^{\left(b\right)}}$
& ${2D^{\left(b\right)}}$ \\ \vspace{0.2em}\\
${5,1}$
& ${\frac{1}{9}E_{\text{lo}}\mu^{\left(b\right)}}$
& ${-2D^{\left(b\right)}}$
& ${2D^{\left(b\right)}}$
& ${-2D^{\left(b\right)}}$
& ${2D^{\left(b\right)}}$ \\ \vspace{0.2em}\\
${5,2}$
& ${\frac{4}{45}E_{\text{lo}}\mu^{\left(b\right)}}$
& ${-2D^{\left(b\right)}}$
& ${-4D^{\left(b\right)}}$
& ${4D^{\left(b\right)}}$
& ${2D^{\left(b\right)}}$ \\ \vspace{0.2em}\\
${6,1}$
& ${\frac{1}{9}E_{\text{lo}}\mu^{\left(a\right)}}$
& ${-2D^{\left(a\right)}}$
& ${2D^{\left(a\right)}}$
& ${-2D^{\left(b\right)}}$
& ${2D^{\left(b\right)}}$ \\ \vspace{0.2em}\\
${7,1}$
& ${\frac{1}{9}E_{\text{lo}}\mu^{\left(b\right)}}$
& ${-2D^{\left(b\right)}}$
& ${2D^{\left(b\right)}}$
& ${-2D^{\left(b\right)}}$
& ${2D^{\left(b\right)}}$ \\ \vspace{0.2em}\\
${7,2}$
& ${\frac{4}{45}E_{\text{lo}}\mu^{\left(b\right)}}$
& ${-2D^{\left(b\right)}}$
& ${-4D^{\left(b\right)}}$
& ${4D^{\left(b\right)}}$
& ${2D^{\left(b\right)}}$ \\ \vspace{0.2em}\\
${8,1}$
& ${\frac{1}{9}E_{\text{lo}}\mu^{\left(a\right)}}$
& ${-2D^{\left(a\right)}}$
& ${2D^{\left(a\right)}}$
& ${-2D^{\left(b\right)}}$
& ${2D^{\left(b\right)}}$ \\ \vspace{0.2em}\\
${9,1}$
& ${\frac{1}{9}E_{\text{lo}}\mu^{\left(a\right)}}$
& ${-2D^{\left(a\right)}}$
& ${-2D^{\left(b\right)}}$
& ${2D^{\left(a\right)}}$
& ${2D^{\left(b\right)}}$ \\ \vspace{0.2em}\\
${10,1}$
& ${\frac{1}{9}E_{\text{lo}}\mu^{\left(a\right)}}$
& ${-2D^{\left(a\right)}}$
& ${-2D^{\left(b\right)}}$
& ${2D^{\left(a\right)}}$
& ${2D^{\left(b\right)}}$ \\ \vspace{0.2em}\\
${11,1}$
& ${\frac{1}{9}E_{\text{lo}}\mu^{\left(b\right)}}$
& ${-2D^{\left(b\right)}}$
& ${-2D^{\left(a\right)}}$
& ${2D^{\left(a\right)}}$
& ${2D^{\left(b\right)}}$ \\ \vspace{0.2em}\\
${12,1}$
& ${\frac{1}{9}E_{\text{lo}}\mu^{\left(b\right)}}$
& ${-2D^{\left(b\right)}}$
& ${2D^{\left(b\right)}}$
& ${-2D^{\left(b\right)}}$
& ${2D^{\left(b\right)}}$ \\ \vspace{0.2em}\\
${12,2}$
& ${\frac{4}{45}E_{\text{lo}}\mu^{\left(b\right)}}$
& ${-2D^{\left(b\right)}}$
& ${-4D^{\left(b\right)}}$
& ${4D^{\left(b\right)}}$
& ${2D^{\left(b\right)}}$ \\ \vspace{0.2em}\\
${13,1}$
& ${\frac{1}{9}E_{\text{lo}}\mu^{\left(a\right)}}$
& ${-2D^{\left(a\right)}}$
& ${2D^{\left(a\right)}}$
& ${-2D^{\left(b\right)}}$
& ${2D^{\left(b\right)}}$ \\ \vspace{0.2em}\\
${14,1}$
& ${\frac{1}{9}E_{\text{lo}}\mu^{\left(a\right)}}$
& ${-2D^{\left(a\right)}}$
& ${2D^{\left(a\right)}}$
& ${-2D^{\left(a\right)}}$
& ${2D^{\left(a\right)}}$ \\ \vspace{0.2em}\\
${14,2}$
& ${\frac{4}{45}E_{\text{lo}}\mu^{\left(a\right)}}$
& ${-2D^{\left(a\right)}}$
& ${-4D^{\left(a\right)}}$
& ${4D^{\left(a\right)}}$
& ${2D^{\left(a\right)}}$ \\ \vspace{0.2em}\\
${15,1}$
& ${\frac{1}{9}E_{\text{lo}}\mu^{\left(b\right)}}$
& ${-2D^{\left(b\right)}}$
& ${2D^{\left(b\right)}}$
& ${-2D^{\left(a\right)}}$
& ${2D^{\left(a\right)}}$ \\ \vspace{0.2em}\\
${16,1}$
& ${\frac{1}{9}E_{\text{lo}}\mu^{\left(a\right)}}$
& ${-2D^{\left(a\right)}}$
& ${2D^{\left(a\right)}}$
& ${-2D^{\left(a\right)}}$
& ${2D^{\left(a\right)}}$ \\ \vspace{0.2em}\\
${16,2}$
& ${\frac{4}{45}E_{\text{lo}}\mu^{\left(a\right)}}$
& ${-2D^{\left(a\right)}}$
& ${-4D^{\left(a\right)}}$
& ${4D^{\left(a\right)}}$
& ${2D^{\left(a\right)}}$ \\ \vspace{0.2em}\\
${17,1}$
& ${\frac{1}{9}E_{\text{lo}}\mu^{\left(b\right)}}$
& ${-2D^{\left(b\right)}}$
& ${2D^{\left(b\right)}}$
& ${-2D^{\left(a\right)}}$
& ${2D^{\left(a\right)}}$ \\ \vspace{0.2em}\\
${18,1}$
& ${\frac{1}{9}E_{\text{lo}}\mu^{\left(b\right)}}$
& ${-2D^{\left(b\right)}}$
& ${2D^{\left(b\right)}}$
& ${-2D^{\left(a\right)}}$
& ${2D^{\left(a\right)}}$ \\ \vspace{0.2em}\\
${19,1}$
& ${\frac{1}{9}E_{\text{lo}}\mu^{\left(a\right)}}$
& ${-2D^{\left(a\right)}}$
& ${2D^{\left(a\right)}}$
& ${-2D^{\left(a\right)}}$
& ${2D^{\left(a\right)}}$ \\ \vspace{0.2em}\\
${19,2}$
& ${\frac{4}{45}E_{\text{lo}}\mu^{\left(a\right)}}$
& ${-2D^{\left(a\right)}}$
& ${-4D^{\left(a\right)}}$
& ${4D^{\left(a\right)}}$
& ${2D^{\left(a\right)}}$ \\ \vspace{0.2em}\\
${20,1}$
& ${\frac{1}{9}E_{\text{lo}}\mu^{\left(b\right)}}$
& ${-2D^{\left(b\right)}}$
& ${2D^{\left(b\right)}}$
& ${-2D^{\left(a\right)}}$
& ${2D^{\left(a\right)}}$ \\ \vspace{0.2em}\\
${21,1}$
& ${\frac{1}{9}E_{\text{lo}}\mu^{\left(a\right)}}$
& ${-2D^{\left(a\right)}}$
& ${2D^{\left(a\right)}}$
& ${-2D^{\left(a\right)}}$
& ${2D^{\left(a\right)}}$ \\ \vspace{0.2em}\\
${21,2}$
& ${\frac{4}{45}E_{\text{lo}}\mu^{\left(a\right)}}$
& ${-2D^{\left(a\right)}}$
& ${-4D^{\left(a\right)}}$
& ${4D^{\left(a\right)}}$
& ${2D^{\left(a\right)}}$ \\ \vspace{0.2em}\\
${22,1}$
& ${\frac{1}{9}E_{\text{lo}}\mu^{\left(b\right)}}$
& ${-2D^{\left(b\right)}}$
& ${2D^{\left(b\right)}}$
& ${-2D^{\left(a\right)}}$
& ${2D^{\left(a\right)}}$ \\ \vspace{0.2em}\\
${23,1}$
& ${\frac{1}{9}E_{\text{lo}}\mu^{\left(a\right)}}$
& ${-2D^{\left(a\right)}}$
& ${2D^{\left(a\right)}}$
& ${-2D^{\left(a\right)}}$
& ${2D^{\left(a\right)}}$ \\ \vspace{0.2em}\\
${23,2}$
& ${\frac{4}{45}E_{\text{lo}}\mu^{\left(a\right)}}$
& ${-2D^{\left(a\right)}}$
& ${-4D^{\left(a\right)}}$
& ${4D^{\left(a\right)}}$
& ${2D^{\left(a\right)}}$ \\ \vspace{0.2em}\\
${24,1}$
& ${\frac{1}{9}E_{\text{lo}}\mu^{\left(b\right)}}$
& ${-2D^{\left(b\right)}}$
& ${-2D^{\left(a\right)}}$
& ${2D^{\left(b\right)}}$
& ${2D^{\left(a\right)}}$ \\ \vspace{0.2em}\\
${25,1}$
& ${\frac{1}{9}E_{\text{lo}}\mu^{\left(a\right)}}$
& ${-2D^{\left(a\right)}}$
& ${-2D^{\left(b\right)}}$
& ${2D^{\left(b\right)}}$
& ${2D^{\left(a\right)}}$ \\ \vspace{0.2em}\\
${26,1}$
& ${\frac{1}{9}E_{\text{lo}}\mu^{\left(b\right)}}$
& ${-2D^{\left(b\right)}}$
& ${-2D^{\left(a\right)}}$
& ${2D^{\left(b\right)}}$
& ${2D^{\left(a\right)}}$ \\ \vspace{0.2em}\\
\end{longtable*}
\end{center}

\begin{center}
\begin{longtable*}{cccccc}
\caption{Orientational molecular constants for pathways in Table~\ref{tab_7}.}
\label{tab_102}\\
\hline\vspace{0.2em}\\
${n,\beta}$ & ${Q^{\text{or}}_{n,\beta,r,q,p}}$ & ${K^{\text{or}}_{n,\beta,r,q,p}}$ & ${A^{\text{or}}_{n,\beta,r,q,p}}$ & ${B^{\text{or}}_{n,\beta,r,q}}$ & ${C^{\text{or}}_{n,\beta,r}}$\\
\vspace{0.2em}\\ \hline
\endfirsthead
\multicolumn{6}{c}{{\bfseries \tablename\ \thetable{} -- continued from previous page}} \\
\hline\vspace{0.2em}\\
${n,\beta}$ & ${Q^{\text{or}}_{n,\beta,r,q,p}}$ & ${K^{\text{or}}_{n,\beta,r,q,p}}$ & ${A^{\text{or}}_{n,\beta,r,q,p}}$ & ${B^{\text{or}}_{n,\beta,r,q}}$ & ${C^{\text{or}}_{n,\beta,r}}$\\
\vspace{0.2em}\\ \hline
\endhead
\hline
\multicolumn{6}{|r|}{{Continued on next page}} \\
\hline
\endfoot
\vspace{0.2em}\\ \hline
\endlastfoot
\vspace{0.2em}\\${27,1}$
& ${\frac{1}{9}E_{\text{lo}}\mu^{\left(b\right)}}$
& ${-2D^{\left(b\right)}}$
& ${-2D^{\left(a\right)}}$
& ${2D^{\left(a\right)}}$
& ${2D^{\left(b\right)}}$ \\ \vspace{0.2em}\\
${28,1}$
& ${\frac{1}{9}E_{\text{lo}}\mu^{\left(a\right)}}$
& ${-2D^{\left(a\right)}}$
& ${-2D^{\left(b\right)}}$
& ${2D^{\left(a\right)}}$
& ${2D^{\left(b\right)}}$ \\ \vspace{0.2em}\\
${29,1}$
& ${\frac{1}{9}E_{\text{lo}}\mu^{\left(b\right)}}$
& ${-2D^{\left(b\right)}}$
& ${-2D^{\left(a\right)}}$
& ${2D^{\left(a\right)}}$
& ${2D^{\left(b\right)}}$ \\ \vspace{0.2em}\\
${30,1}$
& ${\frac{1}{9}E_{\text{lo}}\mu^{\left(b\right)}}$
& ${-2D^{\left(b\right)}}$
& ${2D^{\left(b\right)}}$
& ${-2D^{\left(b\right)}}$
& ${2D^{\left(b\right)}}$ \\ \vspace{0.2em}\\
${30,2}$
& ${\frac{4}{45}E_{\text{lo}}\mu^{\left(b\right)}}$
& ${-2D^{\left(b\right)}}$
& ${-4D^{\left(b\right)}}$
& ${4D^{\left(b\right)}}$
& ${2D^{\left(b\right)}}$ \\ \vspace{0.2em}\\
${31,1}$
& ${\frac{1}{9}E_{\text{lo}}\mu^{\left(b\right)}}$
& ${-2D^{\left(b\right)}}$
& ${2D^{\left(b\right)}}$
& ${-2D^{\left(b\right)}}$
& ${2D^{\left(b\right)}}$ \\ \vspace{0.2em}\\
${31,2}$
& ${\frac{4}{45}E_{\text{lo}}\mu^{\left(b\right)}}$
& ${-2D^{\left(b\right)}}$
& ${-4D^{\left(b\right)}}$
& ${4D^{\left(b\right)}}$
& ${2D^{\left(b\right)}}$ \\ \vspace{0.2em}\\
${32,1}$
& ${\frac{1}{9}E_{\text{lo}}\mu^{\left(a\right)}}$
& ${-2D^{\left(a\right)}}$
& ${2D^{\left(a\right)}}$
& ${-2D^{\left(b\right)}}$
& ${2D^{\left(b\right)}}$ \\ \vspace{0.2em}\\
${33,1}$
& ${\frac{1}{9}E_{\text{lo}}\mu^{\left(a\right)}}$
& ${-2D^{\left(a\right)}}$
& ${2D^{\left(a\right)}}$
& ${-2D^{\left(b\right)}}$
& ${2D^{\left(b\right)}}$ \\ \vspace{0.2em}\\
${34,1}$
& ${\frac{1}{9}E_{\text{lo}}\mu^{\left(b\right)}}$
& ${-2D^{\left(b\right)}}$
& ${2D^{\left(b\right)}}$
& ${-2D^{\left(b\right)}}$
& ${2D^{\left(b\right)}}$ \\ \vspace{0.2em}\\
${34,2}$
& ${\frac{4}{45}E_{\text{lo}}\mu^{\left(b\right)}}$
& ${-2D^{\left(b\right)}}$
& ${-4D^{\left(b\right)}}$
& ${4D^{\left(b\right)}}$
& ${2D^{\left(b\right)}}$ \\ \vspace{0.2em}\\
${35,1}$
& ${\frac{1}{9}E_{\text{lo}}\mu^{\left(a\right)}}$
& ${-2D^{\left(a\right)}}$
& ${2D^{\left(a\right)}}$
& ${-2D^{\left(b\right)}}$
& ${2D^{\left(b\right)}}$ \\ \vspace{0.2em}\\
${36,1}$
& ${\frac{1}{9}E_{\text{lo}}\mu^{\left(b\right)}}$
& ${-2D^{\left(b\right)}}$
& ${2D^{\left(b\right)}}$
& ${-2D^{\left(b\right)}}$
& ${2D^{\left(b\right)}}$ \\ \vspace{0.2em}\\
${36,2}$
& ${\frac{4}{45}E_{\text{lo}}\mu^{\left(b\right)}}$
& ${-2D^{\left(b\right)}}$
& ${-4D^{\left(b\right)}}$
& ${4D^{\left(b\right)}}$
& ${2D^{\left(b\right)}}$ \\ \vspace{0.2em}\\
${37,1}$
& ${\frac{1}{9}E_{\text{lo}}\mu^{\left(a\right)}}$
& ${-2D^{\left(a\right)}}$
& ${2D^{\left(a\right)}}$
& ${-2D^{\left(b\right)}}$
& ${2D^{\left(b\right)}}$ \\ \vspace{0.2em}\\
${38,1}$
& ${\frac{1}{9}E_{\text{lo}}\mu^{\left(b\right)}}$
& ${-2D^{\left(b\right)}}$
& ${2D^{\left(b\right)}}$
& ${-2D^{\left(b\right)}}$
& ${2D^{\left(b\right)}}$ \\ \vspace{0.2em}\\
${38,2}$
& ${\frac{4}{45}E_{\text{lo}}\mu^{\left(b\right)}}$
& ${-2D^{\left(b\right)}}$
& ${-4D^{\left(b\right)}}$
& ${4D^{\left(b\right)}}$
& ${2D^{\left(b\right)}}$ \\ \vspace{0.2em}\\
${39,1}$
& ${\frac{1}{9}E_{\text{lo}}\mu^{\left(a\right)}}$
& ${-2D^{\left(a\right)}}$
& ${2D^{\left(a\right)}}$
& ${-2D^{\left(b\right)}}$
& ${2D^{\left(b\right)}}$ \\ \vspace{0.2em}\\
${40,1}$
& ${\frac{1}{9}E_{\text{lo}}\mu^{\left(a\right)}}$
& ${-2D^{\left(a\right)}}$
& ${2D^{\left(a\right)}}$
& ${-2D^{\left(a\right)}}$
& ${2D^{\left(a\right)}}$ \\ \vspace{0.2em}\\
${40,2}$
& ${\frac{4}{45}E_{\text{lo}}\mu^{\left(a\right)}}$
& ${-2D^{\left(a\right)}}$
& ${-4D^{\left(a\right)}}$
& ${4D^{\left(a\right)}}$
& ${2D^{\left(a\right)}}$ \\ \vspace{0.2em}\\
${41,1}$
& ${\frac{1}{9}E_{\text{lo}}\mu^{\left(b\right)}}$
& ${-2D^{\left(b\right)}}$
& ${2D^{\left(b\right)}}$
& ${-2D^{\left(a\right)}}$
& ${2D^{\left(a\right)}}$ \\ \vspace{0.2em}\\
${42,1}$
& ${\frac{1}{9}E_{\text{lo}}\mu^{\left(a\right)}}$
& ${-2D^{\left(a\right)}}$
& ${2D^{\left(a\right)}}$
& ${-2D^{\left(a\right)}}$
& ${2D^{\left(a\right)}}$ \\ \vspace{0.2em}\\
${42,2}$
& ${\frac{4}{45}E_{\text{lo}}\mu^{\left(a\right)}}$
& ${-2D^{\left(a\right)}}$
& ${-4D^{\left(a\right)}}$
& ${4D^{\left(a\right)}}$
& ${2D^{\left(a\right)}}$ \\ \vspace{0.2em}\\
${43,1}$
& ${\frac{1}{9}E_{\text{lo}}\mu^{\left(b\right)}}$
& ${-2D^{\left(b\right)}}$
& ${2D^{\left(b\right)}}$
& ${-2D^{\left(a\right)}}$
& ${2D^{\left(a\right)}}$ \\ \vspace{0.2em}\\
${44,1}$
& ${\frac{1}{9}E_{\text{lo}}\mu^{\left(b\right)}}$
& ${-2D^{\left(b\right)}}$
& ${2D^{\left(b\right)}}$
& ${-2D^{\left(a\right)}}$
& ${2D^{\left(a\right)}}$ \\ \vspace{0.2em}\\
${45,1}$
& ${\frac{1}{9}E_{\text{lo}}\mu^{\left(a\right)}}$
& ${-2D^{\left(a\right)}}$
& ${2D^{\left(a\right)}}$
& ${-2D^{\left(a\right)}}$
& ${2D^{\left(a\right)}}$ \\ \vspace{0.2em}\\
${45,2}$
& ${\frac{4}{45}E_{\text{lo}}\mu^{\left(a\right)}}$
& ${-2D^{\left(a\right)}}$
& ${-4D^{\left(a\right)}}$
& ${4D^{\left(a\right)}}$
& ${2D^{\left(a\right)}}$ \\ \vspace{0.2em}\\
${46,1}$
& ${\frac{1}{9}E_{\text{lo}}\mu^{\left(b\right)}}$
& ${-2D^{\left(b\right)}}$
& ${2D^{\left(b\right)}}$
& ${-2D^{\left(a\right)}}$
& ${2D^{\left(a\right)}}$ \\ \vspace{0.2em}\\
${47,1}$
& ${\frac{1}{9}E_{\text{lo}}\mu^{\left(a\right)}}$
& ${-2D^{\left(a\right)}}$
& ${2D^{\left(a\right)}}$
& ${-2D^{\left(a\right)}}$
& ${2D^{\left(a\right)}}$ \\ \vspace{0.2em}\\
${47,2}$
& ${\frac{4}{45}E_{\text{lo}}\mu^{\left(a\right)}}$
& ${-2D^{\left(a\right)}}$
& ${-4D^{\left(a\right)}}$
& ${4D^{\left(a\right)}}$
& ${2D^{\left(a\right)}}$ \\ \vspace{0.2em}\\
${48,1}$
& ${\frac{1}{9}E_{\text{lo}}\mu^{\left(a\right)}}$
& ${-2D^{\left(a\right)}}$
& ${-2D^{\left(b\right)}}$
& ${2D^{\left(b\right)}}$
& ${2D^{\left(a\right)}}$ \\ \vspace{0.2em}\\
${49,1}$
& ${\frac{1}{9}E_{\text{lo}}\mu^{\left(a\right)}}$
& ${-2D^{\left(a\right)}}$
& ${-2D^{\left(b\right)}}$
& ${2D^{\left(b\right)}}$
& ${2D^{\left(a\right)}}$ \\ \vspace{0.2em}\\
${50,1}$
& ${\frac{1}{9}E_{\text{lo}}\mu^{\left(b\right)}}$
& ${-2D^{\left(b\right)}}$
& ${-2D^{\left(a\right)}}$
& ${2D^{\left(b\right)}}$
& ${2D^{\left(a\right)}}$ \\ \vspace{0.2em}\\
${51,1}$
& ${\frac{1}{9}E_{\text{lo}}\mu^{\left(b\right)}}$
& ${-2D^{\left(b\right)}}$
& ${2D^{\left(b\right)}}$
& ${-2D^{\left(a\right)}}$
& ${2D^{\left(a\right)}}$ \\ \vspace{0.2em}\\
${52,1}$
& ${\frac{1}{9}E_{\text{lo}}\mu^{\left(a\right)}}$
& ${-2D^{\left(a\right)}}$
& ${2D^{\left(a\right)}}$
& ${-2D^{\left(a\right)}}$
& ${2D^{\left(a\right)}}$ \\ \vspace{0.2em}\\
${52,2}$
& ${\frac{4}{45}E_{\text{lo}}\mu^{\left(a\right)}}$
& ${-2D^{\left(a\right)}}$
& ${-4D^{\left(a\right)}}$
& ${4D^{\left(a\right)}}$
& ${2D^{\left(a\right)}}$\\
\end{longtable*}
\end{center}

\begin{center}
\begin{longtable*}{cccccc}
\caption{Orientational molecular constants for pathways in Table~\ref{tab_8}.}
\label{tab_103}\\
\hline\vspace{0.2em}\\
${n,\beta}$ & ${Q^{\text{or}}_{n,\beta,r,q,p}}$ & ${K^{\text{or}}_{n,\beta,r,q,p}}$ & ${A^{\text{or}}_{n,\beta,r,q,p}}$ & ${B^{\text{or}}_{n,\beta,r,q}}$ & ${C^{\text{or}}_{n,\beta,r}}$\\
\vspace{0.2em}\\ \hline
\endfirsthead
\multicolumn{6}{c}{{\bfseries \tablename\ \thetable{} -- continued from previous page}} \\
\hline\vspace{0.2em}\\
${n,\beta}$ & ${Q^{\text{or}}_{n,\beta,r,q,p}}$ & ${K^{\text{or}}_{n,\beta,r,q,p}}$ & ${A^{\text{or}}_{n,\beta,r,q,p}}$ & ${B^{\text{or}}_{n,\beta,r,q}}$ & ${C^{\text{or}}_{n,\beta,r}}$\\
\vspace{0.2em}\\ \hline
\endhead
\hline
\multicolumn{6}{|r|}{{Continued on next page}} \\
\hline
\endfoot
\vspace{0.2em}\\ \hline
\endlastfoot
\vspace{0.2em}\\${53,1}$
& ${\frac{1}{9}E_{\text{lo}}\mu^{\left(b\right)}}$
& ${-2D^{\left(b\right)}}$
& ${2D^{\left(b\right)}}$
& ${-2D^{\left(b\right)}}$
& ${2D^{\left(b\right)}}$ \\ \vspace{0.2em}\\
${53,2}$
& ${\frac{4}{45}E_{\text{lo}}\mu^{\left(b\right)}}$
& ${-2D^{\left(b\right)}}$
& ${-4D^{\left(b\right)}}$
& ${4D^{\left(b\right)}}$
& ${2D^{\left(b\right)}}$ \\ \vspace{0.2em}\\
${54,1}$
& ${\frac{1}{9}E_{\text{lo}}\mu^{\left(b\right)}}$
& ${-2D^{\left(b\right)}}$
& ${2D^{\left(b\right)}}$
& ${-2D^{\left(b\right)}}$
& ${2D^{\left(b\right)}}$ \\ \vspace{0.2em}\\
${54,2}$
& ${\frac{4}{45}E_{\text{lo}}\mu^{\left(b\right)}}$
& ${-2D^{\left(b\right)}}$
& ${-4D^{\left(b\right)}}$
& ${4D^{\left(b\right)}}$
& ${2D^{\left(b\right)}}$ \\ \vspace{0.2em}\\
${55,1}$
& ${\frac{1}{9}E_{\text{lo}}\mu^{\left(a\right)}}$
& ${-2D^{\left(a\right)}}$
& ${2D^{\left(a\right)}}$
& ${-2D^{\left(b\right)}}$
& ${2D^{\left(b\right)}}$ \\ \vspace{0.2em}\\
${56,1}$
& ${\frac{1}{9}E_{\text{lo}}\mu^{\left(b\right)}}$
& ${-2D^{\left(b\right)}}$
& ${-2D^{\left(a\right)}}$
& ${2D^{\left(a\right)}}$
& ${2D^{\left(b\right)}}$ \\ \vspace{0.2em}\\
${57,1}$
& ${\frac{1}{9}E_{\text{lo}}\mu^{\left(a\right)}}$
& ${-2D^{\left(a\right)}}$
& ${-2D^{\left(b\right)}}$
& ${2D^{\left(a\right)}}$
& ${2D^{\left(b\right)}}$ \\ \vspace{0.2em}\\
${58,1}$
& ${\frac{1}{9}E_{\text{lo}}\mu^{\left(a\right)}}$
& ${-2D^{\left(a\right)}}$
& ${-2D^{\left(b\right)}}$
& ${2D^{\left(a\right)}}$
& ${2D^{\left(b\right)}}$ \\ \vspace{0.2em}\\
${59,1}$
& ${\frac{1}{9}E_{\text{lo}}\mu^{\left(b\right)}}$
& ${-2D^{\left(b\right)}}$
& ${-2D^{\left(a\right)}}$
& ${2D^{\left(a\right)}}$
& ${2D^{\left(b\right)}}$ \\ \vspace{0.2em}\\
${60,1}$
& ${\frac{1}{9}E_{\text{lo}}\mu^{\left(a\right)}}$
& ${-2D^{\left(a\right)}}$
& ${2D^{\left(a\right)}}$
& ${-2D^{\left(a\right)}}$
& ${2D^{\left(a\right)}}$ \\ \vspace{0.2em}\\
${60,2}$
& ${\frac{4}{45}E_{\text{lo}}\mu^{\left(a\right)}}$
& ${-2D^{\left(a\right)}}$
& ${-4D^{\left(a\right)}}$
& ${4D^{\left(a\right)}}$
& ${2D^{\left(a\right)}}$ \\ \vspace{0.2em}\\
${61,1}$
& ${\frac{1}{9}E_{\text{lo}}\mu^{\left(b\right)}}$
& ${-2D^{\left(b\right)}}$
& ${2D^{\left(b\right)}}$
& ${-2D^{\left(a\right)}}$
& ${2D^{\left(a\right)}}$ \\ \vspace{0.2em}\\
${62,1}$
& ${\frac{1}{9}E_{\text{lo}}\mu^{\left(a\right)}}$
& ${-2D^{\left(a\right)}}$
& ${2D^{\left(a\right)}}$
& ${-2D^{\left(a\right)}}$
& ${2D^{\left(a\right)}}$ \\ \vspace{0.2em}\\
${62,2}$
& ${\frac{4}{45}E_{\text{lo}}\mu^{\left(a\right)}}$
& ${-2D^{\left(a\right)}}$
& ${-4D^{\left(a\right)}}$
& ${4D^{\left(a\right)}}$
& ${2D^{\left(a\right)}}$ \\ \vspace{0.2em}\\
${63,1}$
& ${\frac{1}{9}E_{\text{lo}}\mu^{\left(b\right)}}$
& ${-2D^{\left(b\right)}}$
& ${-2D^{\left(a\right)}}$
& ${2D^{\left(b\right)}}$
& ${2D^{\left(a\right)}}$ \\ \vspace{0.2em}\\
${64,1}$
& ${\frac{1}{9}E_{\text{lo}}\mu^{\left(a\right)}}$
& ${-2D^{\left(a\right)}}$
& ${-2D^{\left(b\right)}}$
& ${2D^{\left(b\right)}}$
& ${2D^{\left(a\right)}}$ \\ \vspace{0.2em}\\
${65,1}$
& ${\frac{1}{9}E_{\text{lo}}\mu^{\left(a\right)}}$
& ${-2D^{\left(a\right)}}$
& ${-2D^{\left(b\right)}}$
& ${2D^{\left(b\right)}}$
& ${2D^{\left(a\right)}}$ \\ \vspace{0.2em}\\
${66,1}$
& ${\frac{1}{9}E_{\text{lo}}\mu^{\left(b\right)}}$
& ${-2D^{\left(b\right)}}$
& ${-2D^{\left(a\right)}}$
& ${2D^{\left(b\right)}}$
& ${2D^{\left(a\right)}}$ \\
\end{longtable*}
\end{center}

\end{document}